\documentclass[12pt,letterpaper,a4paper]{article}
\usepackage[includeheadfoot,
marginratio={1:1,2:3},
width=525pt,
height=750pt,]{geometry}

\usepackage{amsmath}
\usepackage{amsfonts}
\usepackage{amssymb}
\usepackage{graphicx}
\usepackage{cancel}
\usepackage{empheq}
\usepackage{color}
\usepackage{hyperref}

\usepackage{graphicx}
\graphicspath{{images/}}

\numberwithin{equation}{section}
\usepackage{float}
\restylefloat{table}
\restylefloat{figure}
\usepackage[utf8]{inputenc}
\usepackage{amsmath, amsthm, amssymb, amsfonts}
\usepackage[font=small, labelfont=bf]{caption}
\usepackage{amsmath, amsthm, amssymb, amsfonts}
\usepackage{multirow}
\usepackage{graphicx}
\usepackage{float}
\usepackage[numbers,sort&compress]{natbib}
\usepackage{enumerate}
\usepackage{amsmath}
\usepackage{amsfonts}
\usepackage{amssymb}
\usepackage{graphicx}
\usepackage{cancel}
\usepackage{empheq}
\usepackage{color}
\usepackage{colortbl}
\definecolor{green2}{cmyk}{0, 1, 0.5, 0.3}
\definecolor{green3}{cmyk}{1, 0.75, 1.0, 0.0}

\definecolor{lightgreen}{cmyk}{0.2, 0, 0.2, 0.2}
\definecolor{lightgray}{cmyk}{0.1,0.2,0,0.1}
\definecolor{lightgray2}{cmyk}{0.4,0.4,0,0.8}
\definecolor{black}{cmyk}{1.0,1.0,1.0,1.0}

\usepackage{hyperref}

\restylefloat{table}
\restylefloat{figure}
\usepackage{longtable}
\usepackage{lipsum} % just for dummy text- not needed for a longtable

\usepackage{amsfonts}
\usepackage{amssymb}
\usepackage{comment}
\usepackage{graphicx}
\usepackage{amsmath}
\usepackage{tabu}
\usepackage{dsfont}

\usepackage{amsthm}
\usepackage{slashed}
\usepackage{setspace}
\usepackage[labelformat=simple]{subcaption}

\usepackage{pgfplots}
\usepackage{tikz}
\usepackage{tikz-cd}
\usetikzlibrary{shapes.misc,shadows,fit,decorations.pathmorphing,positioning,trees,decorations.markings,decorations.pathreplacing,calc,shapes,patterns,arrows,positioning,arrows.meta}
\usepackage{xcolor}
\usepackage{pgfplots}
\usepackage{pgfplotstable}
\usepackage{xcolor}
\usepackage{makecell}
\usepackage{diagbox}
\usepackage{environ}
\usepackage[utf8]{inputenc}
\usepackage{amsmath, amsthm, amssymb, amsfonts}
\usepackage{multirow}
\usepackage{graphicx}
\usepackage{float}
\usepackage[numbers,sort&compress]{natbib}
\usepackage{enumerate}
\usepackage{enumitem}
\usepackage[capitalize,sort]{cleveref}
\usepackage{hhline}
\usepackage{mathtools}
\usepackage{longtable}
\usepackage{makecell}
\usepackage{tabularx}
\usepackage{cellspace}
\usepackage{alphalph}
\usepackage{lscape}
\usepackage{slashed}
\usepackage{setspace}
\usepackage{pifont}

\usepackage{float}

\crefname{figure}{Figure}{Figures}
\crefname{table}{Table}{Tables}

\def\be{\begin{equation}}
\def\ee{\end{equation}}
\def\bea{\begin{eqnarray}}
\def\eea{\end{eqnarray}}
\def\bes{\begin{subequations}}
	\def\ees{\end{subequations}}

\def\ov{\overline}

\def\ov{\overline}
\def\1{{\bf 1}}
\def\2{{\bf 2}}
\def\3{{\bf 3}}
\def\4{{\bf 4}}
\def\6{{\bf 6}}

\newcommand{\nn}{\nonumber}

\newcommand{\beq}{\begin{equation}}
\newcommand{\eeq}{\end{equation}}

\def\ov{\overline}

\allowdisplaybreaks[2]
\numberwithin{equation}{section}

\def\be{\begin{equation}}
\def\ee{\end{equation}}
\def\bea{\begin{eqnarray}}
\def\eea{\end{eqnarray}}
\def\bes{\begin{subequations}}
	\def\ees{\end{subequations}}

\usepackage{multicol}
\usepackage{float}
\usepackage{caption}
\usepackage{setspace}
\allowdisplaybreaks[2]
\numberwithin{equation}{section}

\begin{document}
	{\hfill
		\hfill
		arXiv: 2506.22630}

	\vspace{1.0cm}
	\begin{center}
		{\Large
			%Multi-fibre
            Assisted Fibre Inflation in Perturbative LVS
			}
		\vspace{0.4cm}
	\end{center}

\vspace{0.35cm}

 	\begin{center}
 		George K. Leontaris$^\dagger$ and
 		Pramod Shukla$^\Diamond$ \footnote{Emails:~leonta@uoi.gr, pshukla@jcbose.ac.in}
	\end{center}
 \vspace{0.1cm}

 \begin{center}
  {$^\dagger$ Physics Department, University of Ioannina, University Campus, \\
 Ioannina 45110, Greece}\\
  \vspace{0.3cm}
 {$^\Diamond$ Department of Physical Sciences, Bose Institute,\\
 Unified Academic Campus, EN 80, Sector V, Bidhannagar, Kolkata 700091, India}\\
  \vspace{0.3cm}

 \end{center}

\vspace{1cm}

\abstract{
We propose a multi-field fibre inflation scenario in type IIB perturbative large volume compactifications, showing that the multi-field dynamics suppresses trans-Planckian displacements of the canonical  inflaton. Considering a concrete K3-fibred Calabi-Yau (CY) threefold with $h^{1,1}({\rm CY})=3$ and having certain underlying symmetries, we show that the presence of multi-fibre moduli creates an assisted inflation scenario where multiple moduli collectively help in producing the cosmological observables consistent with the current experimental bounds. We further argue that individual field range excursions $(\Delta\phi_n)$ corresponding to each of the inflaton fields can be estimated as $\Delta\phi_n = \Delta\phi/\sqrt{n}$, where $\Delta\phi$ denotes the effective single-field inflaton range needed to generate the desired cosmological observables, and $n$ is the number of moduli assisting the multi-fibre inflation. We also present various numerical benchmark models consistently producing cosmological observables in light of the recent ACT experiments.}

\clearpage

\tableofcontents

%%%%%%%%%%%%%%%%%%%%%%%%%%%%%%%%%%%%%%%%%%%%%%%%%%%%%%%%%%%%%%
%%%%%%%%%%%%%%%%%%%%%%%%%%%%%%%%%%%%%%%%%%%%%%%%%%%%%%%%%%%%%%%%%%%%%%%%%%%%%%%%%%%%%%%%%%

\section{Introduction}
\label{sec_intro}

In string theory, inflation is frequently driven by scalar moduli fields which arise naturally in the low-energy Effective Field Theories (EFTs) derived from string compactifications.   Inflation is a well-established mechanism in early universe cosmology, as it resolves several long-standing theoretical problems \cite{Starobinsky:1980te,Guth:1980zm,Linde:1981mu}. For instance, inflation provides a compelling solution to the flatness and horizon problems in the standard $\Lambda$CDM  model of cosmology. According to the inflationary scenario,  the universe underwent an exponential expansion, stretching the initial curvature to near zero and thereby explaining its observed spatial flatness. Additionally, because the entire observable universe originated from a tiny, causally connected region before inflation, this rapid expansion accounts for the large-scale uniformity of the cosmic microwave background (CMB) in a natural way and without requiring fine-tuning. Quantum fluctuations during inflation can generate primordial perturbations, which seeded the formation of large-scale structures and the temperature anisotropies observed in the cosmic microwave background. 

Due to all these successful predictions, in the context of plain field theories, a plethora of inflationary models have been proposed over the past four decades. The simplest of those are single-field inflationary models which only depend on a few adjustable parameters. However, with increasingly precise bounds on so-called spectral index ($n_s$)  and the tensor-to-scalar ratio ($r$) from improved cosmological  observations, many minimal field-theoretic inflation models may be now excluded. Yet, even viable  candidates in the context of plain field theory often fail to satisfy string-theoretic consistency requirements, particularly the swampland conjectures \cite{Vafa:2005ui,Ooguri:2006in} and the subsequent implications such as the (refined) de Sitter conjectures \cite{Obied:2018sgi,Garg:2018reu}, issues related to trans-Planckian field ranges \cite{Ooguri:2018wrx,Blumenhagen:2018nts,Grimm:2018ohb,Scalisi:2018eaz,Bedroya:2019tba} etc. (for review see \cite{Palti:2019pca,vanBeest:2021lhn} and reference therein). These constraints which are rooted in ultra-violet (UV) completion, severely limit the space of allowed inflationary potentials in string-derived EFTs. Furthermore, a crucial requirement for building viable low-energy effective string models -particularly those that incorporate inflation with a modulus serving as the inflaton- is the stabilization of all string moduli.
Without stabilization, the moduli remain massless and dynamically unfixed, resulting in undetermined couplings, masses, and other physical parameters in the low-energy effective theory. This would render the theory phenomenologically and cosmologically inconsistent and unsuitable for making meaningful predictions. Consequently, a major task in string cosmology is to identify a moduli potential which, preferably exhibits a de Sitter vacuum and, upon minimization, consistently fixes the moduli at their vacuum expectation values. Furthermore, if the potential has the appropriate shape, one or more moduli can play the role of the inflaton field, thereby realizing the inflationary scenario. Some huge amount of efforts have been devoted to this pursuit over the past two decades, which have led to the development of various inflationary scenarios, e.g. see \cite{Cicoli:2023opf,McAllister:2023vgy} and references therein.

Moduli stabilization has been thoroughly explored in the context of type IIB superstring compactifications resulting in two popular schemes, namely KKLT \cite{Kachru:2003aw} and the Large Volume Scenarios (LVS) \cite{ Balasubramanian:2005zx} offering a robust solution.  In LVS, complex structure moduli and axio-dilaton are first stabilized by fluxes \cite{Dasgupta:1999ss,Gukov:1999ya, Taylor:1999ii,Blumenhagen:2003vr}, while K\"ahler moduli are subsequently fixed via perturbative ($\alpha'$) \cite{Becker:2002nn} and non-perturbative effects \cite{Witten:1996bn,Green:1997di,Blumenhagen:2009qh, Blumenhagen:2008zz, Bianchi:2011qh,Blumenhagen:2012kz}, yielding a viable low-energy vacuum. The essential ingredients for moduli stabilization in these two schemes, namely KKLT and the standard LVS, are the various non-perturbative superpotential contributions \cite{Witten:1996bn,Green:1997di}. However such corrections need not be generically present in a given concrete type IIB CY orientifold setup which depends on the detailed specifics of the underlying CY geometry and the brane-setups/fluxes. For example, some of these requirements are:  satisfying the Witten's unit arithmetic genus condition \cite{Witten:1996bn} in the models based on CY having a rigid divisor, and ~``rigidification" of non-rigid divisors using magnetic fluxes \cite{Bianchi:2011qh,Bianchi:2012pn,Louis:2012nb}, as well as the conflict of chirality and visible sector while using the rigid del-Pezzo divisors \cite{Blumenhagen:2007sm,Blumenhagen:2008zz,Blumenhagen:2009qh,Cvetic:2012ts, Blumenhagen:2012kz}. 

As an alternative to non-perturbative corrections, attempts for moduli stabilization using only the perturbative effects have attracted some significant amount of interests in recent years. The various perturbative $\alpha^\prime$ corrections \cite{Becker:2002nn} including the higher derivative F$^4$-corrections \cite{Ciupke:2015msa}, and a variety of string-loop corrections such as KK/Winding types \cite{Berg:2004ek,vonGersdorff:2005bf, Berg:2005ja, Berg:2005yu, Cicoli:2007xp, Gao:2022uop} and the so-called log-loop type \cite{Antoniadis:2018hqy,Antoniadis:2019rkh,Antoniadis:2020ryh} have been used in this regard. In addition, the non-geometric fluxes arising from the T/S dual completion of the GVW flux superpotential has also received attention in the the context of perturbative moduli stabilization \cite{Aldazabal:2006up,deCarlos:2009fq,deCarlos:2009qm, Blumenhagen:2015kja,Shukla:2016xdy,Plauschinn:2020ram,Damian:2023ote,Shukla:2022srx}. In the perturbative LVS proposal of \cite{Antoniadis:2018hqy,Antoniadis:2019rkh,Antoniadis:2020ryh,Leontaris:2022rzj,Leontaris:2025xit}, it was shown that using logarithmic string loop corrections  (``log-loop" for short) to the K\"ahler potential along with the BBHL correction, one can realize an AdS minimum with exponentially large VEV for the overall volume ${\cal V}$ of the compactifying sixfold background. In turns out that one can have $\langle {\cal V} \rangle \propto e^{c_1/g_s^2}$ where $g_s$ is the string coupling and $c_1 \simeq {\cal O}(1)$ positive constant \cite{Leontaris:2022rzj}. The perturbative LVS framework is quite similar to the standard LVS but does not involve any non-perturbative effects.

In the framework of standard LVS, a variety of  inflationary scenarios and their concrete global embedding have been studied in the last two decades. These inflationary proposals are known as volume modulus inflation \cite{Conlon:2008cj}, Blowup inflaton \cite{Conlon:2005ki,Blanco-Pillado:2009dmu,Cicoli:2017shd}, Fibre Inflation \cite{Cicoli:2008gp,Cicoli:2016chb,Cicoli:2016xae,Cicoli:2017axo, Cicoli:2024bxw}, poly-instanton inflation \cite{Cicoli:2011ct,Blumenhagen:2012ue,Gao:2013hn,Gao:2014fva}, Loop blowup inflation \cite{Bansal:2024uzr}. In this regard, some of these inflationary proposals such as inflation point inflation and fibre inflation have been also realized and studied in the perturbative LVS \cite{Bera:2024ihl,Bera:2024zsk, Hai:2025wvs, Chakraborty:2025yms}. In the meantime, some inflationary models with multi-field analysis have been presented in the context of type IIB superstring compactifications. However, most of these models are effectively single-field models in the sense that only one field is significantly involved during the e-folds are gained \cite{Cicoli:2017axo,Cicoli:2017shd}. Other fields are either just sitting in their respective minima or if axionic, they result in some turnings in the trajectories without helping in generating significant e-folds \cite{Bond:2006nc,Gao:2013hn,Gao:2014fva}. However, in the current work we plan to present a multi-field inflation model in LVS for which multiple fields can ``assist"  in generating the sufficient e-folds, similar to the multi-field proposals of \cite{Liddle:1998jc, Dimopoulos:2005ac}.

Among the various possible string inspired scenarios,  Fibre Inflation \cite{Cicoli:2008gp,Cicoli:2016chb,Cicoli:2016xae,Cicoli:2017axo, Cicoli:2024bxw} offers several attractive features. First of all, it naturally arises in type IIB string compactifications in the context of LVS, providing a robust UV completion.	 The standard Fibre Inflationary scenario predicts a value of $n_s\sim 0.967$ which is in accordance with Planck 2018 data, while $r\lesssim 0.007$ which is compatible with Planck constraints $r<0.06$ \cite{Planck:2018jri,Planck:2018vyg}. Finally, Fibre Inflation admits a global embedding in specific CY manifolds \cite{Cicoli:2016xae, Cicoli:2017axo, Cicoli:2024bxw} and, within the  perturbative LVS, retains the advantages of standard LVS while avoiding non-perturbative effects. This circumvents  the need for an exceptional divisor, relaxing stringent K\"ahler cone constraints on the inflaton field range \cite{Bera:2024ihl}.
			
However, while all these predictions are in agreement with the above mentioned experimental data,  the latest combined results from Planck, ACT, and DESI measure the spectral index to be  $n_s = 0.9743 \pm  0.0034$ \cite{Planck:2018jri,ACT:2025tim,ACT:2025fju,DESI:2024mwx}. Taking these results at face value, may challenge many canonical inflationary models \cite{Frolovsky:2025iao}, and %marginally disfavor single field Fibre Inflation models discussed above. Hence, 
this discrepancy may signal the need for new physics beyond single-field slow-roll inflation. Interestingly, as already discussed above, such a  setup arises naturally in string  theory compactifications, where multiple moduli fields emerge. Hence, while single-field Fibre Inflation offers simplicity, multi-field inflation presents a phenomenologically richer framework. By involving multiple scalar fields in the inflationary dynamics, it ensures greater flexibility and a more robust agreement with current cosmological data.

In this  paper we present a multi-field Fibre Inflation model within the perturbative Large Volume Scenario (LVS) framework of type IIB string theory. We propose an ``assisted inflation'' mechanism where multiple K3-fibre moduli collectively generate sufficient efolds, reducing the need for large field excursions pushing towards the boundaries of the K\"ahler cone  which are typical in single-field models \cite{Cicoli:2017axo,Bera:2024ihl,Cicoli:2024bxw}. Our investigations focus on Calabi-Yau (CY) orientifolds with $h^{1,1} = 3$, demonstrating how symmetries and sub-leading corrections (e.g., string-loop and higher-derivative effects) stabilize the moduli and drive inflation. The main results of the paper are summarized as follows. Each inflaton field $\phi_n$ has to travel for a lower range  $\Delta\phi_n = \Delta\phi/\sqrt{n}$ as compared to what an inflaton in the single-field case would need to travel, alleviating constraints from K\"ahler cone conditions. Moduli stabilization via perturbative effects, i.e., BBHL corrections, and log-loop terms, avoids non-perturbative superpotential terms of the form $A e^{-a T}$ and thus, exceptional divisors are not necessary. Furthermore, we demonstrate that multi-field assisted Fibre Inflation proposal has a global embedding and this can be illustrated with an explicit CY example with toroidal-like volume. 

The layout of the paper is as follows: in section \ref{sec_setup} we review the basics of the LSV scenario and present the various (non-)perturbative contributions to the K\"ahler potential and the superpotential, including the possible quantum corrections to the K\"ahler potential. Furthermore, we outline the key concepts of multi-field inflation and establish the necessary framework for the present work. In Section \ref{sec_fibre-inflation}, we present a two-field analysis for investigating the robustness of the single field fibre inflation model in the perturbative LVS framework, showing that the leading-order effects stabilize the volume modulus which remains seated at its minimum during the inflation while the subleading corrections drive inflation. The analysis is implemented in the context of a threefold example which possesses a toroidal-like volume. A numerical analysis is performed and some benchmark models are presented. 
%ALTERNATIVE: In section 3 we start our investigations with a two-field Fibre inflation model embedded in the framework of perturbative LVS where we show that the overall volume is stabilized via perturbative LVS by the leading order effects while the subleading corrections drive inflation.
We extend our analysis in section \ref{sec_assisted-fibre} where we discuss a three-field model for assisted Fibre Inflation embedded in a specific CY threefold. We discuss how a de Sitter vacuum is achieved using either D-term or T-brane uplifting. 
Finally, we present numerical model with predictions consistent with cosmological observables, identifying parameter regions that produce a spectral index  $n_s = 0.9743\pm0.0034$ matching ACT observations \cite{ACT:2025fju}. In section \ref{sec_conclusions} we present our conclusions, whilst some details of our calculations are collected in the Appendix \ref{sec_appendix}. 

%%%%%%%%%%%%%%%%%%%%%%%%%%%%%%%%%%%%%%%%%%%%%%%%%%%%%%%%%%%%%%%%%%%%%%%%%%%%%%%%%%%%%%%%%%
%%%%%%%%%%%%%%%%%%%%%%%%%%%%%%%%%%%%%%%%%%%%%%%%%%%%%%%%%%%%%%%%%%%%%%%%%%%%%%%%%%%%%%%%%%

\section{Preliminaries}
\label{sec_setup}

The low energy dynamics of the four-dimensional effective supergravity theory arising from the type IIB superstring compactifications on CY orientifolds can be captured by a holomorphic superpotential ($W$) and a real K\"ahler potential ($K$) and the gauge kinetic function $(g)$. These quantities depend on the various chiral coordinates obtained by complexifying  various moduli with a set of RR axions. Let us start by fixing the conventions. We will be using the following definitions of such chiral variables:
\bea
\label{eq:chiral-variables}
& & U^i = v^i - i\, u^i, \qquad S = c_0 + i\, s, \qquad T_\alpha = c_\alpha - i\, \tau_\alpha~,
\eea
where $s$ is the dilaton dependent modulus, $u^i$'s are the complex structure saxions, and $\tau_\alpha$'s are the Einstein frame four-cycle volume moduli defined as $\tau_\alpha = \partial_{t^\alpha} {\cal V} = \frac12 k_{\alpha\beta\gamma} t^\beta t^\gamma$. In addition, the $c_0$ and $c_\alpha$'s are universal RR axion and RR four-form axions, respectively, while the complex structure axions are denoted by $v^i$. Here the indices $\{i, \alpha\}$ are such that $i \in h^{2,1}_-({\rm CY}/{\cal O})$ while $\alpha \in h^{1,1}_+({\rm CY}/{\cal O})$. Moreover, we assume that $h^{1,1} = h^{1,1}_+$ for simplicity, and hence, the so-called odd-moduli $G^a$  are not present in our analysis; we refer interested readers to \cite{Cicoli:2021tzt}.

The F-term contributions to the scalar potential are computed using the following well known formula,
\be
\label{eq:V_gen}
e^{- {K}} \, V = {K}^{{\cal A} \ov {\cal B}} \, (D_{\cal A} W) \, (D_{\ov {\cal B}} \ov{W}) -3 |W|^2 \equiv V_{\rm cs} + V_{\rm k}\,,
\ee
where:
\be
\label{eq:VcsVk}
V_{\rm cs} =  K_{\rm cs}^{i \ov {j}} \, (D_i W) \, (D_{\ov {j}} \ov{W}) \qquad \text{and}\qquad V_{\rm k} =  K^{{A} \ov {B}} \, (D_{A} W) \, (D_{\ov {B}} \ov{W}) -3 |W|^2\,.
\ee
Moduli stabilisation in 4D type IIB effective supergravity models follows a two-step strategy. First, one fixes the complex structure moduli $U^i$ and the axio-dilaton $S$ by the leading order flux superpotential $W_{\rm flux}$ induced by usual S-dual pair of the 3-form fluxes $(F_3, H_3)$ \cite{Gukov:1999ya}. This demands solving the following supersymmetric flatness conditions:
\bea
&& D_i W_{\rm flux} = 0 = D_{\ov {i}} \ov{W}_{\rm flux}, \qquad D_{S} W_{\rm flux} = 0 = D_{\ov {S}} \ov{W}_{\rm flux}.
\label{UStab}
\eea
After supersymmetric stabilization of axio-dilaton and the complex structure moduli, one has $\langle W_{\rm flux} \rangle = W_0$. At this leading order no-scale structure protects the K\"ahler moduli $T_\alpha$ which subsequently remain flat, and as a second step, they can be stabilized via including other sub-leading contributions to the scalar potential, e.g. those induced via the non-perturbative corrections in the holomorphic superpotential $W$ or the other (non-)perturbative corrections arising from the whole series of $\alpha^\prime$ and string-loop ($g_s$) corrections. %\textbf{DC: I will also add another point which starts by "second."}

%%%%%%%%%%%%%%%%%%%%%%%%%%%%%%%%%%%%%%%%%%%%%%%%%%%%%%%%%%%%%%%%%%%%%%%%%%%%%%%%%%%%%%%%%%

\subsection{Two schemes for Large Volume Scenarios}
\subsubsection{Standard LVS}
The LVS scheme of moduli stabilization considers a combination of perturbative $(\alpha^\prime)^3$ corrections \cite{Becker:2002nn} to the K\"ahler potential ($K$) and a non-perturbative contribution to the superpotential $W$ which can be generated by using rigid divisors, such as shrinkable del-Pezzo 4-cycles, or by rigidifying non-rigid divisors using magnetic fluxes \cite{Bianchi:2011qh, Bianchi:2012pn, Louis:2012nb}. The minimal LVS construction includes two K\"ahler moduli corresponding to a so-called Swiss-cheese like volume form of the CY threefold given as\footnote{However, we also note that LVS moduli fixing may be realized by using CY threefolds without having a Swiss-cheese structure; for example see  \cite{AbdusSalam:2020ywo}.}:
\be
{\cal V} = \frac{k_{bbb}}{6} \, (t^b)^3 + \frac{k_{sss}}{6} \, (t^s)^3 \, ,
\ee
where $k_{\alpha\beta\gamma}$ denotes the triple intersection number on the CY threefold, and the 2-cycle volume moduli $t^{\alpha}$ are related to the 4-cycle volume moduli $\tau_\alpha$ via $\tau_\alpha = \partial_{t^\alpha} {\cal V}$. Subsequently one has the following Swiss-cheese like volume form:
\be
{\cal V} =  \gamma_b \, \, \tau_{b}^{3/2} - \gamma_s\, \, \tau_{s}^{3/2} \, ,
\ee
where $\gamma_b$ and $\gamma_s$ are determined through the triple intersection numbers $k_{bbb}$ and $k_{sss}$. Given that the CY threefold has a Swiss-cheese form, one can always find a basis of divisors such that the only non-vanishing intersection numbers are $k_{bbb}$ and $k_{sss}$, which leads to the relation $t^s = - \sqrt{2\tau_s/k_{sss}}$. Here, the minus sign is dictated from the K\"ahler-cone conditions because the so-called `small' divisor $D_s$ in this Swiss-cheese CY is an exceptional 4-cycle. The K\"ahler potential including $\alpha'^3$ corrections takes the form \cite{Becker:2002nn}:
\be
\label{eq:K}
K = -\ln\left[-i\int \Omega\wedge\bar{\Omega}\right]-\ln\left[-\,i\,(S-\bar{S})\right]-2\ln{\cal Y}, \quad {\cal Y} = {\cal V}\,+\frac{\xi}{2}\left(\frac{S-\bar{S}}{2i}\right)^{3/2} = {\cal Y}_0,
\ee
where $\Omega$ denotes the nowhere vanishing holomorphic 3-form which depends on the complex-structure moduli, while the CY volume ${\cal V}$ receives a shift through the $\alpha'^3$ corrections encoded in the parameter $\xi=-\frac{\chi(X)\,\zeta(3)}{2\,(2\pi)^3}$, where $\chi(X)$ is the CY Euler characteristic and $\zeta(3)\simeq 1.202$.

Furthermore, the presence of a `diagonal' del-Pezzo divisor corresponding to the so-called `small' $4$-cycle of the CY threefold induces the superpotential with a non-perturbative effect of the following form:
\be
W= W_0 + A_s\, e^{- i\, a_s\, T_s}\,,
\label{eq:Wnp-n}
\ee
where after fixing $S$ and the $U$-moduli, the flux superpotential can effectively be considered as constant: $W_0=\langle W_{\rm flux}\rangle$. In addition, the pre-factor $A_s$ can generically depend on the complex-structure moduli which after the first-step of the supersymmetric moduli stabilisation can be considered as a parameter. Moreover, without any loss of generality, we consider $W_0$ and $A_s$ to be a real quantity. Subsequently the leading order pieces in the large volume expansion are collected in three types of terms \cite{Balasubramanian:2005zx}:
\be
V \simeq \frac{\beta_{\alpha'}}{{\cal V}^3} + \beta_{\rm np1}\,\frac{\tau_s}{{\cal V}^2}\, e^{- a_s \tau_s} \cos\left(a_s \,c_s\right) \\
+ \beta_{\rm np2}\,\frac{ \sqrt{\tau_s}}{{\cal V}}\, e^{-2 a_s \tau_s},
\label{VlvsSimpl}
\ee
with:
\be
\beta_{\alpha'} = \frac{3 \,\kappa\,\hat\xi |W_0|^2}{4}\,, \quad \beta_{\rm np1} = 4 \, \kappa\, a_s |W_0| |A_s|\,, \quad \beta_{\rm np2} = 4 \, \kappa\, a_s^2 |A_s|^2 \sqrt{2 k_{sss}}, \quad \kappa = \frac{g_s}{2}\, e^{K_{cs}}\,.
\ee
The minimal LVS scheme of moduli stabilization fixes the CY volume ${\cal V}$ along with a small modulus $\tau_s$ controlling the volume of an exceptional del Pezzo divisor. Therefore any LVS models with 3 or more K\"ahler moduli, $h^{1,1}\geq 3$, can generically have flat directions at leading order. These flat directions are promising inflaton candidates with a potential generated at sub-leading order. Based on the geometric nature of the inflaton field and the source of inflaton potential, there are various popular LVS inflationary models such as Blowup inflaton \cite{Conlon:2005ki,Blanco-Pillado:2009dmu,Cicoli:2017shd}, Fibre Inflation \cite{Cicoli:2008gp,Cicoli:2016chb,Cicoli:2016xae,Cicoli:2017axo, Cicoli:2024bxw}, poly-instanton inflation \cite{Cicoli:2011ct,Blumenhagen:2012ue,Gao:2013hn,Gao:2014fva}, and Loop blowup inflation \cite{Bansal:2024uzr}.

%%%%%%%%%%%%%%%%%%%%%%%%%%%%%%%%%%%%%%%%%%%%%%%%%%%%%%%%%%%%%%%%%%%%%%%%%%%%%%%%%%%%%%%%%%

\subsubsection{Perturbative LVS}

With the inclusion of one-loop effects --also known as log-loop corrections-- to the K\"ahler potential on top of the BBHL corrections used in the standard LVS, one arrives at an effectively modified overall volume ${\cal V}$ which we denote as ${\cal Y}$. It takes the following explicit form,
\bea
& & {\cal Y} = {\cal Y}_0 + {\cal Y} _1, %= {\cal V} +  \frac{\xi}{2} \, e^{-\frac{3}{2} \phi} + e^{\frac{1}{2} \phi}\, \left(\sigma + \eta \, \ln{\cal V}\right).
\eea
where ${\cal Y}_0$ denotes the overall volume modified by $\alpha^\prime$ corrections appearing at string tree-level while ${\cal Y}_1$ is induced at string one-loop level as given  below  \cite{Antoniadis:2018hqy,Antoniadis:2018ngr,Antoniadis:2019doc,Antoniadis:2019rkh,Antoniadis:2020ryh,Antoniadis:2020stf},
\bea
& & {\cal Y}_0 = {\cal V} +  \frac{\xi}{2} \, e^{-\frac{3}{2} \phi} = {\cal V} + \frac{\xi}{2}\, \left(\frac{S-\ov{S}}{2\,{\rm i}}\right)^{3/2} \,, \\
& & {\cal Y}_1 = e^{\frac{1}{2} \phi}\, f({\cal V}) = \left(\frac{S-\ov{S}}{2\,{\rm i}}\right)^{-1/2} \left(\sigma + \eta \, \ln{\cal V}\right)\,.\nonumber
\label{eq:defY}
\eea
Here one has the following correlations among the various coefficients, $\xi$, $\sigma$ and $\eta$,
\bea
\label{eq:def-xi-eta}
& & \hskip-1cm  \xi = - \frac{\chi({\rm CY})\, \zeta[3]}{2(2\pi)^3}~, \quad \sigma  = - \frac{\chi({\rm CY})\, \zeta[2]}{2(2\pi)^3} \sigma_0, \quad \eta =  \frac{\chi({\rm CY})\, \zeta[2]}{2(2\pi)^3} \eta_0, \quad \frac{\xi}{\eta} = -\frac{\zeta[3]}{\zeta[2]} \\
& & \hskip-1cm \hat\xi = \frac{\xi}{g_s^{3/2}}~, \qquad \hat\sigma = g_s^{1/2}\, \sigma~,\qquad \hat\eta = g_s^{1/2}\, \eta~, \qquad \frac{\hat\xi}{\hat\eta} = -\frac{\zeta[3]}{\zeta[2]\,g_s^2\, \eta_0}, \quad \frac{\hat\sigma}{\hat\eta} = -\frac{\sigma_0}{\eta_0} ~. \nonumber
\eea
Note that, the three parameters $\xi$, $\sigma$ and $\eta$ do not have any $g_s$ dependence which can be fixed by the $SL(2,\mathbb Z)$ arguments \cite{Leontaris:2022rzj}, however $\sigma$ and $\eta$ can generically depend on the complex structure moduli. In order to keep track of this possibility we introduce $\sigma_0$ and $\eta_0$ in Eq.~(\ref{eq:def-xi-eta}) as complex-structure moduli dependence parameters while still keeping the $SL(2,\mathbb Z)$ motivated factors of the Riemann $\zeta$-functions and the Euler characteristic of the CY. Subsequently we have the following pieces at leading order,
\bea
\label{eq:pheno-potV2}
& & V_{\alpha^\prime +{\rm log} \, g_s}^{(1)} \simeq \frac{3\, \kappa\, \hat\xi}{4\, {\cal V}^3}\, |W_0|^2 + \frac{3 \, \kappa\, (\hat\eta\ln{\cal V} - 4\hat\eta + \hat\sigma)}{2{\cal V}^3}\,|W_0|^2 \equiv V_{\rm pLVS},
\eea
where $\kappa = e^{K_{cs}}\,g_s/2$. This scalar potential results in an exponentially large VEV for the overall volume determined by the following approximate relation:
\bea
\label{eq:pert-LVS}
& & \langle {\cal V} \rangle \simeq e^{\frac{13}{3}-\frac{\hat\xi}{2\, \hat\eta} -\frac{\hat\sigma}{ \hat\eta}} = e^{a/g_s^2 + b}, \qquad a = \frac{\zeta[3]}{2 \zeta[2]\eta_0} \simeq 0.365381/\eta_0, \quad b = \frac{13}{3}+\frac{\sigma_0}{\eta_0}~\cdot
\eea
For natural values $\sigma_0 = -2$ and $\eta_0 = 1$, the numerical estimate for $g_s = 0.2$ gives $\langle {\cal V} \rangle = 95594.5$ while $g_s = 0.1$ leads to $\langle {\cal V} \rangle = 7.615 \cdot 10^{16}$. Given that an exponentially large VEV of the overall volume $\cal V$ is obtained by using only the perturbative effects, this scheme is refereed as ``perturbative LVS". Moreover, similar to the standard LVS case, it corresponds to an AdS minimum.

%%%%%%%%%%%%%%%%%%%%%%%%%%%%%%%%%%%%%%%%%%%%%%%%%%%%%%%%%%%%%%%%%%%%%%%%%%%%%%%%%%%%%%%%%%
\subsection{Sub-leading corrections to scalar potential}
%%%%%%%%%%%%%%%%%%%%%%%%%%%%%%%%%%%%%%%%%%%%%%%%%%%%%%%%%%%%%%%%%%%%%%%%%%%%%%%%%%%%%%%%%%

In the absence of any non-perturbative effects, there can be several types of perturbative effects which can induce useful scalar potential contributions, namely the BBHL's $(\alpha^\prime)^3$ corrections \cite{Becker:2002nn}, the perturbative string-loop effects of \cite{Antoniadis:2018hqy} as well as the higher derivative F$^4$ corrections of \cite{Ciupke:2015msa}. Using the Gukov-Vafa-Witten's (GVW) flux superpotential $W_0$ for stabilizing the complex-structure moduli and the axio-dilaton at their respective supersymmetric minimum, the perturbative scalar potential contributions can be collected in the following form
\bea
\label{eq:masterVgen}
& & \hskip-0.75cm V_{\rm tot} = V_{\rm up} + V_{\rm D} + V_{\alpha^\prime + {\rm log} \, g_s} + V_{g_s}^{\rm KK} + V_{g_s}^{\rm W} +V_{{\rm F}^4} + \dots 
\eea
We now examine each term in greater detail.
\begin{itemize}

\item 
For our purpose, we consider the three popular classes of uplifting schemes by simply characterising them via the following contributions to the scalar potential,
\bea
\label{eq:Vuplift}
& & V_{\rm up} = \frac{{\cal C}_{\rm up}}{{\cal V}^p},
\eea
where $p = 4/3$ for anti-D3 uplifting \cite{Kachru:2003aw,Crino:2020qwk,Cicoli:2017axo,AbdusSalam:2022krp}, and $p = 2$ for D-term uplifting \cite{Burgess:2003ic,Achucarro:2006zf,Braun:2015pza} while  $p = 8/3$ for the T-brane uplifting \cite{Cicoli:2015ylx,Cicoli:2017shd}.

\item 
The most significant contributions which depend on the K\"ahler moduli can arise from the D-term, which appear at  ${\cal O}({\cal V}^{-2})$ in the terms of the simple volume scaling. However, the overall coefficient is flux dependent and can be tuned to have some hierarchy between the contributions which are responsible for the complex-structure moduli stabilization.

To have some D-term contributions, we need to consistently turn-on some worldvolume gauge fluxes on the three stacks of D7-branes. Such fluxes take the form:
\be
{\cal F}_i = \sum_{j=1}^{h^{1,1}} f_{ij}\hat{D}_j + \frac12 \hat{D}_i - \iota_{D_i}^*B \quad\text{with}\quad f_{ij}\in \mathbb{Z} \quad\text{and}\quad i=1,2,3\,,
\ee
where the half-integer contribution is due to Freed-Witten anomaly cancellation \cite{Minasian:1997mm,Freed:1999vc}. After appropriately turning on non-trivial gauge fluxes, one generates the following FI-parameters,
\be
\xi_\alpha = \frac{1}{4\pi {\cal V}}\, \int_{D_\alpha} {\cal F} \wedge J = - \frac{i}{2\pi} \sum_\beta q_{\alpha\beta} \,\partial_{T_\beta}K, \quad {\rm where} \quad q_{\alpha\beta} = \int_{\rm CY} D_\alpha \wedge D_\beta \wedge {\cal F}.
\ee
Subsequently, depending on the non-trivial triple intersection numbers, the D-term induced contributions to the scalar potential can be given as,
\bea
\label{eq:VDup1}
& & \hskip-2cm V_{\rm D} \propto \sum_{\alpha =1}^n \left[\frac{1}{\tau_\alpha} \left(\sum_{\beta \neq \alpha}\, q_{\alpha\beta}\, \frac{\partial K}{\partial \tau_\beta}\right)^2 \right] \simeq \sum_{\alpha =1}^3 \frac{\hat{d}_\alpha}{f_\alpha^{(3)}},
\eea
where $f_\alpha^{(3)}$ denotes some homogeneous cubic polynomial in generic four-cycle volume $\tau_\beta$.

\item
The leading order contributions from the piece $V_{\alpha^\prime + {\rm log} \, g_s}$ appears at ${\cal O}({\cal V}^{-3})$ in the large volume expansion, which can be expressed as follows
\bea
V_{\rm pLVS} \equiv V_{\alpha^\prime +{\rm log} \, g_s} \simeq \frac{3\, \kappa\, \hat\xi}{4\, {\cal V}^3}\, |W_0|^2 + \frac{3 \, \kappa\, (\hat\eta\ln{\cal V} - 4\hat\eta + \hat\sigma)}{2{\cal V}^3}\,|W_0|^2
\eea

\item 
The so-called KK-type and Winding-type string loop effects are encoded as  corrections to the K\"ahler potential, which in the Einstein frame they are written as follows \cite{Berg:2004ek, Berg:2005ja, Berg:2005yu, Berg:2007wt, Cicoli:2007xp},
\bea
\label{eq:KgsE}
& & \hskip-1cm K_{g_s}^{\rm KK} = g_s \sum_\alpha \frac{{\cal C}_\alpha^{\rm KK} \, t^\alpha_\perp}{\cal V} \,, \qquad K_{g_s}^{\rm W} =  \sum_\alpha \frac{{\cal C}_{w_\alpha}}{{\cal V}\, t^\alpha_\cap}\,.
\eea
where ${\cal C}_\alpha^{\rm KK}$ and ${\cal C}_{w_\alpha}$ are some complex structure moduli dependent functions which can generically also depend on the open-string moduli. The two-cycle volume moduli $t^\alpha_{\perp}$ denote the transverse distance among the various stacks of the non-intersecting $D7$-brane and $O7$-planes, whilst  $t^\alpha_{\cap}$ denotes the volume of the curve sitting at the intersection loci of the various non-trivially intersecting stacks of $D7$-branes such that the intersecting curve is non-contractible. These effects induce correction to the scalar potential which take the following form,
\bea
& & V_{g_s}^{\rm KK} = \kappa \, g_s^2 \, \frac{|W_0|^2}{4\,{\cal V}^4} \sum_{\alpha,\beta} {\cal C}_\alpha^{\rm KK} {\cal C}_\beta^{\rm KK} \left(2\,t^\alpha t^\beta - 4\, {\cal V} \,k^{\alpha\beta}\right), \\
& & V_{g_s}^{\rm W} = -2 \kappa \frac{|W_0|^2}{{\cal V}^3} \, \sum_\alpha \frac{{\cal C}_{w_\alpha}}{t^\alpha_\cap}, \nonumber
\eea
Such effects can generically appear at ${\cal O}({\cal V}^{-10/3})$ in the large volume expansion. In fact there can be additional loop corrections motivated by the field theoretic computations \cite{vonGersdorff:2005bf,Gao:2022uop}, however we do not include those corrections in the current analysis.

\item
Finally, the higher derivative F$^4$-corrections cannot be captured by the conventional two-derivative corrections through the K\"ahler potential and the superpotential. These corrections take the following simple form \cite{Ciupke:2015msa},
\bea
& & V_{{\rm F}^4} = -  \frac{\lambda\,\kappa^2\,|W_0|^4}{g_s^{3/2} {\cal V}^4} \Pi_\alpha \, t^\alpha\,.
\eea
We note that such corrections naively appear at ${\cal O}({\cal V}^{-11/3})$ in the large volume expansion.

\end{itemize}

%%%%%%%%%%%%%%%%%%%%%%%%%%%%%%%%%%%%%%%%%%%%%%%%%%%%%%%%%%%%%%%%%%%%%%%%%%%%%%%%%%%%%%%%%%
\subsection{Ingredients for multi-field inflation}
%%%%%%%%%%%%%%%%%%%%%%%%%%%%%%%%%%%%%%%%%%%%%%%%%%%%%%%%%%%%%
To analyze the multi-field evolution of inflationary dynamics, we adopt the e-folding number $N$ 
as the time coordinate, defined by  $dN = H dt$. In this framework, the Einstein-Friedmann equations governing the scalar fields $\Phi^a$ take the following form:

%%%%%%%%%%%%%%%%%%%%%%%%%%%%%
\bea
\label{EOM}
\frac{d^2 \Phi^a}{dN^2}+{\Gamma^a}_{bc} \frac{d\Phi^b}{dN} \frac{d\Phi^c}{dN}+\left(3+ \frac{1}{H} \frac{dH}{dN}\right) \frac{d\Phi^a}{dN}+ \frac{{\cal G}^{ab} \partial_b V}{H^2}=0,
\eea
where the Hubble function is given as
\bea
\label{constran}
H^2= \frac{1}{3}\left(V(\Phi^a)+ \frac{1}{2}H^2 \, {\cal G}_{ab} \frac{d\Phi^a}{dN} \frac{d\Phi^b}{dN} \right).
\eea
Here, the generic (non-flat) field space metric is represented as ${\cal G}_{ab}$ while ${\Gamma^a}_{bc}$'s denote the corresponding Christoffel connections, and the derivatives of the scalar potential are defined as: $\partial_a V = \frac{\partial V}{\partial \Phi^a}$. Furthermore, the field space metric ${\cal G}_{ab}$ for the real moduli is obtained from the K\"ahler moduli space metric $K_{T_\alpha\ov{T}_\beta}$ using the following relation
\bea
\label{eq:fieldspace-metric}
& & K_{T_\alpha\ov{T}_\beta} (\partial_\mu T_\alpha) (\partial^\mu \ov{T}_\beta) = \frac{1}{2} {\cal G}_{ab} (\partial_\mu \Phi^a) \,(\partial^\mu\Phi^b),
\eea
where $\{\Phi^a\}$ forms the new basis of the real fields.
%However, given that the scalar potential does not have any axionic moduli dependence in the current work, all the $C_4$ axion moduli remain flat in perturbative LVS framework, and therefore we restrict the set $\{\Phi^a\}$ to the volume moduli only.
However, since the scalar potential in the present construction has no dependence on the axionic moduli, all $C_4$ axions  remain flat within the perturbative LVS framework. As a result, we restrict the set $ \{\Phi^a\}$ to only the volume moduli.
In addition, given that the overall volume ${\cal V}$ is stabilized at the leading order in perturbative LVS and conversion from two-cycle volume $t^\alpha$ to the four-cycle volume $\tau_\alpha$ does not necessarily lead to a simple enough form for generic CY threefolds, we will work in the basis of real fields $\Phi^a = \{{\cal V}, t^2, .., t^n\}$.
%This basis consists of the overall volume ${\cal V}$ and therefore will be useful in respecting the mass hierarchy between the ${\cal V}$ modulus and the remaining (orthogonal) $t^\alpha$ moduli.
%Now, in the context of studying inflationary aspects, one has to look at the sufficient conditions for realizing the slow-roll inflation which are encoded in a set of slow-roll parameters. We define two such inflationary parameters, namely $\epsilon$ and $\eta$, is the following manner,
This basis includes the overall volume ${\cal V}$, making it well-suited for preserving the mass hierarchy between the modulus ${\cal V}$ and the remaining orthogonal moduli  $t^\alpha$.

To study inflationary dynamics, we must examine the sufficient conditions for slow-roll inflation, which are determined by a set of slow-roll parameters. We define two such parameters, $\epsilon$ and $\eta$,  as follows:
\bea
\label{eq:epH-etaH}
& & \epsilon \equiv - \frac{\dot{H}}{H^2} = - \frac{1}{H} \frac{dH}{dN},\qquad \eta \equiv \frac{\dot{\epsilon}}{\epsilon H} = \frac{1}{\epsilon}\frac{d\epsilon}{dN}.
\eea
Moreover, we find the following useful relation, 
\bea
\label{third}
& & \epsilon = 3 - \frac{V}{H^2} = \frac{1}{2} \, {\cal G}_{ab} \frac{d\Phi^a}{dN} \frac{d\Phi^b}{dN} > 0.
\eea
%Note that, one needs $\epsilon < 1$ as a necessary condition for the inflationary evolution to occur.
%We emphasize that $\epsilon < 1$ is a necessary condition for inflationary dynamics. 
%However, the two inflationary parameters must remain small, $\epsilon \ll 1$ and $\eta \ll 1$,
% to satisfy the slow-roll conditions and  generate sufficient number of $e$-folds during the so-called slow-roll regime.  These generic form of the slow-roll parameters can be correlated with another form which involves the derivatives of the potential given as,
%\bea
%\label{eq:epV-etaV}
%& & \epsilon \simeq \epsilon_V, \qquad \eta \simeq 4 \epsilon_V - 2 \eta_V~.
%\eea
%These generic slow-roll parameters $(\epsilon, \eta)$ can be related to another set of slow-roll parameters $(\epsilon_V, \eta_V)$ defined in the following manner while $\eta_V$ corresponding to the negative most Eigenvalue of matrix $N^a_b$ given as below,
%\bea
%& & \epsilon_V = \frac{{\cal G}^{ab}\, (\partial_a V) (\partial_b V)}{2\,V^2}, \qquad  N^a_b = \frac{{\cal G}^{ac} \left(\partial_c\partial_b V - \Gamma^d_{cb}\partial_d V\right)}{V}, 
%\eea
%These slow-roll parameters are used to determine the cosmological observables  characterizing the inflationary model. 

%--- replaced with:

We emphasize that $\epsilon \ll 1$ is a necessary condition for inflationary dynamics, alongside $\eta \ll 1$, to ensure the slow-roll regime produces a sufficient number of $e$-folds. These generic slow-roll parameters $(\epsilon, \eta)$ can be expressed in terms of potential-dependent parameters $(\epsilon_V, \eta_V)$ as
\bea
\label{eq:epV-etaV}
& & \epsilon \simeq \epsilon_V, \qquad \eta \simeq 4 \epsilon_V - 2 \eta_V~.
\eea
where $\epsilon_V$ and $\eta_V$ are defined via the field-space metric $\mathcal{G}^{ab}$ and the Hessian structure of the scalar potential $V$:
\bea
\epsilon_V &=& \frac{\mathcal{G}^{ab} (\partial_a V) (\partial_b V)}{2 V^2},  \\
N^a_b &=& \frac{\mathcal{G}^{ac} \left(\partial_c \partial_b V - \Gamma^d_{cb} \partial_d V\right)}{V}.\nonumber
\eea
Here, $\eta_V$ corresponds to the most negative eigenvalue of $N^a_b$. These parameters are critical for determining the cosmological observables that characterize the inflationary model.

%---up to here 

Given that one usually starts with a scalar potential which is a function of the moduli fields, it turns out that the following form of the field equations (\ref{EOM}) is useful,
\bea
\label{eq:EOM2}
& & \frac{d^2\Phi^a}{dN^2}+{\Gamma^a}_{bc} \frac{d\Phi^b}{dN} \frac{d\Phi^c}{dN}+\left(3- \frac{1}{2} \, {\cal G}_{ab} \frac{d\Phi^a}{dN} \frac{d\Phi^b}{dN} \right) \left(\frac{d\Phi^a}{dN}+ \frac{{\cal G}^{ab} \partial_b V}{V} \right)=0~,
\eea
where the scalar potential $V$, the metric ${\cal G}_{ab}$ as well as its inverse ${\cal G}^{ab}$ are explicitly known functions of the fields $\Phi^a$.

%Now, the main idea is to solve the second order differential equation in Eq.~(\ref{EOM}) to get 
Now, our primary objective is to solve the second-order differential equation (\ref{EOM}) to determine the evolutionary trajectories of various fields under different initial conditions.  For instance, we can impose the initial conditions 
\bea
& & \Phi^a(0)=\Phi^a_0 \qquad  {\rm and} \qquad \frac{d\Phi^a}{dN}|_{N=0}=0~,
\eea
and numerically evolve the system until the end of inflation.
%After numerically solving the second order differential equations (\ref{eq:EOM2}) one finds the respective trajectories of the fields in terms of the number of $e$-folds $\Phi^a \equiv \Phi^a(N)$ which can be used to determine the evolution of the scalar potential and the slow-roll parameters. Subsequently, the various cosmological observables such as the scalar power spectrum $P_s$, the spectral index $n_s$, the running of spectral index $\alpha_s$, and the tensor-to-scalar ratio $r$, can be determined in terms of the number of e-foldings. 

By numerically solving the second-order differential equations (\ref{eq:EOM2}), we obtain the field trajectories $\Phi^a(N)$ as functions of the number of $e$-folds. These solutions allow us to track the evolution of the scalar potential and the slow-roll parameters. From these, we can derive the cosmological observables, in particular  the scalar power spectrum $P_s$, the spectral index $n_s$, its running $\alpha_s$, and the tensor-to-scalar ratio $r$, all expressed in terms of the $e$-folding evolution.
These observables are defined as
\bea
\label{eq:cosmo-observables}
& & P_s(N) = \frac{V(N)}{24\pi^2\, \epsilon(N)}, \quad n_s(N) = 1 + \frac{1}{P_s(N)} \frac{d}{dN} \,P_s(N), \\
& & \alpha_s(N) = \frac{d}{dN} n_s(N), \qquad r(N) = 16 \epsilon(N). \nonumber
\eea
All the cosmological observables are evaluated at the horizon exit $\Phi^a = \Phi^{a\ast}$ with suitable initial conditions such that one typically has $N(\Phi^{a\ast}) \gtrsim 60$ as an estimate for the sufficient $e$-folds needed to ensure a successful inflationary scenario. 
%However, the number of e-foldings $(N)$ depends on many things including the post-inflationary aspects and can be given as a sum several contributions
However, the total number of 
$e$-folds $(N)$ depends on multiple factors, including post-inflationary dynamics, and can be expressed as a sum of several distinct contributions\cite{Liddle:2003as,Cicoli:2017axo}:
\bea
\label{eq:cosmo-observables1}
& & \hskip-0.5cm N = \int H \, dt \simeq 57 + \frac{1}{4} \ln(r_\ast V_\ast) - \frac{1}{3}\ln\left(\frac{10V_{\rm end}}{m_{\inf}^{3/2}}\right),
\eea
where $V_{\rm end}$ corresponds to the  value of the potential at the end of inflation determined by $\epsilon = 1$ and $m_{\rm inf}$ is the inflaton mass. From the single-field analysis, one usually has $N \simeq 50$ for Fibre inflation \cite{Cicoli:2017axo,Bhattacharya:2020gnk,Cicoli:2020bao}. Following the constraints from the Planck 2018 data, one typically needs the scalar perturbation amplitude to satisfy $P_s \simeq 2.1 \times 10^{-9}$ while the primordial scalar tilt is $n_s=0.9651\pm 0.0044$ and its running turns out to be $\alpha_s = - 0.0041 \pm 0.0067$ \cite{Planck:2018jri,Planck:2018vyg}. In addition, the Atacama Cosmology Telescope (ACT) data gives $n_s= 0.9666 \pm  0.0077$, in agreement with Planck while a new constraint based on a combination of the Planck and ACT data (P-ACT) gives $ n_s = 0.9709 \pm  0.0038$. Furthermore, a combination of Planck, ACT, and DESI (P-ACT-LB) gives $n_s = 0.9743 \pm  0.0034$ and $\alpha_s = 0.0062 \pm 0.0052$ \cite{ACT:2025tim,ACT:2025fju,DESI:2024mwx, Frolovsky:2025iao}.

%%%%%%%%%%%%%%%%%%%%%%%%%%%%%%%%%%%%%%%%%%%%%%%%%%%%%%%%%%%%%%%%%%%%%%%%%%%%%%%%%%%%%%%%%%
%%%%%%%%%%%%%%%%%%%%%%%%%%%%%%%%%%%%%%%%%%%%%%%%%%%%%%%%%%%%%%%%%%%%%%%%%%%%%%%%%%%%%%%%%%

\section{Fibre Inflation in Perturbative LVS}
\label{sec_fibre-inflation}

In this section we present a two-field Fibre inflation model embedded in the framework of perturbative LVS. The main idea is to consider type IIB superstring compactification on a concrete Calabi-Yau threefold with $h^{1,1}({\rm CY})=3$ with multiple K3-fibrations. The overall volume is stabilized via perturbative LVS by the leading order effects while the subleading corrections drive inflation.

\subsection{An explicit CY threefold with $h^{1,1}({\rm CY})=3$}
We start by presenting an explicit CY threefold example which possesses a toroidal-like volume, being of the form: 
\bea 
{\cal V} \propto \sqrt{\tau_1\tau_2\tau_3}~,
\label{Tvol}
\eea 
where $\tau_\alpha$'s are four-cycle volumes corresponding the divisors of the CY threefold. The main motivation for this choice follows from the proposal of \cite{Antoniadis:2018hqy,Antoniadis:2018ngr,Antoniadis:2019doc,Antoniadis:2019rkh,Antoniadis:2020ryh,Antoniadis:2020stf, Antoniadis:2021lhi} where certain  symmetries between the various volume moduli were required for the implementation of the overall mechanism. For this purpose, we have scanned the CY dataset of Kreuzer-Skarke \cite{Kreuzer:2000xy} with $h^{1,1} = 3$ and we have identified at least two geometric configurations ~\cite{Gao:2013pra,Leontaris:2022rzj} capable of producing the volume  form~(\ref{Tvol}).  One such CY threefold corresponding to the polytope Id: 249 in the CY database of \cite{Altman:2014bfa} can be defined by the following toric data:
\begin{center}
\begin{tabular}{|c|ccccccc|}
\hline
\cellcolor[gray]{0.9}Hyp &\cellcolor[gray]{0.9} $x_1$  &\cellcolor[gray]{0.9} $x_2$  &\cellcolor[gray]{0.9} $x_3$  &\cellcolor[gray]{0.9} $x_4$  &\cellcolor[gray]{0.9} $x_5$ & \cellcolor[gray]{0.9}$x_6$  &\cellcolor[gray]{0.9} $x_7$       \\
\hline
\cellcolor[gray]{0.9}4 & 0  & 0 & 1 & 1 & 0 & 0  & 2   \\
\cellcolor[gray]{0.9}4 & 0  & 1 & 0 & 0 & 1 & 0  & 2   \\
\cellcolor[gray]{0.9}4 & 1  & 0 & 0 & 0 & 0 & 1  & 2   \\
\hline
& K3  & K3 & K3 &  K3 & K3 & K3  &  SD  \\
\hline
\end{tabular}
\end{center}
The Hodge numbers are $(h^{2,1}, h^{1,1}) = (115, 3)$, the Euler number is $\chi=-224$ and the Stanley-Reisner ideal is:
\be
{\rm SR} =  \{x_1 x_6, \, x_2 x_5, \, x_3 x_4 x_7 \} \,. \nn
\ee
This CY threefold was also considered for exploring odd-moduli and exchange of non-trivially identical divisors in \cite{Gao:2013pra}. Moreover, a del-Pezzo upgraded version of this example which corresponds to a CY threefold with $h^{1,1}=4$, has been considered  for chiral global embedding of Fibre inflation model in \cite{Cicoli:2017axo}.

The analysis of the divisor topologies using {\it cohomCalg} \cite{Blumenhagen:2010pv, Blumenhagen:2011xn} shows that they can be represented by the following Hodge diamonds:
\bea
K3 &\equiv& \begin{tabular}{ccccc}
    & & 1 & & \\
   & 0 & & 0 & \\
  1 & & 20 & & 1 \\
   & 0 & & 0 & \\
    & & 1 & & \\
  \end{tabular}, \qquad \quad {\rm SD} \equiv \begin{tabular}{ccccc}
    & & 1 & & \\
   & 0 & & 0 & \\
  27 & & 184 & & 27 \\
   & 0 & & 0 & \\
    & & 1 & & \\
  \end{tabular}.
\eea
Considering the basis of smooth divisors $\{D_1, D_2, D_3\}$ and their respective dual basis $\{\hat{D}_1, \hat{D}_2, \hat{D}_3\}$, we get the following intersection polynomial which has just one non-zero classical triple intersection number \footnote{There is another CY threefold in the database of \cite{Altman:2014bfa} which has the intersection polynomial of the form $I_3 = D_1 D_2 D_3$, however that CY threefold (corresponding to the polytope Id: 52) has non-trivial fundamental group.}:
\bea
& & I_3 = 2\, \hat{D}_1\, \hat{D}_2\, \hat{D}_3,
\eea
while the second Chern-class of the CY is given by,
\bea
c_2({\rm CY}) = 5 \hat{D}_3^2+12 \hat{D}_1 \hat{D}_2 + 12 \hat{D}_2 \hat{D}_3+12 \hat{D}_1 \hat{D}_3.
\eea
Subsequently, the second Chern numbers, defined as $\Pi_\alpha = \int_{\rm CY} c_2({\rm CY}) \wedge \hat{D}_\alpha$ corresponding to the divisors $D_\alpha$, are given as,
\bea
& & \Pi_\alpha = 24 \quad \forall \, \alpha \in \{1, 2,..,6\}; \qquad \Pi_7 = 124.
\eea
Moreover, considering the K\"ahler form $J = \sum_{\alpha =1}^3 t^\alpha \hat{D}_\alpha$, the overall volume and the 4-cycle volume moduli can be given as follows:
\bea
& & \hskip-1cm {\cal V} = 2\, t^1\, t^2\, t^3, \qquad \qquad \tau_1 = 2\, t^2 t^3,  \quad  \tau_2 = 2\, t^1 t^3, \quad  \tau_3 = 2 \,t^1 t^2 \,.
\label{Taus}
\eea
Using the above relations between the four- and two-cycle moduli,
the volume can also be expressed in the following form:
\bea
& & \hskip-1cm {\cal V} = 2 \, t^1\, t^2\, t^3 = t^1 \tau_1 = t^2 \tau_2 = t^3 \tau_3 = \frac{1}{\sqrt{2}}\,\sqrt{\tau_1 \, \tau_2\, \tau_3}~.
\eea
This confirms that the volume form ${\cal V}$ is indeed like a toroidal case with an exchange symmetry $1 \leftrightarrow 2 \leftrightarrow 3$ under which all the three K3 divisors which are part of the basis are exchanged. We note that the volume form can also be expressed as,
\bea
\label{eq:t-tau-vol}
{\cal V} = t^1 \, \tau_1 = t^2 \, \tau_2  = t^3 \, \tau_3~,
\eea
which means that the transverse distance for the stacks of $D7$-branes wrapping the divisor $D_1$ is given by $t^1$ and similarly $t^2$ is the transverse distance for $D7$-branes wrapping the divisor $D_2$ and so on. 
The behavior of transverse distances and the intersection properties of K3 divisors on a ${\mathbb T}^2$ closely mirror those observed in the toroidal case. These symmetry structures are compatible with the minimal requirements for producing logarithmic string-loop corrections as elaborated in \cite{Antoniadis:2018hqy,Antoniadis:2018ngr,Antoniadis:2019doc,Antoniadis:2019rkh,Antoniadis:2020ryh,Antoniadis:2020stf,Antoniadis:2021lhi}.
Moreover, using the general K\"ahler metric expression in Eq. (\ref{eq:simpinvKij-1}) and the classical triple intersection numbers, the tree-level metric takes the following form,
\bea
\label{eq:Kij-tree}
& & K_{\alpha\beta}^0 = \frac{1}{4\, {\cal V}^2} \left(
\begin{array}{ccc}
 (t^1)^2 & 0 & 0 \\
 0 & (t^2)^2 & 0 \\
 0 & 0 & (t^3)^2 \\
\end{array}
\right) = \frac{1}{4} \left(
\begin{array}{ccc}
 (\tau_1)^{-2} & 0 & 0 \\
 0 & (\tau_2)^{-2} & 0 \\
 0 & 0 & (\tau_3)^{-2} \\
\end{array}
\right),
\eea
where we have used (\ref{eq:t-tau-vol}) in the second step. Furthermore, the K\"ahler cone for this setup is described by the following conditions,
\bea
\label{KahCone}
\text{K\"ahler cone:} &&  t^1 > 0\,, \quad t^2 > 0\,, \quad t^3 > 0\,.
\eea
The intersection curves between two coordinate divisors are presented in Table \ref{Tab1} where ${\cal H}_g$ denotes a curve with Hodge numbers $h^{0,0} = 1$ and $h^{1,0} = g$, and hence ${\cal H}_1 \equiv {\mathbb T}^2$, while ${\cal H}_0 \equiv {\mathbb P}^1$.
\begin{table}[h]
  \centering
 \begin{tabular}{|c|c|c|c|c|c|c|c|}
\hline
\cellcolor[gray]{0.9}  &\cellcolor[gray]{0.9} $D_1$  &\cellcolor[gray]{0.9} $D_2$  &\cellcolor[gray]{0.9} $D_3$  & \cellcolor[gray]{0.9}$D_4$  & \cellcolor[gray]{0.9}$D_5$ &\cellcolor[gray]{0.9} $D_6$  & \cellcolor[gray]{0.9}$D_7$  \\
    \hline
		\hline
\cellcolor[gray]{0.9}$D_1$ & $\emptyset$  &  ${\mathbb T}^2$      &  ${\mathbb T}^2$        &  ${\mathbb T}^2$   &  ${\mathbb T}^2$  &  $\emptyset$   &  ${\cal H}_9$  \\
\cellcolor[gray]{0.9}$D_2$ &  ${\mathbb T}^2$ & $\emptyset$        &  ${\mathbb T}^2$        &  ${\mathbb T}^2$   &    $\emptyset$   & ${\mathbb T}^2$  & ${\cal H}_9$
\\
\cellcolor[gray]{0.9}$D_3$  &  ${\mathbb T}^2$      &  ${\mathbb T}^2$        & $\emptyset$ & $\emptyset$ &  ${\mathbb T}^2$   &  ${\mathbb T}^2$   &  ${\cal H}_9$
\\
\cellcolor[gray]{0.9}$D_4$  &  ${\mathbb T}^2$      &  ${\mathbb T}^2$        & $\emptyset$ & $\emptyset$ &  ${\mathbb T}^2$   &  ${\mathbb T}^2$   &  ${\cal H}_9$
\\
\cellcolor[gray]{0.9}$D_5$ &  ${\mathbb T}^2$ & $\emptyset$        &  ${\mathbb T}^2$        &  ${\mathbb T}^2$   &    $\emptyset$   & ${\mathbb T}^2$  & ${\cal H}_9$
\\
\cellcolor[gray]{0.9}$D_6$ & $\emptyset$  &  ${\mathbb T}^2$      &  ${\mathbb T}^2$        &  ${\mathbb T}^2$   &  ${\mathbb T}^2$  &  $\emptyset$   &  ${\cal H}_9$  \\
\cellcolor[gray]{0.9}$D_7$ & ${\cal H}_9$  &  ${\cal H}_9$      &  ${\cal H}_9$        &  ${\cal H}_9$   &  ${\cal H}_9$  &  ${\cal H}_9$   &  ${\cal H}_{97}$
\\
    \hline
  \end{tabular}
  \caption{Intersection curves of the two coordinate divisors.}
\label{Tab1}
\end{table}
Subsequently, using the K\"ahler form $J = t^1 D_1 + t^2 D_2 +t^3 D_3$ the corresponding sizes of the curves in Table \ref{Tab1} can be expressed in terms of two-cycle volumes $t^\alpha$ as presented in Table \ref{Tab2}. This also shows, for example, that the curve intersecting at divisors $D_1$ and $D_2$ has a volume along $t^3$, like in the usual toroidal scenarios. Also, this table is symmetrical and lower left entries can be read-off from the right upper sector.
\begin{table}[h]
  \centering
 \begin{tabular}{|c|c|c|c|c|c|c|c|}
\hline
 \cellcolor[gray]{0.9} &\cellcolor[gray]{0.9} $D_1$  &\cellcolor[gray]{0.9} $D_2$  & \cellcolor[gray]{0.9}$D_3$  &\cellcolor[gray]{0.9} $D_4$  &\cellcolor[gray]{0.9} $D_5$ & \cellcolor[gray]{0.9}$D_6$  &\cellcolor[gray]{0.9} $D_7$  \\
    \hline
		\hline
\cellcolor[gray]{0.9}$D_1$  & 0 & 2 ${t^3}$ & 2 ${t^2}$ & 2 ${t^2}$ & 2 ${t^3}$ & 0 & 4 ${t^2}$+4 ${t^3}$ \\
\cellcolor[gray]{0.9}$D_2$  &  & 0 & 2 ${t^1}$ & 2 ${t^1}$ & 0 & 2 ${t^3}$ & 4 ${t^1}$+4 ${t^3}$ \\
\cellcolor[gray]{0.9}$D_3$  & & & 0 & 0 & 2 ${t^1}$ & 2 ${t^2}$ & 4 ${t^1}$+4 ${t^2}$ \\
\cellcolor[gray]{0.9}$D_4$  & & &  & 0 & 2 ${t^1}$ & 2 ${t^2}$ & 4 ${t^1}$+4 ${t^2}$ \\
\cellcolor[gray]{0.9}$D_5$  & && & & 0 & 2 ${t^3}$ & 4 ${t^1}$+4 ${t^3}$ \\
\cellcolor[gray]{0.9}$D_6$  & &&& & & 0 & 4 ${t^2}$+4 ${t^3}$ \\
\cellcolor[gray]{0.9}$D_7$  & &&&&& & 16 $(t^1 + t^2 + t^3)$ \\
    \hline
  \end{tabular}
  \caption{Size of the curves at the intersection locus of the two  divisors presented in Table \ref{Tab1}. }
\label{Tab2}
\end{table}
\noindent
Also, the divisor intersection curves given in table \ref{Tab1} show that the possible $D7$-brane stacks and the $O7$-planes would intersect on a ${\mathbb T}^2$ or on a curve ${\cal H}_9$ defined by $h^{0,0} = 1$ and $h^{1,0} = 9$. 
The behavior of transverse distances and the intersection properties of K3 divisors on a ${\mathbb T}^2$ closely mirror those observed in the toroidal case, though  in the current situation the divisors are K3 -unlike the ${\mathbb T}^4$ divisors of the six-torus. These symmetries are consistent with the basic requirement for generating logarithmic string-loop effects as elaborated in \cite{Antoniadis:2018hqy,Antoniadis:2018ngr,Antoniadis:2019doc,Antoniadis:2019rkh,Antoniadis:2020ryh,Antoniadis:2020stf,Antoniadis:2021lhi}.

%%%%%%%%%%%%%%%%%%%%%%%%%%%%%%%%%%%%%%%%%%%%%%%%%%%%%%%%%%%%%%%%%%%%%%%%%%%%%%%%%%%%%%%%%%

\subsubsection*{Orientifold involution, fluxes and brane setting}
For a given holomorphic involution, one needs to introduce D3/D7-branes and fluxes in order to cancel all the charges. For instance, D7-tadpoles can be canceled by introducing appropriate $N_a$ D7-brane stacks wrapped around suitable divisors (say $D_a$) and their orientifold images ($D_a^\prime$) such that the following relation holds \cite{Blumenhagen:2008zz}:
\bea
\label{eq:D7tadpole}
& & \sum_a\, N_a \left([D_a] + [D_a^\prime] \right) = 8\, [{\rm O7}]\,.
\eea
Moreover, the presence of D7-branes and O7-planes also contributes to the D3-tadpoles, which, in addition, receive contributions from  $H_3$ and $F_3$ fluxes, D3-branes and O3-planes. The D3-tadpole cancellation condition is given as \cite{Blumenhagen:2008zz}:
\be
N_{\rm D3} + \frac{N_{\rm flux}}{2} + N_{\rm gauge} = \frac{N_{\rm O3}}{4} + \frac{\chi({\rm O7})}{12} + \sum_a\, \frac{N_a \left(\chi(D_a) + \chi(D_a^\prime) \right) }{48}\,,
\label{eq:D3tadpole}
\ee
where $N_{\rm flux} = (2\pi)^{-4} \, (\alpha^\prime)^{-2}\int_X H_3 \wedge F_3$ is the contribution from background fluxes and $N_{\rm gauge} = -\sum_a (8 \pi)^{-2} \int_{D_a}\, {\rm tr}\, {\cal F}_a^2$ is due to D7 worldvolume fluxes. However, for the simple case where D7-tadpoles are cancelled by placing 4 D7-branes (plus their images) on top of an O7-plane, (\ref{eq:D3tadpole}) reduces to the following form:
\be
N_{\rm D3} + \frac{N_{\rm flux}}{2} + N_{\rm gauge} =\frac{N_{\rm O3}}{4} + \frac{\chi({\rm O7})}{4}\,.
\label{eq:D3tadpole1}
\ee
For our CY threefold, we note that there are six equivalent reflection involutions corresponding to flipping first six coordinates, i.e. $x_i \to - x_i$ for each $i \in \{1, 2, .., 6\}$. Each of these involutions produces two O7-planes; however, there is limited scope for considering D7-brane stacks wrapping all basis divisors.
%Each of these involutions result in two O7-planes however there is not much scope of considering D7-stacks wrapping all the divisors of the basis. 
Furthermore, the D3-tadpole cancellation conditions impose stringent constraints: the right-hand side of equation~(\ref{eq:D3tadpole1}) evaluates to 12, severely limiting the possible choices for the $F_3$ and $H_3$ flux configurations.
%In addition, the D3 tadpole conditions are quite strict in the sense that the RHS of equation~(\ref{eq:D3tadpole1}) results in 12, leaving very little scope for choosing the $F_3/H_3$ fluxes. 
However considering the involution $x_7 \to - x_7$ leads to  better possibilities for brane setting. This results in only one fixed point set with $\{O7 = D_7\}$ along with no $O3$-planes. Subsequently, one can consider a brane configuration of  three stacks of $D7$-branes wrapping each of the three divisors $\{D_1, D_2, D_3\}$ in the basis,
\bea
& & 8\, [O_7] = 8 \left([D_1] + [D_1^\prime] \right) + 8 \left([D_2] + [D_2^\prime] \right) + 8 \left([D_3] + [D_3^\prime] \right)\,,
\eea
along with the $D3$ tadpole cancellation condition being given as
\bea
\label{eq:ND3-tadpole}
& & N_{\rm D3} + \frac{N_{\rm flux}}{2} + N_{\rm gauge} = 0 + \frac{240}{12} + 8 + 8 + 8 = 44\,,
\eea
which unlike the other six reflection involutions results in some flexibility with turning on background flux as well as the gauge flux. In fact, if the D7-tadpole cancellation condition is satisfied by placing four D7-branes on top of the O7-plane, the string loop corrections to the scalar potential may turn out to be very simple. We shall therefore focus on a slightly more complicate D7-brane setup which gives rise also to winding loop effects. This can be achieved by placing D7-branes not entirely on top of the O7-plane.

We note that the VEV of the flux superpotential $W_0$ gets intertwined with the D3 tadpole charge $Q_{\rm D3}$ after complex-structure moduli stabilization, e.g. see \cite{Denef:2004ze,Cicoli:2024bxw}, 
\bea
\label{eq:W_0-Nflux-tadpole}
& & 2\pi g_s \, |W_0|^2 < Q_{\rm D_3} = 88,
\eea
where $Q_{\rm D_3}$ is twice the right-hand-side (rhs) of (\ref{eq:ND3-tadpole}), for example see \cite{Blumenhagen:2008zz}. In fact, the relation (\ref{eq:ND3-tadpole}) is motivated while looking at the connection the type IIB orientifold ingredients with those of the F-theory compactifications, which demands that rhs of \ref{eq:ND3-tadpole} can take only integral values. 

%%%%%%%%%%%%%%%%%%%%%%%%%%%%%%%%%%%%%%%%%%%%%%%%%%%%%%%%%%%%%%%%%%%%%%%%%%%%%%%%%%%%%%%%%%

\subsection{Various contributions to the scalar potential}

%In contrast to generic type IIB Calabi-Yau orientifold scenarios, our current construction exhibits an absence of several expected corrections. 
Unlike generic type IIB Calabi-Yau orientifold scenarios, our current construction features a vanishing of several expected corrections.
In particular,
\begin{itemize}
\item
Given that there are no rigid divisors present, a priori this setup will not receive non-perturbative superpotential contributions from instanton or gaugino condensation. In fact because of the very same reason, 
this CY could naively be considered not suitable for carrying out phenomenology in the conventional sense. This is because established moduli stabilization mechanisms (specifically KKLT and LVS) rely crucially on non-perturbative superpotential corrections to stabilize K\"ahler  moduli, which are absent in our construction.
%given that both the popular moduli stabilisation schemes (namely KKLT and LVS) which are available, make use of non-perturbative corrections in the superpotential for stabilising the K\"ahler moduli. 

\item 
Let us note that in our present concrete CY construction the choice of orientifold involution which leads to having three stacks of $D7$-branes intersecting at three ${\mathbb T}^2$'s is such that there are no $O3$-planes present, and therefore anti-$D3$ uplifting proposal of \cite{Cicoli:2015ylx,Crino:2020qwk, AbdusSalam:2022krp} is not directly applicable   to our case. However, $D$-term uplifting and the $T$-brane uplifting processes are applicable to this model. 

\item
Moreover,  we note that there are no non-intersecting  $D7$-brane stacks and the $O7$-planes along without $O3$-planes present as well, and therefore this model does not induce the conventional KK-type string-loop corrections to the K\"ahler potential. 

\item 
Furthermore, the three stacks of D7-branes are wrapping a non-spin K3 divisor, i.e. $c_1({\rm K3}) = 0$ and, given that all the intersection numbers for this CY threefolds are even, the pullback of the $B$-field on any divisor $D_\alpha$ does not generate a half-integer flux. Therefore, one can appropriately adjust fluxes such that ${\cal F}_\alpha \in {\mathbb Z}$ for all $\alpha = 1, 2, 3$, have non-trivial D-term contributions. However, such a possibility has been studied in \cite{Bera:2024zsk} leading to an inflection point inflation driven by the overall CY volume while stabilizing the other two moduli via leading order effects. 

For the current work, we are interested in Fibre inflation where the overall volume modulus, ${\cal V}$,  is stabilized by leading-order effects, while inflation is driven by the remaining two moduli through subleading corrections. Without gauge flux ${\cal F}_\alpha = 0$, or having some other suitable flux choice can lead to vanishing of D-term contributions, i.e. $V_{\rm D} = 0$.

\end{itemize}

\noindent
%We note that each of the three $D7$-brane stacks as well as $O7$-plane intersect one another on non-contractible curves (e.g. see Table \ref{Tab1}), and volume of such non-contractible cycles, namely $t^\alpha_\cap$ lying at the intersection locus of any two $D7$-brane stacks or $D7/O7$ configurations can be given as,

We observe that all three $D7$-brane stacks and the $O7$-plane intersect along non-contractible curves (see Table \ref{Tab1}). The volume of these intersection cycles, denoted $t^\alpha_\cap$, corresponds to the loci where any two $D7$-brane stacks or $D7/O7$ configurations meet, and can be expressed as
\bea
& & \int_{CY} J \wedge \hat{D}_1 \wedge \hat{D}_2 = 2t^3, \qquad \int_{CY} J \wedge \hat{D}_2 \wedge \hat{D}_3 = 2t^1, \\
& &  \int_{CY} J \wedge \hat{D}_3 \wedge \hat{D}_1 = 2t^2, \qquad  \int_{CY} J \wedge [O7] \wedge \hat{D}_1= 4 (t^2 + t^3), \nonumber\\
& & \int_{CY} J \wedge [O7] \wedge \hat{D}_2 = 4 (t^1 + t^3), \qquad \int_{CY} J \wedge [O7] \wedge \hat{D}_3= 4 (t^1 + t^2)~, \nonumber
\eea
where the K\"ahler form is taken as $J = t^1 \hat{D}_1 + t^2 \hat{D}_2 + t^3 \hat{D}_3$. Subsequently, one has the winding-type string-loop effects contributing to the scalar potential as below,
\bea
\label{eq:Vgs-Winding-globalmodel}
& & \hskip-1cm V_{g_s}^{\rm W} = -2 \kappa \frac{|W_0|^2}{{\cal V}^3} \, \sum_\alpha \frac{{\cal C}_{w_\alpha}}{t^\alpha_\cap} \\
& & \hskip-0.25cm = - \frac{\kappa\,|W_0|^2}{{\cal V}^3} \left(\frac{{\cal C}_{w_1}}{t^1} + \frac{{\cal C}_{w_2}}{t^2} +\frac{{\cal C}_{w_3}}{t^3} +\frac{{\cal C}_{w_4}}{2(t^1+t^2)} +\frac{{\cal C}_{w_5}}{2(t^2+t^3)} +\frac{{\cal C}_{w_6}}{2(t^3+t^1)} \right)\,, \nonumber
\eea
where ${\cal C}_{w_\alpha}$'s are complex-structure moduli dependent quantities and can be taken as parameter for the moduli dynamics of the sub-leading effects. 

Moreover, notice that although this K3-fibred CY has several properties like a toroidal case, unlike the ${\mathbb T}^4$ case which has a vanishing  $\Pi$ ({\it Chern numbers}), the K3 divisors have $\Pi = 24$, and hence non-trivial higher derivative effects F$^4$-corrections can be induced,  taking the following form in the scalar potential,
\bea
\label{eq:F^4-term-globalmodel}
& & V_{{\rm F}^4} = - \frac{\lambda\,\kappa^2\,|W_0|^4}{g_s^{3/2} {\cal V}^4}\, 24 \, \left(t^1 + t^2 + t^3\right).
\eea
Collecting all the contributions, we have the following scalar potential~\cite{Bera:2024zsk}:
\bea
\label{eq:Vfinal-simp}
& & \hskip-0.75cm V = V_{\rm up} +  \frac{{\cal C}_1}{{\cal V}^3} \left(\hat\xi + 2\,\hat\eta \, \ln{\cal V} - 8\,\hat\eta + 2\,\hat\sigma \right) \\
& & \hskip-0.35cm  - \frac{\kappa\,|W_0|^2}{{\cal V}^3} \left(\frac{{\cal C}_{w_1}}{t^1} + \frac{{\cal C}_{w_2}}{t^2} +\frac{{\cal C}_{w_3}}{t^3} +\frac{{\cal C}_{w_4}}{2(t^1+t^2)} +\frac{{\cal C}_{w_5}}{2(t^2+t^3)} +\frac{{\cal C}_{w_6}}{2(t^3+t^1)} \right) \nonumber\\
& &  \hskip-0.35cm  \, +  \frac{{\cal C}_3}{{\cal V}^4}\,\left(t^1 + t^2 + t^3 \right) + \cdots,\nonumber
\eea
where we used the relations (\ref{eq:t-tau-vol}), and the various coefficients ${\cal C}_i$'s are given as,
\bea
\label{eq:calCis}
& & \hskip-1cm {\cal C}_1 = \frac{3\,\kappa\, |W_0|^2}{4} = \frac34 {\cal C}_2, \quad {\cal C}_3 = - \frac{24\, \lambda\,\kappa^2\, |W_0|^4}{g_s^{3/2}}, \quad |\lambda| =  \, {\cal O}(10^{-4}), \quad \kappa = \frac{g_s}{2}e^{K_{cs}}\cdot
\eea
As ${\cal V} = 2 \,t^1\, t^2\, t^3$, the scalar potential (\ref{eq:Vfinal-simp}) has the symmetry of the permutation group $S_3$ following from the underlying exchange symmetries ($1 \leftrightarrow 2 \leftrightarrow 3$) of the CY threefold. As we have discussed earlier, one can have isotropic moduli stabilization if one appropriately choses the following exchange symmetry among the coefficients,
\bea
\label{eq:exchange-symmetry}
& & t^1 \leftrightarrow t^2 \leftrightarrow t^3, \quad {\cal C}_{w_1} \leftrightarrow {\cal C}_{w_2} \leftrightarrow {\cal C}_{w_3}, \quad {\cal C}_{w_4} \leftrightarrow {\cal C}_{w_5} \leftrightarrow {\cal C}_{w_6}.
\eea
For the current global model candidate, the Euler characteristic is: $\chi({\rm CY}) = -224$, and $\Pi_\alpha = 24\, \, \forall \alpha \in \{1,2,3\}$ corresponding to the three K3 divisors of the underlying CY threefold. Further, using Eq.~(\ref{eq:def-xi-eta}) we have,
\bea
& & \hskip-1cm \hat\xi %= - \frac{\chi({\rm CY})\, \zeta[3]}{2(2\pi)^3\, g_s^{3/2}}
= \frac{14\, \zeta[3]}{\pi^3\, g_s^{3/2}}, \quad \hat\sigma %= \sqrt{g_s}\, \frac{\chi({\rm CY})\, \zeta[2]}{2(2\pi)^3}
= - \frac{14 \sqrt{g_s} \zeta[2] \sigma_0}{\pi^3}, \quad \hat\eta %= \sqrt{g_s}\, \frac{\chi({\rm CY})\, \zeta[2]}{2(2\pi)^3}
= - \frac{14 \sqrt{g_s} \zeta[2] \eta_0}{\pi^3}, \quad \frac{\hat\xi}{\hat\eta} = -\frac{\zeta[3]}{\zeta[2]\, g_s^2 \eta_0}\,.
\eea
In LVS frameworks, the overall volume ${\cal V}$ is fixed by the leading order contribution to the scalar potential and the full exchange symmetry seen from (\ref{eq:exchange-symmetry}) is reduced to ${\mathbb Z}_2$, which helps in assisted fibre inflation with two fields as we will present in the upcoming section.

\subsection{Single field fibre inflation}
The minimal standard LVS scheme of moduli stabilisation stabilizes the CY volume ${\cal V}$ along with a small modulus $\tau_s$ controlling the volume of an exceptional del Pezzo divisor. Therefore any LVS models with 3 or more K\"ahler moduli, $h^{1,1}\geq 3$, can generically have flat directions at leading order. These flat directions are promising inflaton candidates with a potential generated at subleading order. Based on the geometric nature of the inflaton field and the source of inflaton potential, there are basically several classes of interesting inflationary models realized in the LVS framework. In this regard, fibre inflation happens to be one of the most popular inflationary models in LVS \cite{Cicoli:2008gp,Cicoli:2016xae, Cicoli:2017axo,Cicoli:2024bxw}. The minimal design of fibre inflation consists of $3$-fields corresponding to the K\"ahler moduli of a K3-fibred CY threefold with one diagonal del-Pezzo divisor to support LVS. In this class of models, the typical volume form can be considered as ${\cal V} =  \gamma_b \, \, \tau_{b} \, \sqrt{\tau_f} - \gamma_s \, \, \tau_{s}^{3/2}$ such that the overall volume ${\cal V}$ and small divisor volume $\tau_s$ are stabilized via the standard LVS prescription while the volume modulus corresponding to the K3-fibre $\tau_f$ drives the inflation through a sub-leading scalar potential effect encoded as one-loop correction in the K\"ahler potential. It has been also observed that higher derivative F$^4$-corrections can be useful for the purpose of realizing fibre-inflation like models, especially for the cases when KK-type string loop corrections are absent. 

The underlying design of the fibre inflation remains similar in the perturbative LVS, where the overall volume ${\cal V}$ is stabilized perturbatively via using BBHL corrections and the log-loop effects in the K\"ahler potential. In this subsection, we will discuss the realization of an effective single-field  potential which can drive fibre inflation using a combination of string-loop effects of winding-type  and the higher derivative F$^4$-corrections to the scalar potential.

Given a single field scalar potential $V(\phi)$ depending on a canonically normalized field $\phi$, the sufficient condition for having the slow-roll inflation is encoded in a set of two so-called slow-roll parameters $\epsilon_V$, $\eta_V$ and $\xi^{(2)}_V$. These are defined through the derivatives of the inflationary potential as below
\bea
\label{eq:slow-roll-V-def}
& & \epsilon_V \equiv \frac{1}{2} \left(\frac{V_\phi}{V}\right)^2 \ll 1 , \qquad \eta_V \equiv \frac{V_{\phi\phi}}{V} \ll 1, \qquad \xi^{(2)}_V \equiv \frac{V_\phi \, V_{\phi\phi\phi}}{V^2} \ll 1,
\eea
where $V_\phi, V_{\phi\phi}$ and $V_{\phi\phi\phi}$ respectively corresponds to the single, double and triple derivatives of the potential. For a single field inflation, the two sets of slow-roll parameters, namely $(\epsilon, \eta)$ defined in (\ref{eq:epH-etaH}) and $(\epsilon_V, \eta_V)$ can be correlated as $\epsilon \simeq \epsilon_V, \, \eta \simeq -2\, \eta_V + 4\, \epsilon_V$ (e.g. see \cite{Achucarro:2018vey}). For a single field model, the evolutions equation (\ref{eq:EOM2} simplifies as follows.
\bea
\label{eq:single-field-evolution}
& & \hskip-1cm \phi^{\prime\prime} + \left(3 - \frac{1}{2} \phi^{\prime2} \right) \left(\phi^\prime + \frac{V_\phi}{V}\right) = 0, \quad \epsilon \equiv \frac{1}{2} \phi^{\prime2} \simeq \epsilon_V.
\eea
During the slow-roll regime, the evolution equation can be further simplified as $d\phi/dN\simeq -V_\phi/V$. Remarkably, all cosmological observables can be derived directly from derivatives of the scalar potential, effectively circumventing the need to solve the second-order field equations. \footnote{For completion, we will nevertheless present the numerical analysis for the field evolution as well.}. For that, the main cosmological observables, namely the scalar perturbation amplitude $(P_s)$, the spectral index $(n_s)$ and its running $\alpha_s$, and the tensor-to-scalar ratio $(r)$,  are correlated with these slow-roll parameters $\epsilon_V, \eta_V$ and $\xi^{(2)}_V$. It turns out that one typically needs to reproduce the following values for the cosmological observables \cite{Planck:2018jri,Planck:2018vyg},
\bea
\label{eq:cosmo-observables1}
& & P_s \equiv \frac{V^\ast}{24 \pi^2 \, \epsilon_V^\ast} \simeq 2.1 \times 10^{-9}, \\ %\qquad {\rm or} \qquad \frac{V_{\rm inf}^{\ast3}}{V_{\rm inf}^{\prime\ast^2}} \simeq 2.6\times 10^{-7},\\
& & n_s \simeq 1 + 2 \, \eta_V^\ast - 6\, \epsilon_V^\ast = 0.9651\pm 0.0044,\nonumber\\
& & \alpha_s = -24 {\epsilon^\ast_V}^2 + 16 \epsilon_V^\ast \eta_V^\ast - 2 \xi^{(2)\ast}_V = - 0.0041 \pm 0.0067, \nonumber\\
& & r \simeq 16 \epsilon_V^\ast\, < 0.036,\nonumber
\eea
where all the cosmological observables are evaluated at the horizon exit $\phi = \phi^\ast$. However, we also note that a combination of Planck, ACT, and DESI (P-ACT-LB) gives $n_s = 0.9743 \pm  0.0034$ and $\alpha_s = 0.0062 \pm 0.0052$ \cite{ACT:2025tim,ACT:2025fju,DESI:2024mwx, Frolovsky:2025iao}. Further, one typically has $N(\phi^\ast) \gtrsim 60$ as an estimate for the sufficient $e$-folds needed to ensure a successful inflationary scenario. Moreover, the number of e-folding is typically determined by the following relation,
\bea
\label{eq:e-fold-def}
& & N(\phi) = \int_{\phi_{\rm end}}^{\phi_\ast} \, \frac{1}{\sqrt{2 \epsilon_V}}\, d\phi \, \simeq \, \int_{\phi_{\rm end}}^{\phi_\ast} \, \frac{V}{V_\phi}\, d\phi\,,
\eea
where $\phi_{\rm end}$ corresponds to the end of inflation at $\epsilon = 1$, and $\phi_\ast$ is  the point of horizon exit at which the cosmological observables are to be matched with the experimentally observed values. Typically one needs $N(\phi) \gtrsim 60$, however as argued earlier, the number of $e$-folds $(N)$ depends on the post-inflationary aspects as well and can be given as a sum several contributions \cite{Liddle:2003as,Cicoli:2017axo}:
\bea
\label{eq:e-fold-def-postinflation}
& & \hskip-1.5cm N_{\rm tot} \simeq 57 + N_{\rm corr}^{(1)} + N_{\rm corr}^{(2)},\quad {\rm where} \quad N_{\rm corr}^{(1)} = \frac{1}{4} \ln(r_\ast V_\ast) , \quad  N_{\rm corr}^{(2)} = - \frac{1}{3}\ln\left(\frac{10V_{\rm end}}{m_{\inf}^{3/2}}\right).
\eea
Here $V_{\rm end} = V(\phi_{\rm end})$ corresponds to end of inflation determined by $\epsilon = 1$ and $m_{\rm inf}$ is the inflaton mass. For fibre inflation models it has been estimated that typically one needed $N \simeq 50$  \cite{Cicoli:2017axo,Bhattacharya:2020gnk,Cicoli:2020bao}.

For the current CY orientifold model with ${\cal V} = 2 \,t^1 t^2 t^3$, after stabilizing the overall volume modulus via the perturbative LVS process, one is left with two more K\"ahler moduli, say $t^2$ and $t^3$. In order to achieve an effective single-field potential $V(\phi)$ for realizing the minimal fibre inflation scenario, one introduces  appropriate worldvolume gauge fluxes on the suitable four-cycles such that ${\cal F}_i = \sum_{j=1}^{h^{1,1}} f_{ij}\hat{D}_j + \frac12 \hat{D}_i - \iota_{D_i}^*B$ with $f_{ij}\in \mathbb{Z}$ and the half-integer contribution is due to Freed-Witten anomaly cancellation \cite{Minasian:1997mm,Freed:1999vc}. Such a flux can fix another combination of the remaining two moduli. Moreover, a chiral visible sector can be supported on the D7-brane stacks wrapping suitable divisors. Also, given the fact that the three stacks of D7-branes are wrapping a spin divisor K3 with $c_1({\rm K3}) = 0$, and the triple intersections on this CY are even, the pullback of the $B$-field on any divisor $D_\alpha$ does not generate a half-integer flux, and therefore one can appropriately adjust fluxes such that ${\cal F}_\alpha \in {\mathbb Z}$ for all $\alpha \in \{1, 2, 3\}$. However we consider a non-vanishing gauge flux ${\cal F}_1$ only on the worldvolume of the $D_1$ divisor while considering ${\cal F}_2 = 0 = {\cal F}_3$. This way we still keep the underlying symmetry $2 \leftrightarrow 3$ intact. Following the prescription of \cite{Cicoli:2017axo}, the vanishing of FI parameter for such a case leads to $t^2 = t^3$, or equivalently $\tau_2 = \tau_3$. Subsequently the volume  takes the following form,
\bea
\label{}
& & {\cal V} = 2 \, t^1 (t^2)^2 = \frac{\tau_2\, \sqrt{\tau_1}}{\sqrt 2}, \qquad \tau_1 = 2 \,(t^2)^2, \, \, \tau_2 = 2 \, t^1 \,t^2, 
\eea
This leads to the following tree-level K\"ahler metric and its inverse metric,
\bea
\label{eq:Kij-tree1}
& & K_{\alpha\beta}^0 = \frac{1}{4} \left(
\begin{array}{cc}
 (\tau_1)^{-2} & 0 \\
 0 & 2(\tau_2)^{-2} \\
\end{array}
\right),
\eea
However, for our current analysis we work in the basis of real fields which involves the overall volume as it is stabilized at the leading order in the perturbative LVS, and therefore we take it as $\Phi^a = \{{\cal V}, t^2\}$. Subsequently using the leading order tree-level K\"ahler metric (\ref{eq:Kij-tree1}) and the definition in (\ref{eq:fieldspace-metric}), the field space metric ${\cal G}_{ab}$ and its inverse metric ${\cal G}^{ab}$ are given by
\bea
\label{eq:metric-phia1}
& & \hskip-1.5cm {\cal G}_{ab} = \left(\begin{array}{cc}
\frac{1}{{\cal V}^2} &  -\frac{1}{t^2 {\cal V}} \\
&  \\
 -\frac{1}{t^2 {\cal V}} & \frac{3}{(t^2)^2} \\
\end{array}
\right), \qquad {\cal G}^{ab} = \left(\begin{array}{cc}
\frac{3{\cal V}^2}{2} & \frac{{\cal V}t^2}{2} \\
& \\
 \frac{{\cal V}t^2}{2} & \frac{(t^2)^2}{2} \\
\end{array}
\right)
\eea
The main idea for achieving the single field fibre inflation is the following:

\begin{itemize}

\item 
First, using some appropriate uplifting piece, the overall volume ${\cal V}$ can be fixed via perturbative LVS at leading order, resulting in a de-Sitter minimum. 

\item 
Subsequently we will realize inflation effectively driven by a single field, namely $t^2$, while the overall volume modulus ${\cal V}$ being mostly remained in its perturbative LVS minimum.

\end{itemize}

\noindent
Using (\ref{eq:Vfinal-simp}) along with $t^3 = t^2$, and further considering $t^2 \equiv e^{\phi/\sqrt{3}}$ for the canonical field $\phi$ which follows from (\ref{eq:metric-phia}), one gets the following effective single field scalar potential,
\bea
\label{eq:Vinf-single-field}
& & \hskip-0.9cm V \simeq {\cal C}_0 \biggl({\cal R}_{\rm LVS} + {\cal R}_0 e^{-2\gamma\phi} - e^{-\gamma\phi} +{\cal R}_1 e^{\gamma\phi} + {\cal R}_2 e^{2\gamma\phi}\biggr),
\eea
where the exponent $\gamma=1/\sqrt{3}$ and the various quantities ${\cal C}_0$, ${\cal R}_{\rm LVS}$, ${\cal R}_0$, ${\cal R}_1$ and ${\cal R}_2$ are given as:
\bea
\label{eq:Cis-new}
& & \hskip-0.5cm {\cal C}_0 =\frac{{\cal C}_2 \tilde{\cal C}_w}{{\cal V}^3},  \quad {\cal R}_{\rm LVS} = \frac{{\cal C}_{\rm up}}{{\cal C}_0 {\cal V}^p} +  \frac{{\cal C}_1}{{\cal C}_0 {\cal V}^3} \left(\hat\xi + 2\,\hat\eta \, \ln{\cal V} - 8\,\hat\eta + 2\,\hat\sigma \right),\\
& & \hskip-0.5cm {\cal R}_0 = \frac{{\cal C}_3}{2{\cal C}_2\tilde{\cal C}_w}, \quad \frac{{\cal R}_1}{{\cal R}_0} = \frac{4}{{\cal V}}, \quad \frac{{\cal R}_2}{{\cal R}_0} = -\frac{4\,{\cal C}_2 {\cal C}_{w_1}}{{\cal C}_3\,{\cal V}} \biggl[1-\hat{\cal C}_w \left(1+\frac{2\,e^{\sqrt{3}\phi}}{{\cal V}}\right)^{-1}\biggr],\nonumber
\eea
where $\tilde{\cal C}_w =(4{\cal C}_{w_2} + 4{\cal C}_{w_3} + {\cal C}_{w_5})/4$ and $\hat{\cal C}_w = ({\cal C}_{w_4}+{\cal C}_{w_6})/({2{\cal C}_{w_1}})$. Further, we note that ${\cal C}_0, {\cal R}_{\rm LVS}, {\cal R}_0$ and ${\cal R}_1$ depend on the overall volume ${\cal V}$ only, and do not have a dependence on the inflaton $\phi$ while ${\cal R}_2$ depends on both the $({\cal V}, \phi)$ moduli. The parameter ${\cal C}_{\rm up}$ is the uplifting constant which will depend on the choice of uplifting mechanism via the specific values of the $p$ parameter. In fact, the first three terms correspond to Starobinsky type inflationary potential \cite{Starobinsky:1980te} (see  also~\cite{Brinkmann:2023eph}). Therefore we need to examine if this inflation remains robust against the other sub-leading corrections and if there are any knock-on effects on inflation dynamics. Let us emphasize  the following points:
\begin{itemize}
\item 
The first three terms of the inflationary potential in Eq.~(\ref{eq:Vinf-single-field}) determine the minimum while the other two terms create the steepening. 

\item 
After fixing the overall volume ${\cal V}$, the quantities ${\cal R}_1 \ll 1$ and ${\cal R}_2 \ll 1$ serve as parameters which, as seen from (\ref{eq:Cis-new}), are further volume (${\cal V}$) suppressed as compared to ${\cal R}_0$. This is to be exploited in finding a sufficiently long plateau.

\item 
The single field slow-roll parameters do not depend on ${\cal C}_0$ as it is an overall factor seen from (\ref{eq:Vinf-single-field}), hence $n_s$, $r$ and $N$ can be determined purely by three parameters ${\cal R}_{\rm LVS}, {\cal R}_0, {\cal R}_1$ and ${\cal R}_2$, and then ${\cal C}_0$ can be appropriately chosen to match the scalar perturbation amplitude $P_s$.

\item 
Although ${\cal R}_2$ depends on the inflaton $\phi$, this dependence is suppressed by an extra volume factor, and does not spoil the effective leading order inflation for reasonable values of the $\hat{\cal C}_w$ parameter. 

\end{itemize}
\noindent
The above mentioned points will be more explicit when we will discuss the numerical models.

\subsection*{Numerical benchmark models}
We denote a numerical model by ${\cal M}_n^p$ where $n$ corresponds to number of fields considered in the inflationary analysis, and $p$ corresponds to the uplifting parameter with $p = 4/3$ for anti-D3 uplifting \cite{Kachru:2003aw,Crino:2020qwk,Cicoli:2017axo,AbdusSalam:2022krp}, and $p = 2$ for D-term uplifting \cite{Burgess:2003ic,Achucarro:2006zf,Braun:2015pza} while  $p = 8/3$ for the T-brane uplifting \cite{Cicoli:2015ylx,Cicoli:2017shd}. We present two sets of benchmark models, first one corresponding to the PLANCK data \cite{Planck:2018jri, Planck:2018vyg} while the other one corresponds to the recent ACT, DESI results \cite{ACT:2025tim,ACT:2025fju,DESI:2024mwx, Frolovsky:2025iao}.

\subsubsection*{Class-I: Standard cosmological observables}

We consider the following set of parameters for the winding-type string loop corrections,
\bea
\label{eq:same-parameters}
& & \hskip-1cm {\cal C}_{w_1} = 0.0006, \quad {\cal C}_{w_2} = - 0.0008 = {\cal C}_{w_3}, \quad {\cal C}_{w_4} = -0.02 = {\cal C}_{w_6} , \quad {\cal C}_{w_5} = 0.4.
\eea
which result in $\tilde{\cal C}_{w} = 0.0984$ and $\hat{\cal C}_w = 100/3$. The additional parameters are considered as in table \ref{tab_single-field-models1} which results in a non-supersymmetric de-Sitter in the perturbative LVS. Subsequently, considering the overall volume to be remained seated at its perturbative LVS minimum with $\langle {\cal V} \rangle$, we present a single field inflationary analysis for a set of numerical models with various cosmological observables as presented in table \ref{tab_single-field-models2}.

\begin{table}[H]
\centering
\begin{tabular}{|c|c|c|c|c|c|c|c|| c| c| c|}
\hline
%& & & & & & & & & & \\
${\cal M}_n^p $ & $p$ & $\chi$ & $\eta_0$ & $\sigma_0$ & $g_s$ & $W_0$ & $\lambda$ & ${\cal C}_{\rm up}$ & $\langle{\cal V}\rangle$ & $\langle t^2 \rangle$ \\
%& & & & & & & & & & \\
\hline
\hline
%& & & & & & & & & & \\
${\cal M}_1^{4/3}$ & 4/3 & -224 & 2 & 0 & 0.275 & 5 & -0.0001 & $2.27\cdot 10^{-5}$ & 1037.06 & 0.587 \\
%& & & & & & & & & & \\
${\cal M}_1^{2}$ & 2 & -224 & 2 & 0 & 0.30 & 5.2 & -0.0001 & $4.69\cdot 10^{-3}$ & 1056.87 & 0.602 \\
%& & & & & & & & & & \\
${\cal M}_1^{8/3}$ & 8/3 & -224 & 6 & -4 & 0.295 & 4.83 & -0.0001 & $3.58$ & 1104.66 & 0.524 \\
%& & & & & & & & & & \\
\hline
\end{tabular}
\caption{Single-field models ${\cal M}_1^p$: Moduli stabilization and uplifting}
\label{tab_single-field-models1}
\end{table}

\begin{table}[H]
\centering
\begin{tabular}{|c|c|c|c||c|c|c||c|c|c|c|}
\hline
%& & & & & & & & & & \\
${\cal M}_n^p $ & $\langle \phi \rangle$ & $\phi_{\rm end}$ & $\phi^\ast$ & $N$ & $N_{\rm corr}^{(1)}$ & $N_{\rm corr}^{(2)}$ & $10^9 P_s^\ast$ & $n_s^\ast$ & $10^4 \alpha_s^\ast$ & $10^3 r^\ast$ \\
& & & & (\ref{eq:e-fold-def}) & (\ref{eq:e-fold-def-postinflation}) & (\ref{eq:e-fold-def-postinflation}) & & & & \\
\hline
\hline
%& & & & & & & & & & \\
${\cal M}_1^{4/3}$ & -0.9231 & 0.1117 & 5.52 & 51.2 & -6.59 & 0.77 & 2.08 & 0.969 & -3.78 & 8.27 \\
%& & & & & & & & & & \\
${\cal M}_1^{2}$ & -0.8799 & 0.1550 & 5.55 & 50.5 & -6.69 & 0.78 & 2.17 & 0.969 & -3.73 & 8.61 \\
%& & & & & & & & & & \\
${\cal M}_1^{8/3}$ & -1.1206 & -0.0861 & 5.27 & 50.7 & -6.77 & 0.79 & 2.13 & 0.967 & -6.40 & 7.41 \\
%& & & & & & & & & & \\
\hline
\end{tabular}
\caption{Single-field models ${\cal M}_1^p$: Cosmological observables}
\label{tab_single-field-models2}
\end{table}

\subsubsection*{Class-II: ACTivated cosmological observables}

In order to distinguish the ACTivated models from the standard ones, we denote them by ${\cal M}_{n}^{\prime p}$. Since the winding loop correction parameters ${\cal C}_{w_1}$, ${\cal C}_{w_4}$ and ${\cal C}_{w_6}$ create the steepening part, we find that larger values of the spectral index parameter $n_s$ as motivated by the ACT experiments, can be easily accommodated in such models. For that purpose, we consider the following parameters
\bea
\label{eq:same-parameters}
& & \hskip-1cm {\cal C}_{w_1} = 0.001, \quad {\cal C}_{w_2} = - 0.0008 = {\cal C}_{w_3}, \quad {\cal C}_{w_4} = -0.1 = {\cal C}_{w_6} , \quad {\cal C}_{w_5} = 0.37.
\eea

\begin{table}[H]
\centering
\begin{tabular}{|c|c|c|c|c|c|c|c|| c| c| c|}
\hline
%& & & & & & & & & & \\
${\cal M}_n^{\prime p} $ & $p$ & $\chi$ & $\eta_0$ & $\sigma_0$ & $g_s$ & $W_0$ & $\lambda$ & ${\cal C}_{\rm up}$ & $\langle{\cal V}\rangle$ & $\langle t^2 \rangle$ \\
%& & & & & & & & & & \\
\hline
\hline
%& & & & & & & & & & \\
${\cal M}_1^{\prime 4/3}$ & 4/3 & -224 & 2 & 0 & 0.280 & 5 & -0.00017 & $2.533\cdot 10^{-5}$ & 986.80 & 1.0524 \\
%& & & & & & & & & & \\
${\cal M}_1^{\prime 2}$ & 2 & -224 & 2 & 0 & 0.298 & 5 & -0.00017 & $4.05\cdot 10^{-3}$ & 1120.42 & 1.0216 \\
%& & & & & & & & & & \\
${\cal M}_1^{\prime 8/3}$ & 8/3 & -224 & 6 & -4 & 0.295 & 5 & -0.00017 & $3.868$ & 1120.32 & 1.0267 \\
%& & & & & & & & & & \\
\hline
\end{tabular}
\caption{ACTivated single-field models ${\cal M}_1^{\prime p}$: Moduli stabilization and uplifting}
\label{tab_single-field-models-ACT1}
\end{table}

\begin{table}[H]
\centering
\begin{tabular}{|c|c|c|c||c|c|c||c|c|c|c|}
\hline
%& & & & & & & & & & \\
${\cal M}_n^p $ & $\langle \phi \rangle$ & $\phi_{\rm end}$ & $\phi^\ast$ & $N$ & $N_{\rm corr}^{(1)}$ & $N_{\rm corr}^{(2)}$ & $10^9 P_s^\ast$ & $n_s^\ast$ & $10^4 \alpha_s^\ast$ & $10^3 r^\ast$ \\
& & & & (\ref{eq:e-fold-def}) & (\ref{eq:e-fold-def-postinflation}) & (\ref{eq:e-fold-def-postinflation}) & & & & \\
\hline
\hline
%& & & & & & & & & & \\
${\cal M}_1^{\prime 4/3}$ & 0.0885 & 1.1405 & 5.83 & 50.61 & -6.93 & 0.92 & 2.13 & 0.974 & 9.37 & 5.39 \\
%& & & & & & & & & & \\
${\cal M}_1^{\prime 2}$ & 0.0370 & 1.0857 & 5.72 & 50.58 & -7.08 & 0.93 & 2.14 & 0.975 & 1.68 & 3.96 \\
%& & & & & & & & & & \\
${\cal M}_1^{\prime 8/3}$ & 0.0456 & 1.0945 & 5.75 & 51.41 & -7.09 & 0.93 & 2.11 & 0.975 & 2.75 & 3.96 \\
%& & & & & & & & & & \\
\hline
\end{tabular}
\caption{ACTivated single-field models ${\cal M}_1^{\prime p}$: Cosmological observables}
\label{tab_single-field-models-ACT2}
\end{table}

\begin{figure}[H]
\centering
\includegraphics[width=7.15cm]{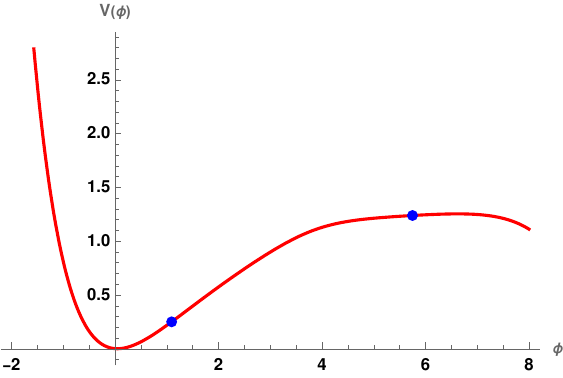}
\caption{Plot of $V(\phi)\cdot10^{10}$ with a display of the horizon exit $\phi^\ast$ and $\phi_{\rm end}$} %corresponding to the end of inflation, i.e. $\epsilon = 1$.}
\label{fig_pot-single-field}
\end{figure}

\begin{figure}[H]
\centering
\includegraphics[width=13.9cm]{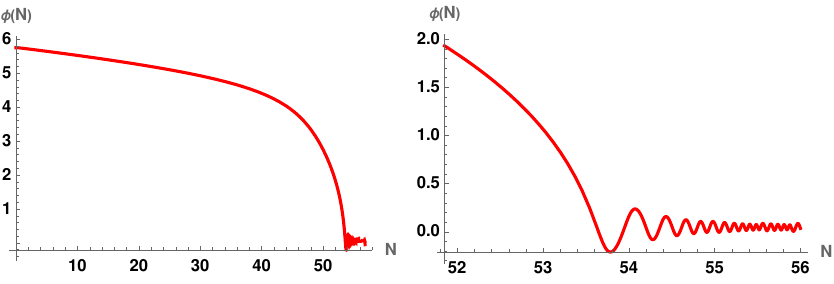}
\caption{Evolution of the inflaton $\phi(N)$ showing $\Delta \phi \simeq 6$ needed for inflation}
\label{fig_phi-single-field}
\end{figure}

\begin{figure}[H]
\centering
\includegraphics[width=13.9cm]{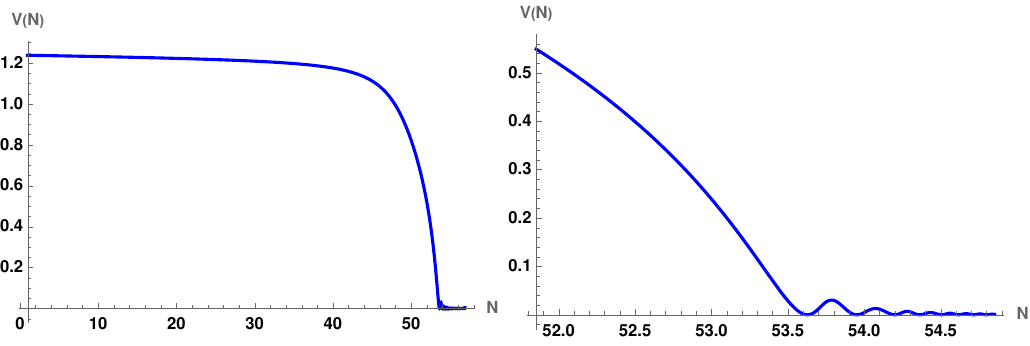}
\caption{Evolution of the scalar potential $V(N)\cdot10^{10}$ plotted for the number of $e$-folds}
\label{fig_pot-N-single-field}
\end{figure}

\begin{figure}[H]
\centering
\includegraphics[width=13.9cm]{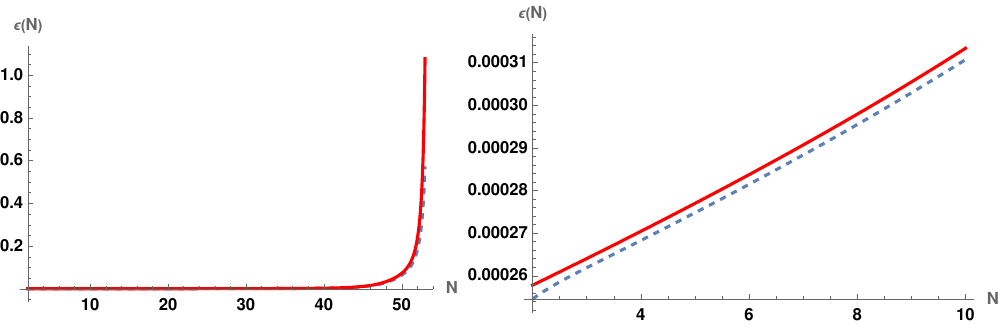}
\caption{Evolution of slow-roll parameter $\epsilon$ along with $\epsilon_V$ represented with dotted lines.}
\label{fig_epsilon-single-field}
\end{figure}

\begin{figure}[H]
\centering
\includegraphics[width=13.9cm]{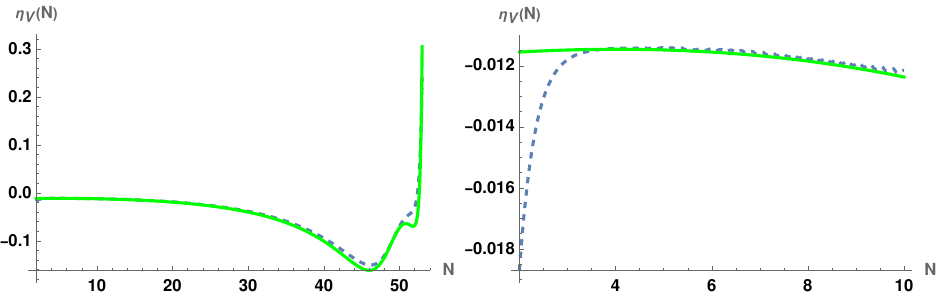}
\caption{Evolution of slow-roll parameter $\eta_V$ with two definitions (\ref{eq:epH-etaH}) and (\ref{eq:epV-etaV}).}
\label{fig_eta-single-field}
\end{figure}

\begin{figure}[H]
\centering
\includegraphics[width=13.9cm]{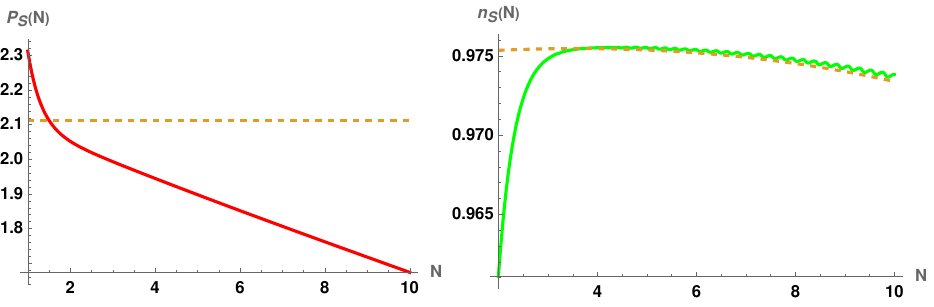}
\caption{Evolution of $P_s(N)\cdot10^9$ and $n_s(N)$ with dotted plots for $P_s = 2.1\cdot10^{-9}$ and $n_s$ using (\ref{eq:cosmo-observables1}).}
\label{fig_Ps-ns-single-field}
\end{figure}

\begin{figure}[H]
\centering
\includegraphics[width=13.9cm]{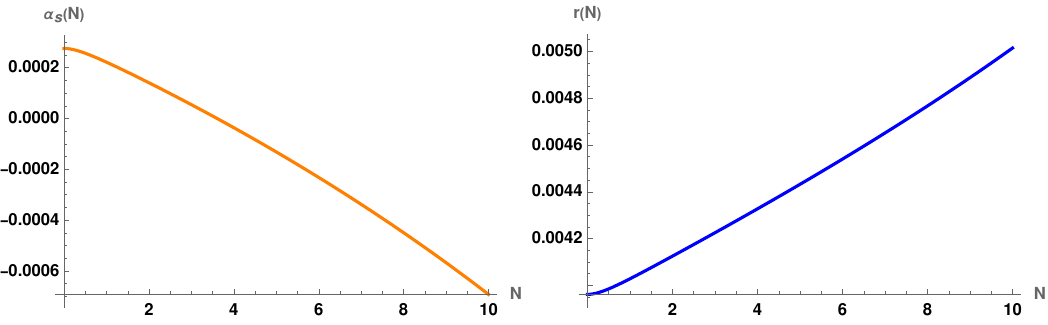}
\caption{Evolution of $\alpha_s(N)$ and  $r(N)$ %where the solid plots follow from the definition (\ref{eq:cosmo-observables}) while the dotted plot corresponds to their 
using definitions in (\ref{eq:cosmo-observables1}) and $\phi\equiv\phi(N)$.}
\label{fig_r-single-field}
\end{figure}

\noindent
Finally, for the graphical analysis of the field evolution, let us consider the ACTivated model ${\cal M}_1^{\prime 8/3}$ described by the parameters given in table \ref{tab_single-field-models-ACT1}. The single field scalar potential $V(\phi)$ is given by (\ref{eq:Vinf-single-field}) with numerical parameters given as below
\bea
\label{eq:Vnum-single-field}
& & {\cal C}_0 = 2.38378\cdot10^{-10}, \quad {\cal R}_{\rm LVS} = 0.480076, \quad {\cal R}_0 = 0.516494,\\
& & {\cal R}_1 = 0.00184409, \quad {\cal R}_2 = -0.0000196392 + \frac{2.20022}{1120.32+2e^{\sqrt{3}\phi}}. \nonumber
\eea
Using this inflationary potential, the motion of the inflaton can be analyzed by the solving the evolution equation (\ref{eq:single-field-evolution}) for the initial conditions: $\phi^\prime(0) = 0$ and $\phi(0) = 5.75$ as mentioned in table \ref{tab_single-field-models-ACT2}. The corresponding inflationary details for this model are presented in figures \ref{fig_pot-single-field} - \ref{fig_r-single-field}. During the inflationary period, i.e. $\epsilon \leq 1$, the total number of $e$-folds  turns out to be around 53. However as the initial conditions are assumed such that the initial inflaton velocity is set to zero, it usually takes 1-3 $e$-folds to get down on the correct inflationary track.

\noindent
The main features of the single-field analysis can be summarized along the following points:
\begin{itemize}

\item
Fibre inflation can be successfully realized in perturbative LVS framework with ${\cal V} \simeq 10^3$ and satisfying the other requirements such as having sufficient number of $e$-folds with inflaton shifts within the K\"ahler cone conditions as well as having the experimentally consistent values for all the cosmological observables, in particular the scalar perturbation amplitude $P_s$. 

\item
One can easily realize the numerical models with larger spectral index $n_s$ as motivated by the recent ACT experiments \cite{ACT:2025tim,ACT:2025fju,DESI:2024mwx, Frolovsky:2025iao}. However, we observe that larger $n_s$ values result in a slightly lower values for the tensor-to-scalar ratio $r$.

\item
The main point we want to emphasize is that the total inflaton shift one needs for driving a successful single-field fibre inflation turns out to be $\Delta\phi = \phi^\ast - \langle \phi \rangle \simeq 6$ M$_p$. However, for ACTivated models, we observe that one needs slightly less inflaton shifts, $\Delta\phi$ for realizing the cosmological inflationary requirements.

\item 
For demonstrating the numerics through graphics, we consider one particular model, namely the ACTivated ${\cal M}_1^{(8/3)}$. The inflationary potential is plotted in figure \ref{fig_pot-single-field}. Subsequently, the evolution of inflaton field $\phi(N)$ is determined by solving (\ref{eq:single-field-evolution}) which is plotted in figure \ref{fig_phi-single-field}. This shows a nice attractor behavior towards the minimum, and this feature is also seen for the inflationary potential $V(N)$ in figure \ref{fig_pot-N-single-field}. Subsequently all the cosmological observables are plotted in figures \ref{fig_epsilon-single-field} - \ref{fig_r-single-field}.
    
\end{itemize}

\subsection{Two-field inflationary analysis}
In this subsection, we will present the two-field inflationary analysis for the benchmark models presented in the single-field case. For the single-field analysis, we assumed that the overall volume modulus ${\cal V}$ sits at its perturbative LVS minimum while the $t^2$ modulus effectively drives the inflation. However, it is necessary to test the validity of such an assumption by considering a two-field analysis.

For this purpose, we consider the inflationary dynamics using the field basis $\Phi^a = \{{\cal V}, t^2\}$ and show that inflation is mostly driven by a single field while the overall volume modulus ${\cal V}$ remains seated in its perturbative LVS minimum. Now imposing $t^2 = t^3$ in the scalar potential (\ref{eq:Vfinal-simp}), the effective two-field potential takes the following form,
\bea
\label{eq:Vfinal-simp1}
& & \hskip-1cm V({\cal V}, t^2) \simeq \frac{{\cal C}_{\rm up}}{{\cal V}^p} +  \frac{{\cal C}_1}{{\cal V}^3} \left(\hat\xi + 2\,\hat\eta \, \ln{\cal V} - 8\,\hat\eta + 2\,\hat\sigma \right) \\
& & \hskip-0.35cm - \frac{\,{\cal C}_2}{{\cal V}^3} \biggl( \frac{2\,{\cal C}_{w_1}\,(t^2)^2}{{\cal V}} +  \frac{4({\cal C}_{w_2}+{\cal C}_{w_3})+{\cal C}_{w_5}}{4\,t^2} + \frac{({\cal C}_{w_4}+{\cal C}_{w_6})\,(t^2)^2}{{\cal V} + 2 (t^2)^3}\biggr)+  \frac{{\cal C}_3}{{\cal V}^3}\,\left(\frac{1}{2 \, (t^2)^2} + \frac{2\,t^2}{\cal V} \right). \nonumber
\eea
Subsequently, using (\ref{eq:fieldspace-metric} and (\ref{eq:metric-phia1}), the field equations for the fields $\{{\cal V}, t^2\}$ arising from Eq.~(\ref{eq:EOM2}) are explicitly given as below,
\bea
\label{eq:Explicit-EOMs1}
& & \hskip-1cm {\cal V}^{\prime\prime} = \frac{{\cal V}^{\prime2}}{{\cal V}} - \left(3- \epsilon\right) \left({\cal V}^{\prime}+\frac{3{\cal V}^2}{2V} \partial_{\cal V} V +\frac{{\cal V} \, t^2}{2V}\partial_{t^2} V \right),\\
& & \hskip-1cm (t^2)^{\prime\prime} = \frac{(t^2)^{\prime2}}{t^2} - \left(3- \epsilon\right) \left((t^2)^{\prime}+\frac{{\cal V} \, t^2}{2V}\partial_{{\cal V}} V +\frac{(t^2)^2}{2V} \partial_{t^2} V \right),\nonumber
\eea
where $^\prime$ denotes derivatives w.r.t.~the $e$-folding $N$, i.e. ${\cal V}^\prime = \frac{d{\cal V}}{dN}$ etc. In addition, the $\epsilon$ parameter and the non-zero Christoffel connections, corresponding to the field space metric ${\cal G}_{ab}$ as defined in Eq.~(\ref{eq:metric-phia1}), are given as follows
\bea
& & \epsilon = \frac{1}{2} \left(\frac{{\cal V}^{\prime2}}{{\cal V}^2} - \frac{2{\cal V}^\prime \, (t^2)^{\prime}}{{\cal V} \, t^2} + \frac{3(t^2)^{\prime2}}{(t^2)^2} \right), \qquad \Gamma_{11}^1 = - \frac{1}{\cal V}, \quad \Gamma_{22}^2 = - \frac{1}{t^2}.
\eea

\subsubsection*{Computation of the effective inflaton shift $\Delta\phi$}
Before discussing  the numerical models, let us have some estimates about the effective inflaton shift needed for having a successful fibre inflation model. Using the two-field K\"ahler metric (\ref{eq:metric-phia1}) one can determine the form of two canonical fields such that the metric ${\cal G}_{ab} = \delta_{ab}$, i.e. a diagonal flat space metric. Moreover, given that the overall volume modulus ${\cal V}$ is stabilized at the leading order in perturbative LVS, we make one of the canonical fields  to be determined entirely by ${\cal V}$ while the other field corresponds to a direction orthogonal to the overall volume ${\cal V}$. This can be achieved by the following canonical fields $\{\phi^1, \phi^2\}$,
\bea
\label{eq:canonical-two-field-def}
& & {\cal V} = \frac{1}{\sqrt{2}}\, e^{\sqrt{\frac{3}{2}} \phi^1}, \qquad t^2 = \frac{1}{\sqrt2} e^{\frac{1}{\sqrt6} \phi^1 + \frac{1}{\sqrt3}\phi^2}.
\eea
In fact, it turns out that the $\phi^2$ direction is along the ratio of two moduli such that $\frac{\tau_1}{\tau_2} = \frac{t^2}{t^1} = e^{\sqrt3 \phi^2}$. Subsequently, the overall inflaton shift needed for realizing successful inflationary predictions can be determined by the flat-space distance formula given as,
\bea
\label{eq:shiftphi-two-field}
& & \Delta \phi = \sqrt{\left(\phi^{1\ast} - \langle\phi^1\rangle\right)^2 + \left(\phi^{2\ast} - \langle\phi^2\rangle\right)^2},
\eea
where $\phi^{a\ast}$ corresponds to the horizon exit while $\langle \phi^a\rangle$ denotes the moduli VEVs at the LVS minimum. Let us note that one can choose any canonical basis, and this inflaton shift $\Delta\phi$ will be invariant. However, the individual fields may result in different shifts in a different choice of basis. As mentioned earlier, this choice (\ref{eq:canonical-two-field-def}) is made only to keep one field along the overall volume, which being heavier, receives a very small shift during the inflation, while the other modulus, which is orthogonal to the overall volume, turns out to be the actual inflaton field.

\subsection*{Numerical benchmark models}
Similarly to the single-field analysis, we will now numerically solve the two-field equations given in (\ref{eq:Explicit-EOMs1}) and present the numerical benchmark models. As the various plots for this two-field analysis will be similar to those of the single-field and three-field cases, we postpone the graphical analysis and the subsequent discussions for the next section.

\subsubsection*{Class-I: Standard cosmological observables}

Similar to the single-field case, we consider the following set of parameters,
\bea
\label{eq:same-parameters-two-fields}
& & \hskip-1cm \chi = -224, \quad \lambda = -0.0001, \quad {\cal C}_{w_1} = 0.0008,\\
& & \hskip-1cm {\cal C}_{w_2} = - 0.0008 = {\cal C}_{w_3}, \quad {\cal C}_{w_4} = -0.02 = {\cal C}_{w_6} , \quad {\cal C}_{w_5} = 0.4. \nonumber
\eea
Considering appropriate set of other parameters along with (\ref{eq:same-parameters-two-fields}) leads to moduli stabilization resulting in perturbative LVS with de-Sitter uplifting. Three benchmark models are presented in table \ref{tab_two-field-models1} followed by the respective cosmological observables being presented in table \ref{tab_two-field-models2}.

\begin{table}[H]
\centering
\begin{tabular}{|c|c|c|c|c|c||c|c|c|c|c|}
\hline
%& & & & & & & & & & \\
${\cal M}_n^p $ & $p$ &  $\eta_0$ & $\sigma_0$ & $g_s$ & $W_0$ & ${\cal C}_{\rm up}$ & $\langle{\cal V}\rangle$ & $\langle t^2 \rangle$ & $\langle\phi^1\rangle$ & $\langle\phi^2\rangle$\\
%& & & & & & & & & & \\
\hline
\hline
%& & & & & & & & & & \\
${\cal M}_2^{4/3}$ & 4/3 & 2 & 0 & 0.275 & 5.0 & $2.271\cdot 10^{-5}$ & 1036.31 & 0.581 & 5.952 & -4.549 \\
%& & & & & & & & & & \\
${\cal M}_2^{2}$ & 2 &  2 & 0 & 0.3 & 5.6 & $5.383\cdot 10^{-3}$ & 1066.71 & 0.698 & 5.976 & -4.249\\
%& & & & & & & & & & \\
${\cal M}_2^{8/3}$ & 8/3 & 6 & -4 & 0.3 & 5.7 & $5.2058$ & 1087.37 & 0.723 & 5.992 & -4.199\\
%& & & & & & & & & & \\
\hline
\end{tabular}
\caption{Two-field models ${\cal M}_2^p$: Moduli stabilization and uplifting}
\label{tab_two-field-models1}
\end{table}

\begin{table}[H]
\centering
\begin{tabular}{|c|c|c||c|c|c|c||c|c|c|c|c|}
\hline
%& & & & & & & & & & & \\
${\cal M}_n^p $ & ${\cal V}^\ast$ & $(t^2)^\ast$ & $\phi^{1\ast}$ & $\phi^{2\ast}$ & $\Delta\phi$ & $N$ & $N^\ast$ & $10^9 P_s^\ast$ & $n_s^\ast$ & $10^4\alpha_s$& $10^3 r^\ast$ \\
%& & & &  & & & & & & & \\
\hline
\hline
%& & & & & & & & & & & \\
${\cal M}_2^{4/3}$ & 1194.98 & 22.09 & 6.07 & 1.67 & 6.22 & 60.3 & 10 & 2.12 & 0.964 & -7.06 & 6.62 \\
%& & & & & & & & & & & \\
${\cal M}_2^{2}$ & 1334.45 & 22.14 & 6.16 & 1.61 & 5.86 & 54.0 & 4 & 2.09 & 0.966 & -5.12 & 6.48 \\
%& & & & & & & & & & & \\
${\cal M}_2^{8/3}$ & 1397.71 & 23.68 & 6.20 & 1.70 & 5.90 & 58.0 & 8 & 2.07 & 0.966 & -6.78 & 6.14 \\
%& & & & & & & & & & & \\
\hline
\end{tabular}
\caption{Two-field models ${\cal M}_2^p$: Cosmological observables}
\label{tab_two-field-models2}
\end{table}

\subsubsection*{Class-II: ACTivated cosmological observables}

For the two-field ACTivated models, we consider the following set of parameters
\bea
\label{eq:same-parameters-two-fields-ACT}
& & \hskip-1cm \chi = -224, \quad \lambda = -0.00017, \quad {\cal C}_{w_1} = 0.001,\\
& & \hskip-1cm {\cal C}_{w_2} = - 0.0008 = {\cal C}_{w_3}, \quad {\cal C}_{w_4} = -0.1 = {\cal C}_{w_6} , \quad W_0 = 5. \nonumber
\eea

\begin{table}[H]
\centering
\begin{tabular}{|c|c|c|c|c|c||c|c|c|c|c|}
\hline
%& & & & & & & & & & \\
${\cal M}_n^{\prime p} $ & $p$ &  $\eta_0$ & $\sigma_0$ & $g_s$ & ${\cal C}_{w_5}$ & ${\cal C}_{\rm up}$ & $\langle{\cal V}\rangle$ & $\langle t^2 \rangle$ & $\langle\phi^1\rangle$ & $\langle\phi^2\rangle$\\
%& & & & & & & & & & \\
\hline
\hline
%& & & & & & & & & & \\
${\cal M}_2^{\prime 4/3}$ & 4/3 & 2 & 0 & 0.280 & 0.355 & $2.520\cdot 10^{-5}$ & 989.74 & 1.096 & 5.915 & -3.423 \\
%& & & & & & & & & & \\
${\cal M}_2^{\prime 2}$ & 2 &  2 & 0 & 0.298 & 0.33 & $4.015\cdot 10^{-3}$ & 1129.05 & 1.144 & 6.022 & -3.425\\
%& & & & & & & & & & \\
${\cal M}_2^{\prime 8/3}$ & 8/3 & 6 & -4 & 0.295 & 0.33 & $3.86422$ & 1123.23 & 1.150 & 6.018 & -3.413\\
%& & & & & & & & & & \\
\hline
\end{tabular}
\caption{ACTivated two-field models ${\cal M}_2^{\prime p}$: Moduli stabilization and uplifting}
\label{tab_two-field-models1-ACT}
\end{table}

\begin{table}[H]
\centering
\begin{tabular}{|c|c|c||c|c|c|c||c|c|c|c|c|}
\hline
% & & & & & & & & & & & \\
${\cal M}_n^p $ & ${\cal V}^\ast$ & $(t^2)^\ast$ & $\phi^{1\ast}$ & $\phi^{2\ast}$ & $\Delta\phi$ & $N$ & $N^\ast$ & $10^9 P_s^\ast$ & $n_s^\ast$ & $10^4\alpha_s$& $10^3 r^\ast$ \\
% & & & &  & & & & & & & \\
\hline
\hline
% & & & & & & & & & & & \\
${\cal M}_2^{\prime4/3}$ & 1062.27 & 27.70 & 5.973 & 2.13 & 5.55 & 54.5 &  4.5 & 2.12 & 0.979 & 0.974 & 4.47 \\
% & & & & & & & & & & & \\
${\cal M}_2^{\prime2}$ & 1257.77 & 25.22 & 6.110 & 1.87 & 5.30 & 53.4 & 3.5 & 2.11 & 0.976 & 1.026 & 2.72 \\
% & & & & & & & & & & & \\
${\cal M}_2^{\prime8/3}$ & 1258.18 & 25.66 & 6.111 & 1.90 & 5.30 & 55.0 & 5.0 & 2.09 & 0.976 & -5.738 & 2.73 \\
%& & & & & & & & & & & \\
\hline
\end{tabular}
\caption{ACTivated two-field models ${\cal M}_2^{\prime p}$: Cosmological observables}
\label{tab_two-field-models2-ACT}
\end{table}

%%%%%%%%%%%%%%%%%%%%%%%%%%%%%%%%%%%%%%%%%%%%%%%%%%%%%%%%%%%%%%%%%%%%%%%%%%%%%%%%%%%%%%%%%%

\section{Assisted Fibre Inflation in Perturbative LVS}
\label{sec_assisted-fibre}
%%%%%%%%%%%%%%%%%%%%%%%%%%%%%%%%%%%%%%%%%%%%%%%%%%%%%%%%%%%%%%%%%%%%%%%%%%%%%%%%%%%%%%%%%%

From the detailed analysis presented in the previous section, we find that standard fibre inflation models typically need $\Delta\phi \simeq 6$M$_p$ while the ACTivated ones require  relatively smaller values of the effective inflaton shifts for successfully realizing the cosmological observables. In this section we present the three-field evolution analysis, where the overall volume ${\cal V}$ remains at the perturbative LVS minimum while the other two moduli assist in driving the multi-field fibre inflation. 

\subsection{Ingredients for three-field evolution}
In our approach, we adopt the real field basis $\Phi^a = \{{\cal V}, t^2, t^3\}$ and employ the symmetry of the CY threefold so that $t^2$ and $t^3$ drive inflation, while ${\cal V}$ stays largely fixed at its perturbative LVS minimum.
Using the leading-order tree-level K\"ahler metric (\ref{eq:Kij-tree}), the field space metric and its inverse metric are given by,
\bea
\label{eq:metric-phia}
& & \hskip-1.5cm {\cal G}_{ab} = \left(\begin{array}{ccc}
\frac{1}{{\cal V}^2} &  -\frac{1}{2 t^2 {\cal V}} & -\frac{1}{2 t^3 {\cal V}} \\
& & \\
 -\frac{1}{2 t^2 {\cal V}} & \frac{1}{(t^2)^2} & \frac{1}{2 t^2 t^3} \\
& & \\
 -\frac{1}{2 t^3 {\cal V}} &  \frac{1}{2 t^2 t^3} & \frac{1}{(t^3)^2} \\
\end{array}
\right), \qquad {\cal G}^{ab} = \left(\begin{array}{ccc}
\frac{3{\cal V}^2}{2} & \frac{{\cal V}t^2}{2} & \frac{{\cal V}t^3}{2} \\
& & \\
 \frac{{\cal V}t^2}{2} & \frac{3 (t^2)^2}{2} & -\frac{t^2t^3}{2} \\
& & \\
 \frac{{\cal V}t^3}{2} &  -\frac{t^2t^3}{2}  & \frac{3 (t^3)^2}{2} \\
\end{array}
\right)
\eea
Subsequently, the non-vanishing Christoffel connections $\Gamma^a_{bc}$ are given as below,
\bea
\label{eq:Christofell}
& & \Gamma_{11}^1 = - \frac{1}{\cal V}, \qquad \Gamma_{22}^2 = - \frac{1}{t^2}, \qquad \Gamma_{33}^3 = - \frac{1}{t^3}.
\eea
The scalar potential presented in (\ref{eq:Vfinal-simp1}) takes the following form,
\bea
\label{eq:Vfinal-simp3}
& & \hskip-0.75cm V \equiv V({\cal V}, t^2, t^3) = \frac{{\cal C}_{\rm up}}{{\cal V}^p} +  \frac{{\cal C}_1}{{\cal V}^3} \left(\hat\xi + 2\,\hat\eta \, \ln{\cal V} - 8\,\hat\eta + 2\,\hat\sigma \right) \\
& & \hskip-0.35cm - \frac{{\cal C}_2}{{\cal V}^3} \biggl(2\,{\cal C}_{w_1} \frac{t^2 \,t^3}{{\cal V}} +  \frac{{\cal C}_{w_2}}{t^2} + \frac{{\cal C}_{w_3}}{t^3} + \frac{{\cal C}_{w_4}\,t^2 t^3}{{\cal V} + 2 (t^2)^2 t^3} +  \frac{{\cal C}_{w_5}}{2(t^2+t^3)} + \frac{{\cal C}_{w_6}\,t^2 t^3}{{\cal V} + 2 t^2 (t^3)^2} \biggr)\nonumber\\
& &  \hskip-0.35cm  \, +  \frac{{\cal C}_3}{{\cal V}^3}\,\left(\frac{1}{2 \, t^2 \, t^3} + \frac{t^2}{\cal V} + \frac{t^3}{\cal V} \right) + \cdots,\nonumber
\eea
We now present benchmark numerical models that stabilize the moduli in a tachyon-free de Sitter vacuum within the perturbative LVS framework.
%Now we present a couple of benchmark numerical models stabilizing the moduli in tachyon-free de-Sitter minimum in perturbative LVS.
We can observe that the exchange symmetry is maintained among the $t^2$ and $t^3$ moduli which follows from the symmetry of underlying CY threefold. This symmetry will be exploited for creating an assisted track for driving inflation. As earlier, we will consider the three popular classes of uplifting schemes by simply characterizing them via the contributions to the scalar potential correspond to $p = 4/3$ for anti-D3 uplifting \cite{Kachru:2003aw,Crino:2020qwk,Cicoli:2017axo,AbdusSalam:2022krp}, and $p = 2$ for D-term uplifting \cite{Burgess:2003ic,Achucarro:2006zf,Braun:2015pza} while  $p = 8/3$ for the T-brane uplifting \cite{Cicoli:2015ylx,Cicoli:2017shd}.

Using the metric and Christoffel connections derived above, the field equations for ${{\cal V}, t^2, t^3}$ obtained from Eq.~(\ref{eq:EOM2}) take the following explicit form:
%Using these pieces of information about the metric and Christoffel connections, the field equations for the three fields, namely $\{{\cal V}, t^2, t^3\}$, arising from Eq.~(\ref{eq:EOM2}) are explicitly given as below,
\bea
\label{eq:Explicit-EOMs}
& & \hskip-1cm {\cal V}^{\prime\prime} = \frac{{\cal V}^{\prime2}}{{\cal V}} - \left(3- \epsilon\right) \left({\cal V}^{\prime}+\frac{3{\cal V}^2}{2V} \partial_{\cal V} V +\frac{{\cal V} \, t^2}{2V}\partial_{t^2} V +\frac{{\cal V} \, t^3}{2V}\partial_{t^3} V \right),\\
& & \hskip-1cm (t^2)^{\prime\prime} = \frac{(t^2)^{\prime2}}{t^2} - \left(3- \epsilon\right) \left((t^2)^{\prime}+\frac{{\cal V} \, t^2}{2V}\partial_{{\cal V}} V +\frac{3(t^2)^2}{2V} \partial_{t^2} V -\frac{t^2 t^3}{2V}\partial_{t^3} V\right),\nonumber\\
& & \hskip-1cm (t^3)^{\prime\prime} = \frac{(t^3)^{\prime2}}{t^3} - \left(3- \epsilon\right) \left((t^3)^{\prime}+\frac{{\cal V} \, t^3}{2V}\partial_{{\cal V}} V -\frac{t^2 t^3}{2V}\partial_{t^2} V +\frac{3(t^3)^2}{2V} \partial_{t^3} V \right).\nonumber
\eea
Here as earlier, the prime $^\prime$ denotes derivatives w.r.t.~the number of $e$-folds $N$, i.e. ${\cal V}^\prime = \frac{d{\cal V}}{dN}$ etc. and now the inflationary parameter $\epsilon$ takes the following explicit form,
\bea
& & \hskip-1cm \epsilon = \frac{1}{2} \left(\frac{{\cal V}^{\prime2}}{{\cal V}^2} + \frac{(t^2)^{\prime2}}{(t^2)^2}  + \frac{(t^3)^{\prime2}}{(t^3)^2} - \frac{{\cal V}^\prime \, (t^2)^{\prime}}{{\cal V} \, t^2} - \frac{{\cal V}^\prime \, (t^3)^{\prime}}{{\cal V} \, t^3} + \frac{(t^2)^\prime \, (t^3)^{\prime}}{t^2\, t^3} \right).
\eea
Note the the underlying symmetry $2 \leftrightarrow 3$ is reflected in the set of three-field equations as well. 

\subsubsection*{Computation of the effective inflaton shift $\Delta\phi$}

%Using a set of canonical fields $\{\phi^a\}$ taking part in the three-field inflationary dynamics, the effective inflaton shift required to  realize successful inflationary predictions can be determined by the flat-space distance formula between the two points in the three-dimensional moduli space given as,
The effective inflaton shift needed for phenomenologically viable inflation can be computed as the geodesic distance in the flat three-dimensional field space spanned by canonical fields ${\phi^a}$ between the initial and final field configurations:
\bea
\label{eq:shiftphi-three-field}
& & \Delta \phi = \sqrt{\left(\phi^{1\ast} - \langle\phi^1\rangle\right)^2 + \left(\phi^{2\ast} - \langle\phi^2\rangle\right)^2 + \left(\phi^{3\ast} - \langle\phi^3\rangle\right)^2}~.
\eea
Here $\phi^{a\ast}$ corresponds to the horizon exit while $\langle \phi^a\rangle$ denotes the moduli VEVs at the LVS minimum. As argued earlier, for any choice of canonical field basis, the inflaton shift $\Delta\phi$ remains invariant. However, the individual fields may result in different shifts in a different choice of basis. 
%As an example, we demonstrate the argument by considering three choice of canonical bases relevant for different purposes.
To exemplify this, we analyze three canonical bases, each serving a distinct purpose.
\begin{itemize}

\item 
Considering the tree-level K\"ahler metric arising from the volume form ${\cal V} = 2 t^1 t^2 t^3$, one obtains the following basis of canonical normalized fields $\{\varphi^a\}$ corresponding to the $4$-cycle volume moduli $\{\tau_1, \tau_2, \tau_3\}$,
\bea
\label{eq:cononical-varphi0}
& & \varphi^a = \frac{1}{\sqrt{2}} \ln \tau_a, \qquad \forall \, a \in \{1, 2, 3\}.
\eea
%This canonical basis $\varphi^a$ respects the exchange symmetry $1 \leftrightarrow 2 \leftrightarrow 3$, however it does not have the overall volume ${\cal V}$ as a singled our direction. 
This canonical basis $\varphi^a$ respects the exchange symmetry $1 \leftrightarrow 2 \leftrightarrow 3$, but it does not treat the overall volume $\mathcal{V}$ as a singled-out direction.
%This is needed for manifestation of the mass hierarchy between the overall volume and the remaining two volume moduli.
This is required to establish the mass hierarchy between the overall volume modulus and the other two volume moduli.

\item 
Following the conventions of \cite{Antoniadis:2020stf}, one can define another canonical basis $\{\psi^a\}$, given as below
%now we define the another canonical basis of fields $\phi^\alpha$ for exploring inflationary aspects given as below,
\bea
\label{eq:cononical-varphi1}
& & \psi^1 = \frac{1}{\sqrt{3}} \left(\varphi^1+ \varphi^2 + \varphi^3 \right) = \sqrt{\frac{2}{3}} \ln(\sqrt{2}\,{\cal V}) , \\
& & \psi^2 = \frac{1}{\sqrt{2}} \left(\varphi^1- \varphi^2 \right), \quad \psi^3 = \frac{1}{\sqrt{6}} \left(\varphi^1+ \varphi^2  -2 \varphi^3 \right)\,, \nonumber
\eea
which is equivalent to
\bea
& & \hskip-1cm {\cal V} = \frac{1}{\sqrt2}e^{\sqrt{\frac{3}{2}} \psi^1}, \quad t^2 = \frac{1}{\sqrt2}e^{\frac{1}{\sqrt6} \psi^1+\psi^2-\frac{1}{\sqrt3}\psi^3}, \quad t^3 = \frac{1}{\sqrt2}e^{\frac{1}{\sqrt6}\psi^1+\frac{2}{\sqrt3}\psi^3}.
\eea
This basis of canonical fields involves the overall volume through $\psi^1$, however the exchange symmetry $2 \leftrightarrow 3$ between the remaining two fields $\psi^2$ and $\psi^3$ is absent. 

\item 
In order to keep the symmetry $2 \leftrightarrow 3$ intact while still keeping the overall volume in the basis, one can choose the canonical fields such that
\bea
\label{eq:cononical-varphi3}
& & \phi^1 = \frac{1}{\sqrt{3}} \left(\varphi^1+ \varphi^2 + \varphi^3 \right) = \sqrt{\frac{2}{3}} \ln(\sqrt{2}\,{\cal V}) , \\
& & \phi^2 = \frac{1}{6} \left(2\sqrt{3}\, \varphi^1 + (3 -\sqrt{3})\varphi^2-(3+\sqrt{3} \phi^3) \right), \nonumber\\
& & \phi^3 = \frac{1}{6} \left(2\sqrt{3}\, \varphi^1 - (3 +\sqrt{3})\varphi^2 + (3 - \sqrt{3}) \phi^3 \right),\nonumber
\eea
which is equivalent to
\bea
\label{eq:cononical-varphi4}
& & \hskip-1cm {\cal V} = \frac{1}{\sqrt2}e^{\sqrt{\frac{3}{2}} \phi^1}, \quad t^2 = \frac{1}{\sqrt2}e^{\frac{1}{\sqrt6} \left(\phi^1 -\left(1 -{\sqrt3}\right)\phi^2+\left(1+{\sqrt3}\right)\right)\phi^3}, \\
& & \hskip-1cm t^3 = \frac{1}{\sqrt2}e^{\frac{1}{\sqrt6} \left(\phi^1 +\left(1 +{\sqrt3}\right)\phi^2 -\left(1 - {\sqrt3}\right)\right)\phi^3}. \nonumber
\eea
This basis indeed has the exchange symmetry $2 \leftrightarrow 3$ between the remaining two fields $\phi^2$ and $\phi^3$ while $\phi^1$ is determined by the overall volume and always remains the heavier than the other two fields. We will work in this basis for our numerical analysis.
    
\end{itemize}

\subsection{Numerical benchmark models}
%For appropriately chosen initial conditions one can numerically solve the second order differential equations (\ref{eq:Explicit-EOMs}), and subsequently can realize the desired cosmological observables for similar benchmark models as presented in the previous section. 
With properly selected initial conditions, we numerically solve the coupled second-order differential equations (\ref{eq:Explicit-EOMs}) to obtain cosmological observables consistent with benchmark models from the previous section. As mentioned earlier we will work with the basis $\Phi^a=\{{\cal V}, t^2, t^3\}$ and will subsequently use the canonical basis presented in (\ref{eq:cononical-varphi4}).

\subsubsection{Class-I: Standard cosmological observables}

In order to realize the standard cosmological observables consistent with the PLANCK 2018 data \cite{Planck:2018jri,Planck:2018vyg}, similar to the previous single- and two-field analysis, we consider the following numerical models ${\cal M}_3^p$ where the first block collects the model dependent parameters, the second block has information on the moduli stabilization and de Sitter realization while the third block presents the inflationary results.

\subsubsection*{Model ${\cal M}_3^{4/3}$:}
\bea
\label{eq:model-A1}
& & \hskip-1.5cm p = 4/3, \qquad \chi({\rm CY}) = -224, \qquad \eta_0 = 2, \qquad \sigma_0 = 0, \qquad g_s = 0.275,  \\
& & \hskip-1.5cm |W_0| = 5, \qquad {\cal C}_{w_1} = 0.0008, \qquad  {\cal C}_{w_2} = -0.0008, \qquad  {\cal C}_{w_3} = -0.0008, \nonumber\\
& & \hskip-1.5cm {\cal C}_{w_4} = -0.02, \qquad {\cal C}_{w_5} = 0.4, \qquad  {\cal C}_{w_6} = -0.02, \qquad \lambda = - 0.0001; \nonumber\\
& & \nonumber\\
& & \hskip-1.5cm {\cal C}_{\rm up} = 2.27101\cdot10^{-5}, \qquad \langle {\cal V} \rangle = 1036.31, \qquad \langle t^2 \rangle = 0.581, \qquad \langle t^3 \rangle = 0.581, \nonumber\\
& & \hskip-1.5cm \langle \phi^1 \rangle = 5.952, \quad \langle \phi^2 \rangle = - 3.21655 , \quad \langle \phi^3 \rangle = -3.21655, \nonumber\\
& & \nonumber\\
& & \hskip-1.5cm {\cal V}^\ast = 1195.03, \qquad (t^2)^\ast = 22.198, \quad  (t^3)^\ast = 22.198, \nonumber\\
& & \hskip-1.5cm \phi^{1\ast} = 6.069, \quad \phi^{2\ast} = 1.187, \quad \phi^{3\ast} = 1.187, \quad \Delta\phi = 6.23, \quad N = 60.2,\nonumber\\
& & \hskip-1.5cm N^\ast = 10.2, \quad P_s^\ast = 2.13 \cdot 10^{-9}, \quad n_s^\ast = 0.964, \quad \alpha_s^\ast = - 6.98\cdot10^{-4}, \quad r^\ast = 6.58\cdot10^{-3}.\nonumber
\eea

\subsubsection*{Model ${\cal M}_3^{2}$:}
\bea
\label{eq:model-B1}
& & \hskip-1.5cm p = 2, \qquad \chi({\rm CY}) = -224, \qquad \eta_0 = 2, \qquad \sigma_0 = 0, \qquad g_s = 0.3,   \\
& & \hskip-1.5cm |W_0| = 5.6, \qquad {\cal C}_{w_1} = 0.0008, \qquad  {\cal C}_{w_2} = -0.0008, \qquad  {\cal C}_{w_3} = -0.0008,    \nonumber\\
& & \hskip-1.5cm {\cal C}_{w_4} = -0.02, \qquad {\cal C}_{w_5} = 0.4, \qquad  {\cal C}_{w_6} = -0.02, \qquad \lambda = - 0.0001; \nonumber\\
& & \nonumber\\
& & \hskip-1.5cm {\cal C}_{\rm up} = 5.38229 \cdot10^{-3} \qquad \langle {\cal V} \rangle = 1066.7, \qquad \langle t^2 \rangle = 0.698, \qquad \langle t^3 \rangle = 0.698, \nonumber\\
& & \hskip-1.5cm \langle \phi^1 \rangle = 5.97586, \quad \langle \phi^2 \rangle = -3.00448  , \quad \langle \phi^3 \rangle = -3.00448 , \nonumber\\
& & \nonumber\\
& & \hskip-1.5cm {\cal V}^\ast = 1334.76, \qquad (t^2)^\ast \rangle = 22.349, \quad  (t^3)^\ast = 22.349, \nonumber\\
& & \hskip-1.5cm \phi^{1\ast} = 6.1589, \quad \phi^{2\ast} = 1.5 , \quad \phi^{3\ast} = 1.5, \quad \Delta\phi = 5.88, \quad N = 54.5,\nonumber\\
& & \hskip-1.5cm N^\ast = 4.5, \quad P_s^\ast = 2.10 \cdot 10^{-9}, \quad n_s^\ast = 0.966, \quad \alpha_s^\ast = -6.30\cdot 10^{-4}, \quad r^\ast = 6.46\cdot10^{-3}.\nonumber
\eea

\subsubsection*{Model ${\cal M}_3^{8/3}$:}
\bea
\label{eq:model-B1}
& & \hskip-1.5cm p = 8/3, \qquad \chi({\rm CY}) = -224, \qquad \eta_0 = 6, \qquad \sigma_0 = -4, \qquad g_s = 0.3, \\
& & \hskip-1.5cm  |W_0| = 5.7, \qquad {\cal C}_{w_1} = 0.0008, \qquad  {\cal C}_{w_2} = -0.0008, \qquad  {\cal C}_{w_3} = -0.0008,    \nonumber\\
& & \hskip-1.5cm {\cal C}_{w_4} = -0.02, \qquad {\cal C}_{w_5} = 0.4, \qquad  {\cal C}_{w_6} = -0.02, \qquad \lambda = - 0.0001; \nonumber\\
& & \nonumber\\
& & \hskip-1.5cm {\cal C}_{\rm up} = 5.20572, %\quad \langle V \rangle =  9.26 \cdot 10^{-23}, 
\quad \langle {\cal V} \rangle = 1087.37, \quad \langle t^2 \rangle = 0.723, \quad \langle t^3 \rangle = 0.723, \nonumber\\
& & \hskip-1.5cm \langle \phi^1 \rangle = 5.99152, \quad \langle \phi^2 \rangle = -2.96905  , \quad \langle \phi^3 \rangle = -2.96905 , \nonumber\\
& & \nonumber\\
& & \hskip-1.5cm {\cal V}^\ast = 1398.06, \qquad (t^2)^\ast \rangle = 23.836, \quad  (t^3)^\ast = 23.836, \nonumber\\
& & \hskip-1.5cm \phi^{1\ast} = 6.19673, \quad \phi^{2\ast} = 1.21 , \quad \phi^{3\ast} = 1.21, \quad \Delta\phi = 5.91, \quad N = 58.4,\nonumber\\
& & \hskip-1.5cm N^\ast = 8.4, \quad P_s^\ast = 2.07\cdot 10^{-9}, \quad n_s^\ast = 0.966, \quad \alpha_s^\ast = -6.753\cdot10^{-4}, \quad r^\ast = 6.13\cdot10^{-3}.\nonumber
\eea

\subsubsection{Class-II: ACTivated cosmological observables}
We find that the numerical models with slightly larger values of the spectral index parameter $n_s = 0.9743 \pm 0.0034$ as suggested by the ACT observations \cite{ACT:2025tim,ACT:2025fju,DESI:2024mwx, Frolovsky:2025iao} can be easily embedded with a slight change of the numerical values of the parameter space. 

\subsubsection*{Model ${\cal M}_3^{\prime4/3}$:}
\bea
\label{eq:model-M3-1-ACT}
& & \hskip-1.5cm p = 4/3, \qquad \chi({\rm CY}) = -224, \qquad \eta_0 = 2, \qquad \sigma_0 = 0, \qquad g_s = 0.28,  \\
& & \hskip-1.5cm |W_0| = 5, \qquad {\cal C}_{w_1} = 0.001, \qquad  {\cal C}_{w_2} = -0.0008, \qquad  {\cal C}_{w_3} = -0.0008, \nonumber\\
& & \hskip-1.5cm {\cal C}_{w_4} = -0.1, \qquad {\cal C}_{w_5} = 0.355, \qquad  {\cal C}_{w_6} = -0.1, \qquad \lambda = - 0.00017; \nonumber\\
& & \nonumber\\
& & \hskip-1.5cm {\cal C}_{\rm up} = 2.51989\cdot10^{-5}, \qquad \langle {\cal V} \rangle = 989.736, \qquad \langle t^2 \rangle = 1.0964, \qquad \langle t^3 \rangle = 1.0964, \nonumber\\
& & \hskip-1.5cm \langle \phi^1 \rangle = 5.91471, \quad \langle \phi^2 \rangle = -2.42019  , \quad \langle \phi^3 \rangle = - 2.42019 , \nonumber\\
& & \nonumber\\
& & \hskip-1.5cm {\cal V}^\ast = 1062.25, \qquad (t^2)^\ast \rangle = 27.5612, \quad  (t^3)^\ast = 27.5612, \nonumber\\
& & \hskip-1.5cm \phi^{1\ast} = 5.97244, \quad \phi^{2\ast} = 1.5 , \quad \phi^{3\ast} = 1.5, \quad \Delta\phi = 5.55, \quad N = 54.1,\nonumber\\
& & \hskip-1.5cm N^\ast = 4.1, \quad P_s^\ast = 2.12 \cdot 10^{-9}, \quad n_s^\ast = 0.9798, \quad \alpha_s^\ast = 2.157\cdot10^{-4}, \quad r^\ast = 4.46\cdot10^{-3}.\nonumber
\eea

\subsubsection*{Model ${\cal M}_3^{\prime2}$:}
\bea
\label{eq:model-M3-2-ACT}
& & \hskip-1.5cm p = 2, \qquad \chi({\rm CY}) = -224, \qquad \eta_0 = 2, \qquad \sigma_0 = 0, \qquad g_s = 0.298,   \\
& & \hskip-1.5cm |W_0| = 5, \qquad {\cal C}_{w_1} = 0.001, \qquad  {\cal C}_{w_2} = -0.0008, \qquad  {\cal C}_{w_3} = -0.0008,    \nonumber\\
& & \hskip-1.5cm {\cal C}_{w_4} = -0.1, \qquad {\cal C}_{w_5} = 0.33, \qquad  {\cal C}_{w_6} = -0.1, \qquad \lambda = - 0.00017; \nonumber\\
& & \nonumber\\
& & \hskip-1.5cm {\cal C}_{\rm up} = 5.38229\cdot10^{-3} \qquad \langle {\cal V} \rangle = 1129.05, \qquad \langle t^2 \rangle = 1.14437, \qquad \langle t^3 \rangle = 1.14437, \nonumber\\
& & \hskip-1.5cm \langle \phi^1 \rangle = 6.02224, \quad \langle \phi^2 \rangle = -2.42149  , \quad \langle \phi^3 \rangle = -2.42149 , \nonumber\\
& & \nonumber\\
& & \hskip-1.5cm {\cal V}^\ast = 1257.93, \qquad (t^2)^\ast \rangle = 25.798, \quad  (t^3)^\ast = 25.798, \nonumber\\
& & \hskip-1.5cm \phi^{1\ast} = 6.1105, \quad \phi^{2\ast} = 1.35, \quad \phi^{3\ast} = 1.35, \quad \Delta\phi = 5.36, \quad N = 55.6,\nonumber\\
& & \hskip-1.5cm N^\ast = 5.6, \qquad P_s^\ast = 2.12 \cdot 10^{-9}, \quad n_s^\ast = 0.976, \quad \alpha_s^\ast = -5.808 \cdot10^{-4}, \quad r^\ast = 2.71\cdot10^{-3}.\nonumber
\eea

\subsubsection*{Model ${\cal M}_3^{\prime8/3}$:}
\begin{figure}[H]
\centering
\includegraphics[width=16.5cm]{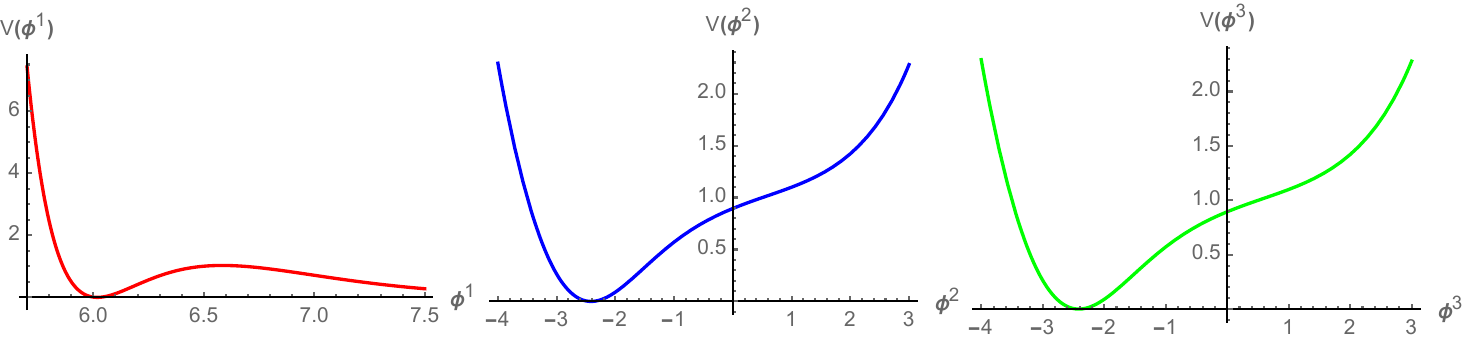}
\caption{One dimensional plot of $V\cdot10^{10}$ while keeping the other two moduli at their minimum}
\label{fig_pot-three-field}
\end{figure}

\bea
\label{eq:model-M3-3-ACT}
& & \hskip-1.5cm p = 8/3, \qquad \chi({\rm CY}) = -224, \qquad \eta_0 = 6, \qquad \sigma_0 = -4, \qquad g_s = 0.295, \\
& & \hskip-1.5cm  |W_0| = 5, \qquad {\cal C}_{w_1} = 0.001, \qquad  {\cal C}_{w_2} = -0.0008, \qquad  {\cal C}_{w_3} = -0.0008,    \nonumber\\
& & \hskip-1.5cm {\cal C}_{w_4} = -0.1, \qquad {\cal C}_{w_5} = 0.33, \qquad  {\cal C}_{w_6} = -0.1, \qquad \lambda = - 0.00017; \nonumber\\
& & \nonumber\\
& & \hskip-1.5cm {\cal C}_{\rm up} = 5.32455, %\quad \langle V \rangle =  9.26 \cdot 10^{-23}, 
\quad \langle {\cal V} \rangle = 1123.23, \quad \langle t^2 \rangle = 1.14996, \quad \langle t^3 \rangle = 1.14996, \nonumber\\
& & \hskip-1.5cm \langle \phi^1 \rangle = 6.01802, \quad \langle \phi^2 \rangle = -2.41341  , \quad \langle \phi^3 \rangle = -2.41341 , \nonumber\\
& & \nonumber\\
& & \hskip-1.5cm {\cal V}^\ast = 1258.22, \qquad (t^2)^\ast \rangle = 25.8, \quad  (t^3)^\ast = 25.8, \nonumber\\
& & \hskip-1.5cm \phi^{1\ast} = 6.11069, \quad \phi^{2\ast} = 1.35 , \quad \phi^{3\ast} = 1.35, \quad \Delta\phi = 5.32, \quad N = 55.5,\nonumber\\
& & \hskip-1.5cm N^\ast = 5.5, \qquad P_s^\ast =  2.095\cdot 10^{-9}, \quad n_s^\ast = 0.9763, \quad \alpha_s^\ast = -5.763\cdot10^{-4}, \quad r^\ast = 2.73\cdot10^{-3}.\nonumber
\eea
The scalar potential $V(\phi^a)$ can be plotted for each of the three moduli while keeping the other two fixed at their minimum. This is shown in figure \ref{fig_pot-three-field}.

\noindent

\begin{figure}[H]
\centering
\includegraphics[width=13.5cm]{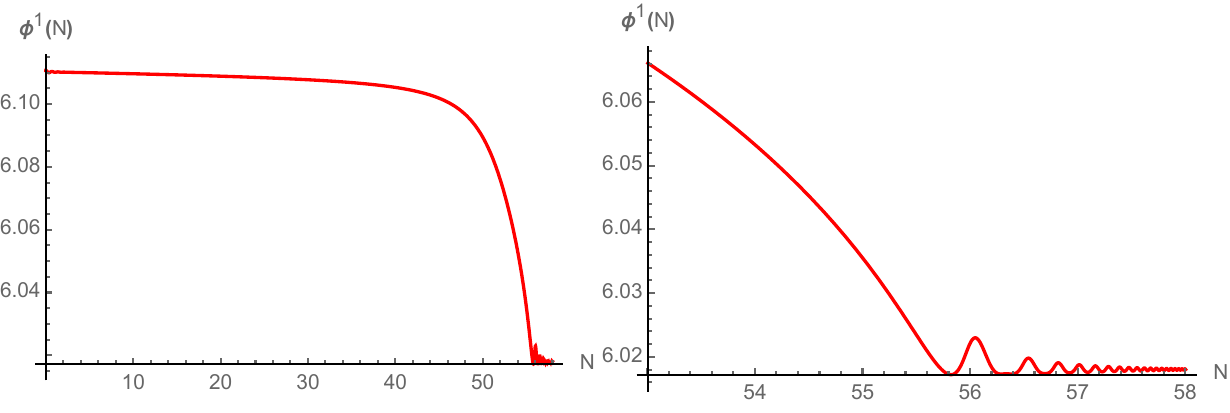}
\caption{Evolution of $\phi^1(N)$}
\label{fig_phi1-three-field}
\end{figure}

\begin{figure}[H]
\centering
\includegraphics[width=13.5cm]{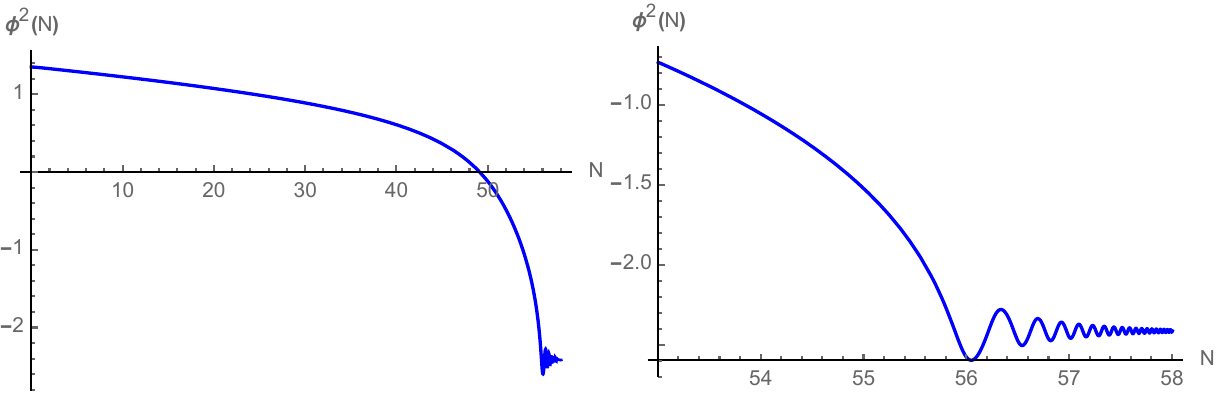}
\caption{Evolution of $\phi^2(N)$}
\label{fig_phi2-three-field}
\end{figure}

\begin{figure}[H]
\centering
\includegraphics[width=13.5cm]{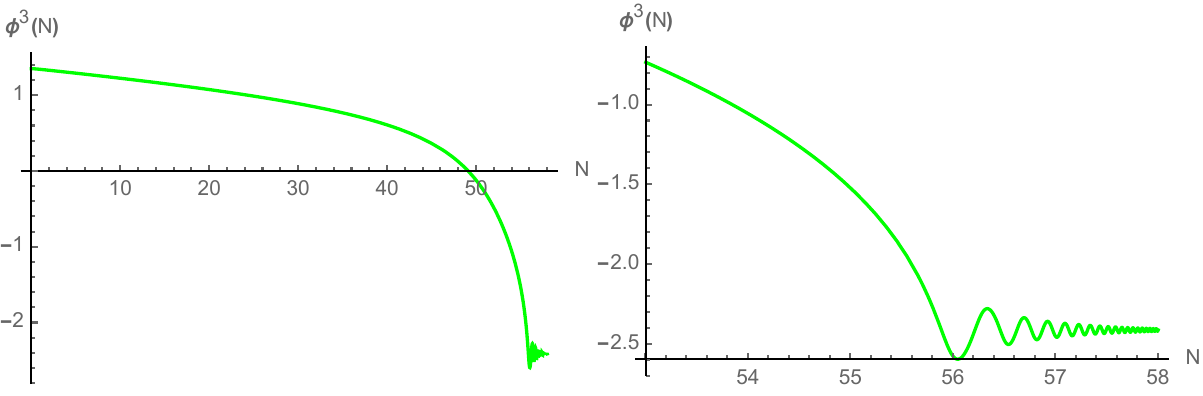}
\caption{Evolution of $\phi^3(N)$}
\label{fig_phi3-three-field}
\end{figure}

\begin{figure}[H]
\centering
\includegraphics[width=13.5cm]{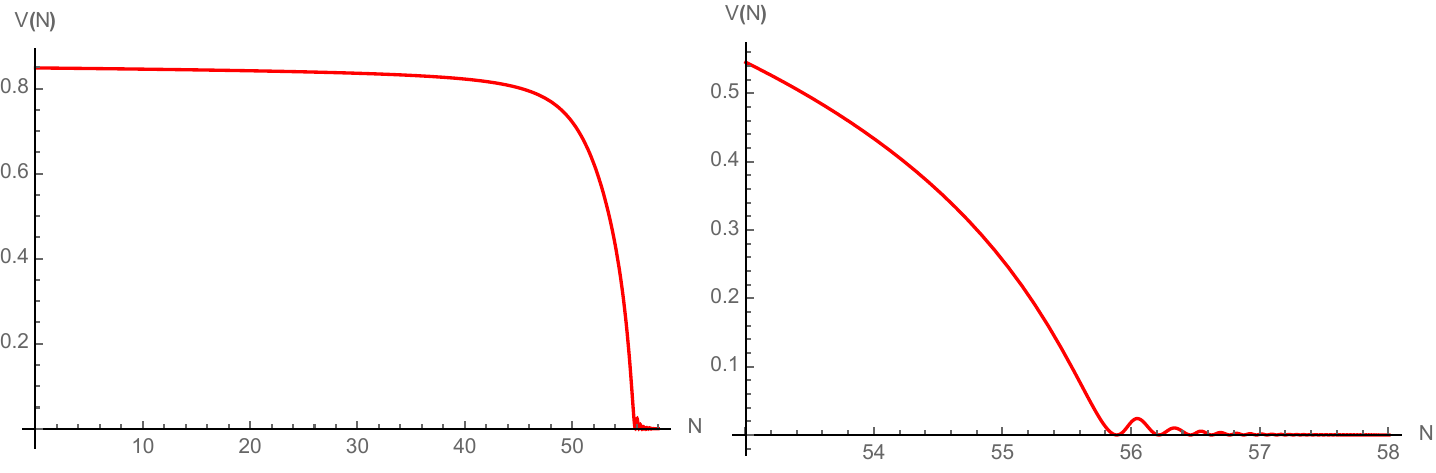}
\caption{Evolution of the scalar potential $V(N)\cdot10^{10}$ plotted for the number of efoldings}
\label{fig_pot-N-three-field}
\end{figure}

\noindent
After solving the evolution equations (\ref{eq:EOM2}) the canonical field evolutions for $\phi^1$, $\phi^2$ and $\phi^3$ are shown in figure {\ref{fig_phi1-three-field}}, figure {\ref{fig_phi2-three-field}} and figure \ref{fig_phi3-three-field} respectively. As anticipated we observe  that the two fields $\phi^2$ and $\phi^3$ have identical evolution as well as minimum due to the underlying symmetry of the CY threefold itself. Moreover, using the evolution of various fields $\phi^a(N)$ the full scalar potential (\ref{eq:Vfinal-simp3}) evolves as shown in figure \ref{fig_pot-N-three-field}.

\subsubsection*{Evolution of cosmological observables}
\begin{figure}[H]
\centering
\includegraphics[width=14cm]{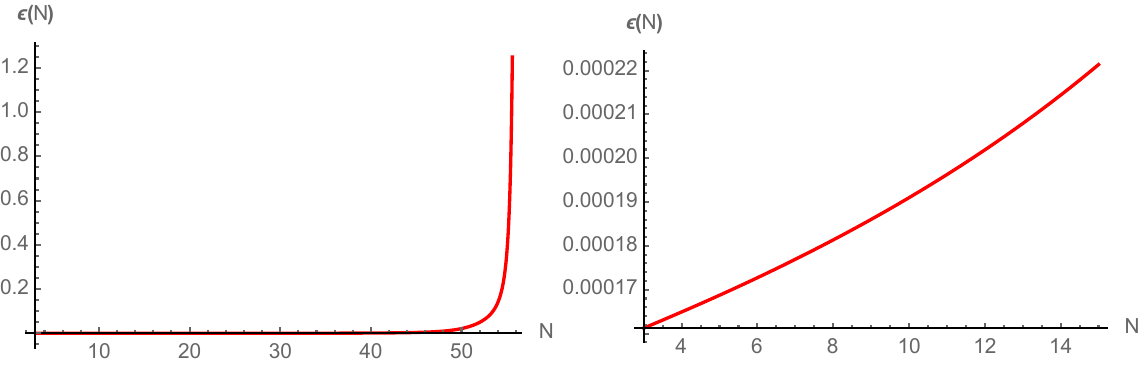}
\caption{Evolution of slow-roll parameter $\epsilon(N)$}
\label{fig_epsilon-three-field}
\end{figure}

\begin{figure}[H]
\centering
\includegraphics[width=14cm]{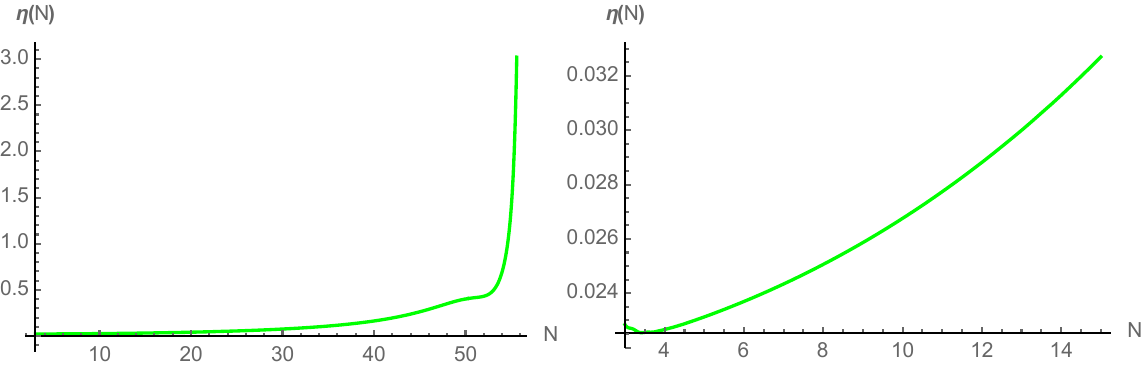}
\caption{Evolution of slow-roll parameter $\eta(N)$}
\label{fig_eta-three-field}
\end{figure}

\begin{figure}[H]
\centering
\includegraphics[width=14cm]{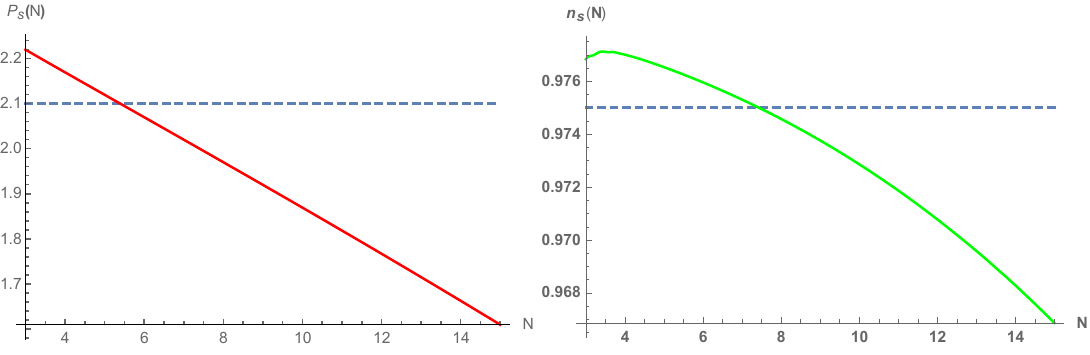}
\caption{Evolution of  $P_s(N)\cdot10^9$ and $n_s(N)$ with dashed lines for $P_s = 2.1 \cdot 10^{-9}$ and $n_s = 0.975$.}
\label{fig_Ps-ns-three-field}
\end{figure}

\begin{figure}[H]
\centering
\includegraphics[width=14cm]{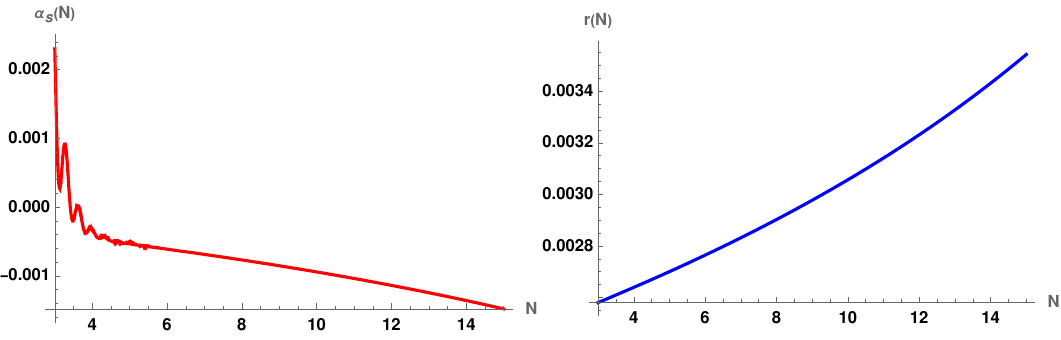}
\caption{Evolution of $\alpha_s(N)$ and  $r(N)$ %where the solid plots follow from the definition (\ref{eq:cosmo-observables}) while the dotted plot corresponds to their 
using definitions in (\ref{eq:cosmo-observables1}) and $\phi\equiv\phi(N)$.}
\label{fig_r-three-field}
\end{figure}

Using the evolution of canonical fields one can compute the evolution of slow-roll parameters $\epsilon(N)$ and $\eta(N)$ are presented in figure \ref{fig_epsilon-three-field} and figure \ref{fig_eta-three-field} respectively. Subsequently, the evolution of various cosmological observables such as scalar power spectrum amplitude $P_s(N)$ and the spectral index $n_s(N)$ is presented in figure \ref{fig_Ps-ns-three-field} while the tensor-to-scalar ratio $r(N)$ is presented in figure \ref{fig_r-three-field}. Further the horizon exit corresponds to $N^\ast = 5.5$ at which the various observations are matched with their respective experimentally consistent values.

\subsection{Demonstrating the assisted nature of inflation}
\begin{figure}[H]
\centering
\includegraphics[width=16.5cm]{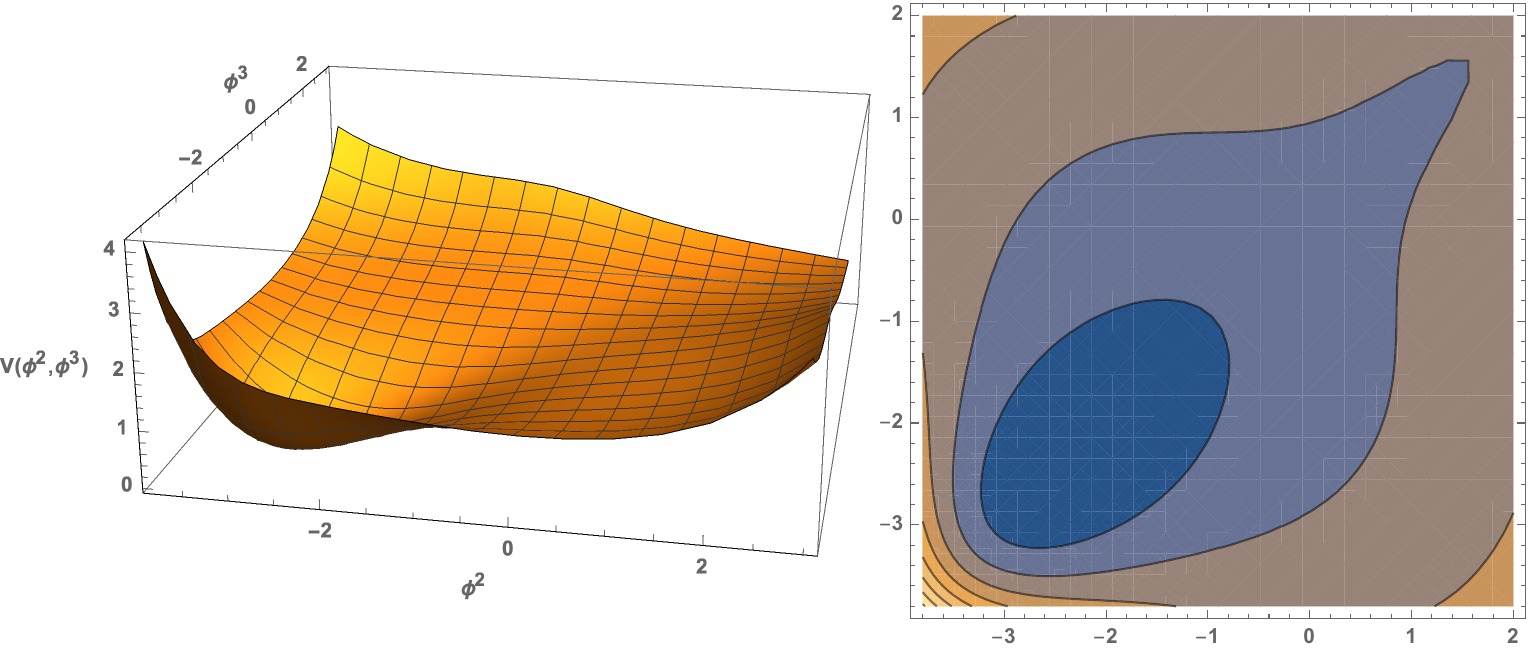}
\caption{Assisted inflationary track in the $(\phi^2, \phi^3)$ plane while keeping $\phi^1$ fixed at its minimum.}
\label{fig_pot-3D-three-field}
\end{figure}

For the chosen ACTivated model ${\cal M}_3^{\prime8/3}$, after fixing the overall volume in its perturbative LVS minimum, the effective two-dimensional scalar potential is plotted in figure \ref{fig_pot-3D-three-field}. Both the plots, namely the standard 3D plot and the contour plot, show that there is a ``diagonal" direction in the $(\phi^2, \phi^3)$ plane which serves as a flat inflationary track for driving the assisted fibre inflation. 

\noindent
While following the ``diagonal" track inflaton needs to move a relatively smaller distance as compared to the sum of two individual directions. This can be estimated by the following shifts in the individual canonical fields during the entire inflationary process. From (\ref{eq:model-M3-1-ACT}) we find that
\bea
& & \Delta\phi^1 \simeq 0.0926, \quad \Delta\phi^2 \simeq 3.763 \simeq \Delta\phi^3 \qquad \Longrightarrow \qquad \Delta\phi \simeq 5.32,
\eea
which shows that the overall volume gets only a negligible shift whilst  the other two fields assist during the inflation. Note that we have exploited the underlying exchange symmetry $2 \leftrightarrow 3$. This clearly follows from the relation $\Delta\phi^n \simeq \Delta\phi/\sqrt{n}$ for $n = 2$. Recall that, in the single- and two-field versions of this model, the total inflaton shift was $\Delta\phi \simeq 5.3$. 

Finally, let us recall that there can be different choices of the canonical fields leading to ${\cal G}_{ab} = \delta_{ab}$ as given by $\{\varphi^a\}$ in (\ref{eq:cononical-varphi0}) and $\{\psi^a\}$ in (\ref{eq:cononical-varphi1}). These lead to the following shifts in the individual inflaton directions
\bea
& & \Delta\varphi^1 \simeq 4.39907, \quad \Delta\varphi^2 \simeq 2.11929, \quad \Delta\varphi^3 \simeq 2.11929, \qquad \Longrightarrow \, \, \Delta\varphi \simeq 5.32, 
\eea
and
\bea
& & \Delta\psi^1 \simeq 0.0927, \quad \Delta\psi^2 \simeq 4.61, \quad \Delta\psi^3 \simeq 2.66, \qquad \Longrightarrow \, \, \Delta\psi \simeq 5.32, 
\eea
which clearly show that assisted two-field inflation needs smaller field excursion in the individual directions to achieve the same ``effective" inflation shift $\Delta\phi = \Delta\psi=\Delta\varphi$ needed for consistently producing the cosmological observables. Note that the canonical basis $\{\varphi^a\}$ as defined in (\ref{eq:cononical-varphi0}) does not involve the overall volume modulus and therefore the three moduli are significantly shifted from the minimum during inflation.

\subsection{Stability of assisted fibre inflation models}

\begin{figure}[H]
\centering
\includegraphics[width=16.5cm]{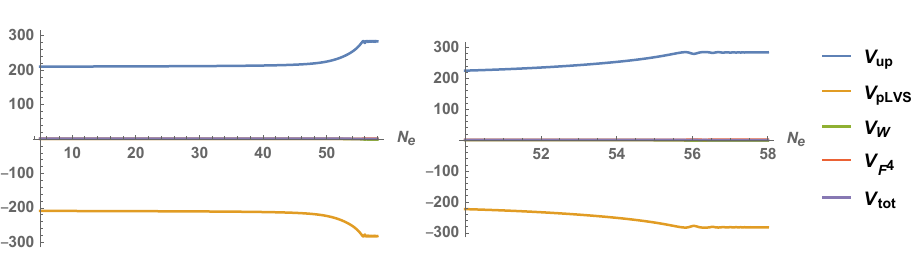}
\caption{Evolution of scalar potential ($V\cdot 10^{10}$) contributions showing the dominance of $V_{\rm up}$ and $V_{\rm pLVS}$}
\label{fig_V-scales1-three-field}
\end{figure}

Let us discuss the stability of the numerical benchmark model ${\cal M}_3^{\prime8/3}$ given in (\ref{eq:model-M3-3-ACT}) by considering the various contributions to the scalar potential. Similar estimates should hold for the other models as well. It turns out that at the LVS minimum, the individual scalar potential contributions are,
\bea
\label{eq:various-V-vevs1}
& & \hskip-1cm \langle V_{\rm up} \rangle = 2.83449 \cdot 10^{-8}, \quad \langle V_{\alpha^\prime} \rangle = 6.61081 \cdot 10^{-9}, \quad \langle V_{\rm logloop} \rangle = -3.48662\cdot 10^{-8},\\
& & \hskip-1cm \langle V_{g_s}^W \rangle = -1.82453 \cdot 10^{-10}, \quad \langle V_{{\rm F}^4} \rangle = 9.28818\cdot 10^{-11}. \nonumber
\eea
This apparently shows that there is no clean hierarchy among the various individual contributions to the scalar potential, in the sense of their origin from $\alpha^\prime$-series or string-loop $g_s$ series. However, we note that the hierarchy is maintained for the purpose of the so-called iterative moduli stabilization. For example, the pieces in the first line of (\ref{eq:various-V-vevs1} correspond to the stabilization of the overall volume modulus using the leading-order effects in perturbative LVS. From the figure \ref{fig_V-scales1-three-field}, we see that $V_{\rm pLVS} = V_{\alpha^\prime} + V_{\rm logloop}$ piece is negative during the inflationary evolution of the volume modulus, and an approximately equal positive contribution from the uplifting piece compensates for this to give a nearly flat de Sitter minimum. 

\noindent
%From (\ref{eq:various-V-vevs1}) we also see that the BBHL's $\alpha^\prime$-corrections, which are tree-level effects in string-loop series, are slightly smaller than the log-loop effects, which are $g_s$ corrections at one-loop level. Also, the winding-type one-loop corrections are smaller than the BBHL as well as log-loop effects. However, given the nature of the model, which synthesizes several corrections from different perturbative series expansions for stabilizing the same set of moduli and building a large field inflationary model, the presence of a significant hierarchy among various individual scalar potential contributions arising from different series is anyway not likely to be anticipated/present.
%
Equation (\ref{eq:various-V-vevs1}) reveals two hierarchies: (i) BBHL's tree-level $\alpha'$-corrections are subdominant to the one-loop $g_s$ corrections (log-loop effects), and (ii) Winding-type one-loop corrections are further suppressed relative to both BBHL and log-loop terms. Crucially, however, our model synthesizes distinct perturbative expansions, i.e., $\alpha'$, $g_s$, and winding ones, to stabilize moduli and achieve large-field inflation. This multi-source approach inherently disfavors large hierarchies between individual potential contributions, a consistency we observe numerically.

\begin{figure}[H]
\centering
\includegraphics[width=16.5cm]{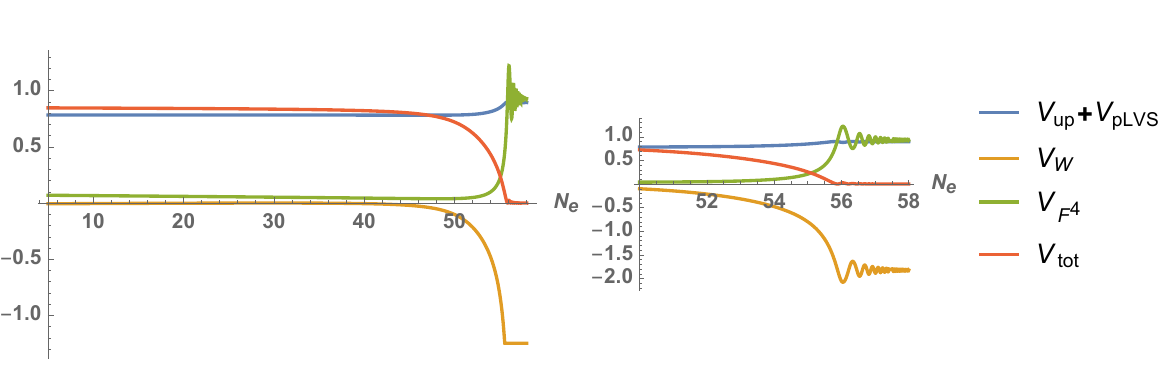}
\caption{Evolution of various pieces of the scalar potential ($V\cdot 10^{10}$) contributions  }
\label{fig_V-scalesc-2-three-field}
\end{figure}

However, one indeed finds that $\left|\langle V_{\rm pLVS} \rangle \right| > \left|\langle V_{g_s}^W \rangle\right| >  \left|\langle V_{{\rm F}^4} \rangle \right|$, and subsequently the winding-type string-loop effects and the higher derivative F$^4$ corrections are used to stabilize the remaining two moduli. These two sub-leading contributions are nearly the same and opposite as nicely seen from the evolution presented in the figure \ref{fig_V-scalesc-2-three-field}. In addition, the following eigenvalues of the mass-squared matrix indeed ensure a mass hierarchy among the various moduli 
\bea
& & m_a^2 = \{4.76\cdot 10^{-9}, \, \, 1.91\cdot 10^{-10}, \, \, 6.27 \cdot 10^{-11}\},
\eea
where the heaviest eigenstate corresponds to the overall volume modulus ${\cal V}$ while the other ones are some combinations of all the three moduli $\{{\cal V}, t^2, t^3\}$. A clean mass hierarchy between the overall volume mode, e.g. see (\ref{eq:model-M3-3-ACT}), and the two inflaton fields ensures that there is only a very small shift in ${\cal V}$ modulus during inflation and the three-field evolution is effectively a two-field assisted inflationary dynamics as we have discussed.

\begin{figure}[H]
\centering
\includegraphics[width=16.5cm]{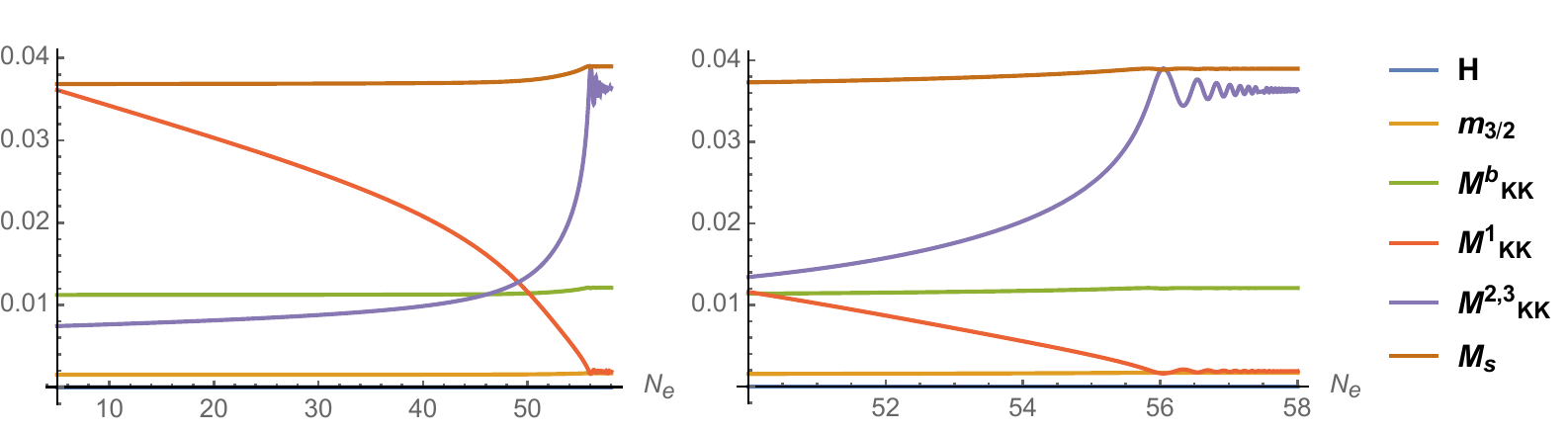}
\caption{Evolutions of various mass scales during inflationary dynamics}
\label{fig_scales-three-field}
\end{figure}

\noindent
Furthermore, the following mass hierarchy should be respected during evolution,
\bea
\label{eq:mass-hierarchy}
& & H < m_{3/2} < M_{\rm KK} < M_s < M_p,
\eea
where $H$ is the Hubble scale and the gravitino mass is denoted as $m_{3/2}$. The string mass $M_s$ and the various KK scales are defined as below 
\bea
\label{eq:KKscales}
& & M_s = \frac{M_p}{\sqrt{\alpha^\prime}}, \quad M_{\rm KK}^a = \frac{M_p}{R_a} = \frac{M_s}{\tilde{R}_a}, \quad m_{3/2} = e^{\frac{1}{2} {\cal K}} |W_0| = \frac{\sqrt{g_s}\, |W_0|}{\sqrt{2} \, {\cal V}},
\eea
In our conventions string length $\ell_s$ and the $\alpha^\prime$ parameter are connected as $\ell_s = 2\pi \sqrt{\alpha^\prime}$ which results in $M_s = \frac{g_s^{1/4} \sqrt\pi}{\sqrt{\cal V}} M_p$, e.g. see \cite{Conlon:2006gv}. Further $R_a = \tilde{R}_a \sqrt{\alpha^\prime}$ where $\tilde{R}_a$ is the size of the relevant length corresponding to a particular KK mode such that $\tilde{R}_a = (t^a)^{1/2}$ for two-cycle volumes, $\tilde{R}_a = (\tau_a)^{1/4}$ for four-cycle volumes and $\tilde{R}_L = {\cal V}^{1/6}$  corresponds to the bulk modulus ($t^b \simeq{\cal V}^{1/3}$ or $\tau_b \simeq {\cal V}^{2/3}$) which we denote as $M_{\rm KK}^b$. Usually these are the lightest KK modes. The evolution of various mass scales are plotted in figure \ref{fig_scales-three-field} where apart from the bulk KK mode, we consider $M_{\rm KK}^1$ corresponding to $t^1$ modulus and $M_{\rm KK}^{2,3}$ corresponding to $t^2$ and $t^3$ moduli. 

The evolution of various scales as presented in figure \ref{fig_scales-three-field} shows that mass-hierarchy (\ref{eq:mass-hierarchy}) is respected throughout the inflationary regime, i.e. till $\epsilon \leq 1$ corresponding to $N\simeq55.5$. However, we also observe that the heaviest KK scales $M_{\rm KK}^{2,3}$ become comparable to the string mass towards the minimum, after the end of inflation. Such an issue is likely to be present for smaller VEVs of the volume moduli, in particular for models with $\langle t^\alpha \rangle < 1$ or $\langle\tau_\alpha \rangle < 1$. Moreover, towards the horizon exit when the inflatons $t^2$ and $t^3$ are shifted significantly away from their minimum, the $t^1$ modulus tends towards smaller size as the overall volume ${\cal V} = 2\, t^1 t^2 t^3$ remains nearly fixed during the entire inflationary process. Subsequently, the corresponding KK scale, namely $M_{\rm KK}^1$, increases and becomes closer to the string mass scale $M_s$ towards the horizon exit $N^\ast = 5.5$. However, as long as the volume moduli do not enter in the regime where they take too smaller values, the supergravity approximations should remain (marginally) valid. It is also worth noting that the string mass scale being comparable to (one of the) KK mass may not be an immediate concern as argued in \cite{Dienes:2002ze}.

Finally, as mentioned earlier, the flux superpotential parameter $W_0$ needs to satisfy the constraint in Eq.~(\ref{eq:W_0-Nflux-tadpole}) which is intertwined with the $D3$ tadpole charge $Q_{D3}$ as $2\pi g_s |W_0|^2 < Q_{\rm tot} = 88$. For our models $W_0 \simeq 5 $ and $g_s \simeq 0.3$, and therefore this tadpole relation is respected.

%%%%%%%%%%%%%%%%%%%%%%%%%%%%%%%%%%%%%%%%%%%%%%%%%%%%%%%%%%%%%%%%%%%%%%%%%%%%%%%%%%%%%%%%%%
%%%%%%%%%%%%%%%%%%%%%%%%%%%%%%%%%%%%%%%%%%%%%%%%%%%%%%%%%%%%%%%%%%%%%%%%%%%%%%%%%%%%%%%%%%

\section{Summary and Conclusions}
\label{sec_conclusions}
In this article, we have proposed a multi-field fibre inflation scenario within type IIB superstring compactifications, specifically in the perturbative Large Volume Scenario (pLVS). We have discussed the possible challenges in single-field inflation models, such as trans-Planckian field displacements and tensions with recent cosmological data from ACT and DESI \cite{ACT:2025tim,ACT:2025fju,DESI:2024mwx, Frolovsky:2025iao}. 

We considered the type IIB superstring compactifications using an orientifold of a K3-fibred Calabi Yau threefold with toroidal-like volume ${\cal V} = 2\, t^1 t^2 t^3$. Following the two-step moduli stabilization strategy, the complex-structure moduli and the axio-dilaton are fixed in the supersymmetric minimum by using the superpotential induced by the $F_3/H_3$ fluxes. The second step used a combination of the so-called {\it log-loop} effects \cite{Antoniadis:2018hqy} and the BBHL corrections \cite{Becker:2002nn} leading to an exponentially large volume AdS minimum \cite{Antoniadis:2018hqy,Leontaris:2022rzj}. Subsequently, using an appropriate de Sitter uplifting mechanism along with a set of additional contributions to the scalar potential arising from the KK- and winding-type string-loop effects \cite{Berg:2005ja,Berg:2007wt,vonGersdorff:2005bf,Cicoli:2007xp} and the higher-derivative F$^4$-corrections \cite{Ciupke:2015msa} can facilitate the slow-roll inflationary dynamics. The present work can be summarized along the following steps:

\begin{itemize}
\item 
{\bf Step 1:} For achieving an effective single-field inflationary model, we fixed a combination of the volume moduli, namely $t^2 = t^3$ by turning on appropriate fluxes. Subsequently, the model simplifies to a two-field case with ${\cal V} = 2 \,t^1 (t^2)^2$, and after fixing the overall volume ${\cal V}$ in perturbative LVS, one gets an effective single-field fibre inflation \cite{Bera:2024ihl}. This model has similar cosmological predictions as the standard fibre inflation \cite{Cicoli:2008gp, Cicoli:2016chb, Cicoli:2017shd, Cicoli:2024bxw} which faces some strong challenge in the form of the inflaton field range bounds arising from the K\"ahler cone conditions \cite{Cicoli:2018tcq}. However, unlike these models, the current setup does not typically face this challenge due to the absence of the rigid 4-cycle as argued in \cite{Bera:2024ihl}.

In this work, we started with revisiting the single-field analysis presented in \cite{Bera:2024ihl} by numerically tracking the evolution of the inflaton field, and subsequently producing the cosmological observables. In addition, we presented additional numerical benchmark models consistent with recent ACT observations. We  found that one typically needed field excursions $\Delta\phi \simeq 6$ M$_p$ in order to consistently produce the cosmological observables. We also observed that the analogous ACTivated models needed slightly smaller field excursions.

\item 
{\bf Step 2:} There are several types of corrections arising from different series expansions which are being used for moduli stabilization and driving an effective single-field inflation. Therefore, in order to check the robustness of the assumption made in the single-field approximations, we performed a detailed numerical analysis for a two-field model, with $\{{\cal V}, t^2\}$ moduli, and  tracked their evolution with respect to the number of e-folds. We discussed the viability of standard and ACTivated models by consistently reproducing the cosmological observables. 

\item 
{\bf Step 3:} From the single-field and two-field analysis, we observed that one needed $\Delta\phi \simeq 6$ M$_p$ for a successful inflationary dynamics, while the overall volume remained seated at its perturbative LVS minimum. This means that even the two-field analysis in {\bf Step 2} is effectively not a multi-field inflation. 

For having a genuine multi-field inflation on top of having the overall volume being fixed at the leading order in perturbative LVS, first we switched-off the fluxes turned-on in {\bf Step 1} so that to relax the constraint $t^2 = t^3$ and analyze the generic three-field potential  (\ref{eq:Vfinal-simp3}). After performing a detailed numerical analysis and exploiting the underlying exchange symmetry following from the Calabi Yau threefold geometry, we find an assisted behavior of the two individual inflaton candidates. It turns out that they ``assist" during the inflation and the need of traveling large distance of order $6$ M$_p$ (as in the single-field case) is reduced such that the individual fields travel only around $4$ M$_p$ for generating the cosmological observables. We have subsequently shown that the canonical inflaton field range for individual inflatons scale as $\Delta\phi_n=\Delta\phi/\sqrt{n}$, where $n$ is the number of assisting moduli fields. 

\end{itemize}

\noindent
For our numerical benchmark models we used multiple uplifting mechanisms with an uplifting term: $V_{\rm up} \propto {\cal V}^{-p}$ where $p = 4/3$ for anti-D3 uplifting, $p = 2$ for D-term uplifting and $p = 8/3$ for T-brane uplifting. We showed that irrespective of the uplifting schemes, one can produce the single- and multi-field fibre inflation models with $g_s\simeq 0.3$, $W_0\simeq 5$, and ${\cal V} \simeq 10^3$ along with correct cosmological predictions. In typical fibre inflation models based on Swiss-Cheese Calabi Yau, the K\"ahler cone conditions require that ${\cal V} \gtrsim 10^4$ to have a sufficiently large inflaton field range \cite{Cicoli:2017axo}, and subsequently one needs a larger value of the parameter $W_0$ in order to match with the correct scalar perturbation amplitude. However, we also note that increasing $W_0$ too much may enhance the gravitino mass such that it becomes heavier than the lightest bulk KK mode. Attempts to realize fibre inflation in Swiss-Cheese-based models with $W_0\simeq 5$ and ${\cal V} \simeq 10^3$ may result in pushing one or several K\"ahler moduli close to the boundary of the K\"ahler cone, which has its own new challenges regarding EFT control \cite{Cicoli:2018tcq}. We also note that the VEV of $W_0$ is intertwined with the total D3 tadpole charge $Q_{D3}$ allowed by the chosen Calabi-Yau orientifold \cite{Denef:2004ze,Bera:2024ihl,Cicoli:2024bxw}. The K\"ahler cone constraints are typically absent or milder in the perturbative LVS model \cite{Bera:2024ihl} and the problems with large $W_0 \simeq {\cal O}(10)$ could be resolved by considering the appropriate Calabi Yau threefold with larger $h^{1,1}$ \cite{Shukla:2022dhz, Crino:2022zjk}. So it is very well possible that these challenges could be just because of the simple choice of Calabi Yau threefolds. However, multi-field models will be natural and inevitable for setups with large $h^{1,1}$, since finding an effective single-field potential in the presence of a huge variety of scalar potential contributions and maintaining the needed iterative hierarchy would be equally challenging.

In our systematic analysis, we have shown that the minimal assisted fibre inflation proposal can be embedded with a range of possible uplifting schemes in perturbative LVS. We performed a detailed numerical analysis and provided several numerical benchmarks models consistent with Planck and ACT data, including the spectral tilt $n_s$, the running of spectral tilt $\alpha_s$ and the tensor-to-scalar ratio $r$. We also found that typically the tensor-to-scalar ratio $r$ reduces for higher values of $n_s$ in ACTivated models while the values of $\alpha_s$ can be found to be positive or negative both, with a slight change of model dependent parameter. Thus, as anticipated, multi-field inflation model is more flexible in accommodating the recent cosmological predictions.
 
Finally, we discussed the viability and robustness of the assisted fibre inflation where we presented a detailed analysis of the evolution of various scalar potenial pieces during the entire inflationary period. We also discussed the issue of hierarchy among the various mass scales as in (\ref{eq:mass-hierarchy}) where we noted that the heaviest KK modes approach the string scale towards the end of inflation. This suggests that the model maybe near the edge of what is allowed in a consistent quantum gravity theory.  Although this does not automatically imply that the EFT breaks down \cite{Dienes:2002ze}, a careful analysis of the backreaction, field excursions, and UV completions is necessary to determine the robustness of this (or any other large field) inflationary model. Alternatively, it may require more fine tuning of our parameter space to achieve appropriate VEVs for the volume moduli such that all the KK scales remain significantly below the string mass scale without lowering (one of) them below the gravitino mass. Furthermore, it would be interesting to investigate if assisted fibre inflation could be realized in the presence of del-Pezzo divisors using K3-fibred Calabi Yau threefolds with Swiss-Cheese structure. Also it would be interesting to extend the assisted fibre inflation proposal to check the relation $\Delta\phi_n=\Delta\phi/\sqrt{n}$ for models with more than two inflatons. We hope to present these (and related) issues in a future work \cite{Leontaris:2025xxx}.

%%%%%%%%%%%%%%%%%%%%%%%%%%%%%%%%%%%%%%%%%%%%%%%%%%%%%%%%%%%%%%%%%%%%%%%%%%%%%%%%%%%%%%%%%%%
%%%%%%%%%%%%%%%%%%%%%%%%%%%%%%%%%%%%%%%%%%%%%%%%%%%%%%%%%%%%%%%%%%%%%%%%%%%%%%%%%%%%%%%%%%%

\section*{Acknowledgments}
We are grateful to Ignatios Antoniadis and Roberto Valandro for useful discussions. We also thank Manas Kumar Sinha for the initial collaboration on the project last year during his visit for the master thesis work at Bose Institute. PS would like to thank the {\it Department of Science and Technology (DST), India} for the kind support. The authors would also like to thank the organizers of the conference ``\href{https://www.swamplandia.com/home}{Lotus and Swamplandia 2025}" for the kind hospitality and support.

%%%%%%%%%%%%%%%%%%%%%%%%%%%%%%%%%%%%%%%%%%%%%%%%%%%%%%%%%%%%%%%%%%%%%%%%%%%%%%%%%%%%%%%%%%%
%%%%%%%%%%%%%%%%%%%%%%%%%%%%%%%%%%%%%%%%%%%%%%%%%%%%%%%%%%%%%%%%%%%%%%%%%%%%%%%%%%%%%%%%%%%

%\newpage
\appendix

%%%%%%%%%%%%%%%%%%%%%%%%%%%%%%%%%%%%%%%%%%%%%%%%%%%%%%%%%%%%%%%%%%%%%%%%%%%%%%%%%%%%%%%%%%%
%%%%%%%%%%%%%%%%%%%%%%%%%%%%%%%%%%%%%%%%%%%%%%%%%%%%%%%%%%%%%%%%%%%%%%%%%%%%%%%%%%%%%%%%%%%

\section{Deriving the scalar potential for perturbative LVS}
\label{sec_appendix}
The K\"ahler derivatives can be subsequently found to take the following form,
\bea
\label{eq:derK}
& & K_S = \frac{{\rm i}}{2\, s\, {\cal Y}} \left({\cal V} + 2 \,\hat\xi \right) = - K_{\ov S}, \quad K_{T_\alpha} = - \frac{{\rm i} \, t^\alpha}{2 \, {\cal Y}} \left(1+ \frac{\partial {\cal Y}_1}{\partial {\cal V}} \right) = - K_{\ov{T}_\alpha}
\eea
Further, it turns out that the K\"ahler metric and its inverse generically admit the following explicit forms,
\bea
\label{eq:simpinvKij-1}
& & \hskip-1cm K_{S \ov{S}} = {\cal P}_1, \qquad K_{T_\alpha\, \ov{S}} = t^\alpha\, {\cal P}_2 = K_{S\,\ov{T}_\alpha}, \qquad  K_{T_\alpha \, \ov{T}_\beta} = (t^\alpha \, t^\beta)\, {\cal P}_3 \, - k^{\alpha\beta}\, {\cal P}_4~, \\
& & \hskip-1cm  K^{S \ov{S}} = \tilde{\cal P}_1, \qquad K^{T_\alpha\, \ov{S}} = k_\alpha\, \tilde{\cal P}_2 = K^{S\,\ov{T}_\alpha}, \qquad  K^{T_\alpha \, \ov{T}_\beta} = (k_\alpha \, k_\beta)\, \tilde{\cal P}_3 \, - k_{\alpha\beta}\, \tilde{\cal P}_4~, \nonumber
\eea
where the two sets of functions ${\cal P}_i$ and $\tilde{\cal P}_i$'s are
\bea
\label{eq:Pis}
& & {\cal P}_1 = \frac{1}{8\,s^2\, {\cal Y}^2}\, \left[{\cal V} ({\cal Y} + {\cal V}) - 4 \hat\xi ({\cal Y} - {\cal V}) + 4 \hat\xi^2\right], \\
& & {\cal P}_2 = -\frac{1}{8\, s\, {\cal Y}^2} \left[\frac{3}{2}\, \hat\xi - s^{-\frac{1}{2}}\, (\sigma+\eta \ln {\cal V}) + \, s^{-\frac{1}{2}} \, ({\cal V} + 2\hat\xi) \frac{\eta}{{\cal V}} \right], \nonumber\\
& & {\cal P}_3 = \frac{1}{8\, {\cal Y}^2} \left[\left(1 + s^{-\frac{1}{2}} \, \frac{\eta}{{\cal V}}\right)^2 +  {\cal Y} \,  s^{-\frac{1}{2}} \, \frac{\eta}{{\cal V}^2}\right], \nonumber\\
& & {\cal P}_4 = \frac{1}{4\, {\cal Y}} \left[1 + s^{-\frac{1}{2}} \, \frac{\eta}{{\cal V}} \right]~, \nonumber
\eea
and
\bea
\label{eq:tildePis}
& & \tilde{\cal P}_1 = \frac{{\cal P}_4-6 {\cal P}_3 {\cal V}}{{\cal P}_1 {\cal P}_4 + 6 {\cal P}_2^2 {\cal V}-6 {\cal P}_1 {\cal P}_3 {\cal V}}\,, \qquad \tilde{\cal P}_2 = \frac{{\cal P}_2}{{\cal P}_1 {\cal P}_4 + 6 {\cal P}_2^2 {\cal V}-6 {\cal P}_1 {\cal P}_3 {\cal V}}\,, \\
& & \tilde{\cal P}_3 = \frac{{\cal P}_2^2-{\cal P}_1 {\cal P}_3}{{\cal P}_4\left({\cal P}_1 {\cal P}_4 + 6 {\cal P}_2^2 {\cal V}-6 {\cal P}_1 {\cal P}_3 {\cal V}\right)}\,, \qquad \tilde{\cal P}_4 = ({\cal P}_4)^{-1}. \nonumber
\eea
This subsequently leads to the following form of the scalar potential,
\bea
\label{eq:masterV}
&&\hskip-1cm  V_{\alpha^\prime + {\rm log} \, g_s} = \frac{\kappa}{{\cal Y}^2} \biggl[\frac{3\, {\cal V} }{2\, {\cal Y}^2} \left(1 + \frac{\partial {\cal Y}_1}{\partial{\cal V}}\right)^2 \left(6 {\cal V} \tilde{\cal P}_3 - \tilde{\cal P}_4 \right) - 3 \biggr]\, |W_0|^2 = V_{\alpha^\prime +{\rm log} \, g_s}^{(1)} + \cdots~,
\eea
where we have the following pieces at leading order which we call as $V_{\rm pLVS}$,
\bea
\label{eq:pheno-potV2}
& & \hskip-1cm V_{\rm pLVS} \equiv V_{\alpha^\prime +{\rm log} \, g_s}^{(1)} \simeq \frac{3\, \kappa\, \hat\xi}{4\, {\cal V}^3}\, |W_0|^2 + \frac{3 \, \kappa\, (\hat\eta\ln{\cal V} - 4\hat\eta + \hat\sigma)}{2{\cal V}^3}\,|W_0|^2, \qquad \kappa = \frac{g_s}{2}\,e^{K_{cs}}.
\eea

%%%%%%%%%%%%%%%%%%%%%%%%%%%%%%%%%%%%%%%%%%%%%%%%%%%%%%%%%%%%%%%%%%%%%%%%%%%%%%%%%%%%%%%%%%%
%%%%%%%%%%%%%%%%%%%%%%%%%%%%%%%%%%%%%%%%%%%%%%%%%%%%%%%%%%%%%%%%%%%%%%%%%%%%%%%%%%%%%%%%%%%

%\newpage
\bibliographystyle{JHEP}
\bibliography{reference}

%\newpage
%\bibliographystyle{utphys}
%\bibliography{reference}

\end{document}